\documentclass[11pt,letterpaper]{article}

\usepackage{latexsym,multirow}
\usepackage{amssymb,amsmath, bm}
\usepackage{graphicx}
\usepackage[color,all,import,arrow]{xy}
\usepackage{enumerate}

\usepackage{natbib}
\usepackage[pdftex, bookmarksopen=true, bookmarksnumbered=true,
pdfstartview=FitH, breaklinks=true, urlbordercolor={0 1 0}, citebordercolor={0 0 1}]{hyperref}
\usepackage{colortbl}
\usepackage{subfigure}

\usepackage{tikz}
\usetikzlibrary{er,positioning,shadows,arrows,shapes,snakes,automata,backgrounds,petri,fit}

\usepackage{dcolumn}
\newcolumntype{.}{D{.}{.}{-1}}
\newcolumntype{d}[1]{D{.}{.}{#1}}
\usepackage{theorem}
\theoremstyle{plain}
\theoremheaderfont{\scshape}
\newtheorem{assumption}{Assumption}

\newtheorem{proposition}{Proposition}
\newtheorem{theorem}{Theorem}

\newcommand{\ind}{\mbox{$\perp\!\!\!\perp$}}

\usepackage{rotating}

\usepackage[compact]{titlesec}

\allowdisplaybreaks

\newcommand\spacingset[1]{\renewcommand{\baselinestretch}%
{#1}\small\normalsize}

\newcommand{\blind}{0}

\newcommand*{\QEDB}{\hfill\ensuremath{\square}}

\newcommand{\cov}{\textnormal{cov}}
\newcommand{\mT}{\mathcal{T}}
\newcommand{\ACE}{\textsc{ACE}}

\newcommand{\ACER}{\textsc{ACER}}

\def\ind{\begin{picture}(9,8)
         \put(0,0){\line(1,0){9}}
         \put(3,0){\line(0,1){8}}
         \put(6,0){\line(0,1){8}}
         \end{picture}
        }

\def\d{{ \mathrm{d}  }}
 \def\Pr{{ \textnormal{Pr}  }}

\begin{document} 

\newcommand{\tit}{An instrumental variable method for  point processes: generalised Wald estimation based on deconvolution}
%
%
\spacingset{1.25}

\if0\blind

{\title{\bf\tit}

\author{Zhichao Jiang\thanks{School of Mathematics,
      Sun Yat-sen University, Guangzhou,  Guangdong 510275, China.  Email:
      \href{mailto:jiangzhch7@mail.sysu.edu.cn}{jiangzhch7@mail.sysu.edu.cn} } \textsuperscript{\textsection}
    \and Shizhe Chen\thanks{Department of Statistics, University of California, Davis, California 95616, U.S.A. Email: \href{mailto:szdchen@ucdavis.edu}{szdchen@ucdavis.edu}} \textsuperscript{\textsection} 
  \and Peng Ding\thanks{Department of Statistics, University of California, Berkeley, California 94720, U.S.A. Email: \href{mailto:pengdingpku@berkeley.edu}{pengdingpku@berkeley.edu}} 
}

\date{
\today
}

\maketitle

}\fi

\if1\blind
\title{\bf \tit}

\maketitle
\fi

\pdfbookmark[1]{Title Page}{Title Page}

\thispagestyle{empty}
\setcounter{page}{0}

\begingroup\renewcommand\thefootnote{\textsection}
\footnotetext{Equal contribution}
\endgroup
         
\begin{abstract}
Point processes are probabilistic tools for modeling event data. While there exists a fast-growing literature studying the relationships between point processes, it remains unexplored how such relationships connect to causal effects. In the presence of unmeasured confounders, parameters from point process models do not necessarily have causal interpretations. We propose an instrumental variable method for causal inference with point process treatment and outcome. We define causal quantities based on potential outcomes and establish nonparametric identification results with a binary instrumental variable.
We extend the traditional Wald estimation to deal with point process treatment and outcome, showing that it should be performed after a Fourier transform of the intention-to-treat effects on the treatment and outcome and thus takes the form of deconvolution. We term this as the generalised Wald estimation and propose an estimation strategy based on well-established deconvolution methods.

\end{abstract}


\clearpage
\spacingset{1.5}

\section{Introduction}

Point processes have long been used for modeling event data. 
The past decade has witnessed a surge of interest in point process models in many fields including neuroscience, finance, and social sciences. In this paper, we consider the analysis of neural data as a concrete motivation.
Modern technologies allow neuroscientists to simultaneously record neural spike trains, i.e., arrays of timestamps when neurons fire, across the brain. With these data, one can hope to peek into the mechanisms of neural computing. The nature of these scientific questions is the inference of causal effects.

Current technologies, however, present a major challenge for causal inference with neural data. Except for experiments on very simple animals, even state-of-the-art technologies can record only a very small fraction of neurons in chosen regions in the nerve systems, leaving the vast majority unobserved.
The unmeasured neural activities inevitably lead to the issue of unmeasured confounding; that is, unmeasured activities might be the common causes of observed neural activities. As a result, any relationship inferred based on the partially observed system might not reflect the true causal relationship, but rather a spurious association.

Fortunately, advances in optogenetics create new opportunities to address unmeasured confounding effects.
Neuroscientists are able to instigate neural activities in a living brain via optical stimulation, which alters the activity of any chosen neuron with high spatial and temporal precision  \citep{mardinly2018precise, carrillo2019controlling}.
From a causal inference perspective, such interventions can serve as {\it instrumental variables}  for inferring the causal relationship between neurons, as they affect the outcome neuron only through the treatment neuron while introducing exogenous variation in the treatment neuron.

Instrumental variable methods are powerful tools for inferring causal effects in the presence of unmeasured confounding between the treatment and the outcome. 
In a seminal paper, \citet{angrist1996identification} clarify the role of a binary instrumental variable in identifying the causal effect of a binary treatment for an unmeasured subgroup, known as the complier average causal effect. They propose two crucial identification assumptions, monotonicity and exclusion restriction. Under these assumptions, they show that the complier average causal effect is identified by the Wald estimator \citep{wald1940fitting,ridder2007data} that equals the ratio of the differences in means of the outcome and the treatment when the instrumental variable changes from $0$ to $1$.

With an instrumental variable, most existing work considers non-dynamic settings and the instrumental variable methods in survival analysis mainly focus on a scalar treatment and a non-recurrent outcome \citep[e.g.,][]{li2015instrumental, martinussen2017instrumental,richardson2017nonparametric,jiang2018twostep}. To the best of our knowledge, there is no formal instrumental variable framework for point processes that addresses nonparametric identification.

We propose an instrumental variable method for causal inference when both the treatment and the outcome take the form of point processes. We define several causal quantities for the effect of the treatment on the outcome over time. Using a binary instrumental variable, we establish the nonparametric identification of causal effects allowing for the unmeasured treatment-outcome confounding. The identification assumptions hold as long as the impact of the unmeasured confounders on the outcome is additive. Our identification result implies that the causal effects can be obtained by solving a convolution equation.
This extends the Wald estimation in traditional instrumental variable method to take the form of deconvolution, leading to the proposed generalised Wald estimation.
We also examine several commonly-used models under our framework, studying the identification of the causal effects and the causal interpretation of the model parameters with a binary instrumental variable. When the unmeasured confounders are additive on the outcome, the causal effects are identifiable without any distributional assumptions on the confounders based on the proposed generalised Wald estimation.
Our finding justifies the identifiability of many commonly-used models such as the Hawkes process, broadening their applicability with fewer assumptions.

We use the following notation. Let $A\, \ind \, B \mid C$ denote the conditional independence of $A$ and $B$ given $C$.
Let  $\mathbb{R}$ denote the set of real numbers and $\mathcal{B}(\mathbb{R})$ denote the Borel $\sigma$-algebra of the whole real line. Let  $L^1(\mathbb{R})$ denote the set of functions $f(x)$ such that  $\int_{-\infty}^{\infty} |f(x)|\d x < \infty $.  Unless specified otherwise, we assume all functions used in this paper belong to $L^{1}(\mathbb{R})$.
 Let $\Psi$ denote the Fourier transform, i.e., for any $f(x) \in L^1(\mathbb{R})$ and $\nu \in \mathbb{R}$, define 
 \begin{eqnarray*}
(\Psi f)(\nu) \ = \ \int_{-\infty}^{\infty} f(x) e^{- \text{i} 2\pi \nu x}\d x,
\end{eqnarray*}
where $\text{i} = \sqrt{-1}$. Let $\Psi^{-1}$ denote the inverse Fourier transform.

\section{An instrumental variable framework for point processes}\label{sec:setup}

\subsection{A brief review of the binary instrumental variable model}

We begin by reviewing the binary instrumental variable model in the context of noncompliance \citep{angrist1996identification}.
For unit $i$, let $Z_i$ be the binary treatment assigned, $N_i$ the actual treatment received, and $Y_i$ the outcome of interest. 
Let $N_{iz}$ be the potential value of the treatment receipt if the assigned treatment condition is $z$, $Y_{izn}$ the potential value of the outcome if the assigned treatment is $z$ and the actually received treatment is $n$. The joint values of $N_{i1}$ and $N_{i0}$ define the unmeasured compliance type
$U_i = ( N_{i1}, N_{i0} ) $. Units with $( N_{i1}=1, N_{i0}=0)$ are compliers who take the treatment assigned, units with $( N_{i1}=1, N_{i0}=1)$ are always-takers who always take treatment $1$, units with $( N_{i1}=0, N_{i0}=0)$ are never-takers who always take treatment $0$, and units with $(N_{i1}=0,N_{i0}=1)$ are defiers who take the treatment opposite to the assigned.

\citet{angrist1996identification} invoke three assumptions: (1) \textit{exclusion restriction} that the treatment assigned affects the outcome only through the treatment received, i.e., $Y_{izn}=Y_{iz^\prime n}$ for all $z,z^\prime,n$; (2) \textit{randomization} that $Z_i$ is independent of $N_{iz}$ and $Y_{izn}$ for $z, n=0,1$; (3) \textit{monotonicity} that the assigned treatment does not negatively affect the treatment receipt for all units, i.e., $N_{i1} \geq N_{i0}$. Exclusion restriction simplifies $Y_{izn}$ to $Y_{in}$. Randomization rules out the confounding between the treatment assignment and the treatment receipt as well as the confounding between the treatment assignment and the outcome. Monotonicity rules out defiers. Under these assumptions, \citet{angrist1996identification} introduce the complier average causal effect as the average effect of the treatment receipt on the outcome for compliers,
$\textsc{CACE}\ =\ E  (Y_{i1}-Y_{i0} \mid N_{i1}=1,N_{i0}=0),$
and show that it is identified by
\begin{eqnarray}\label{eq::identify-cace}
\textsc{CACE}\ =\ \frac{E (Y_i \mid Z_i=1)-E (Y_i \mid Z_i=0)}{E (N_i \mid Z_i=1)-E (N_i \mid Z_i=0)}.
\end{eqnarray}
In this model, the treatment assignment $Z_i$ is the instrumental variable.
The expression in~\eqref{eq::identify-cace} suggests the Wald estimator \citep{wald1940fitting} for the CACE, i.e., the ratio of the differences in means of the outcome and the treatment receipt when the treatment assigned changes from $0$ to $1$. 

\citet{angrist1996identification} identify only the treatment effect in the complier subpopulation. For extrapolation to the whole population, we can invoke the {\it homogeneity} assumption \citep[cf.,][]{heckman1996identification,chen2009identifiability} that the treatment effect is the same across compliance groups:
\begin{eqnarray}
\label{eqn::homogeneity}
E  ( Y_{i1}-Y_{i0} \mid N_{i1},N_{i0} )\ = \ E  ( Y_{i1}-Y_{i0} ).
\end{eqnarray}
Under the assumption in \eqref{eqn::homogeneity}, the treatment effect in the whole population equals the CACE.

\subsection{Notation and basic assumptions with a point process treatment}
\begin{figure}[ht!]
	\centering
	\begin{tikzpicture}[node distance=1.3cm,>=stealth',bend angle=45,auto]
	\tikzstyle{iv}=[draw=black!20,fill=black!20,minimum size=8mm]
	\tikzstyle{latent}=[circle,draw=black,fill=black!20,minimum size=10mm,dashed]
	\tikzstyle{obs}=[circle,draw=black!20,	fill=black!20,minimum size=10mm]

	\begin{scope}
	\node [obs] (N){$ N_i$};
	\node [iv] (Z) [left = 1 cm of N] {$Z_i$}
	edge [->, thick, color=black, thick] (N);
	\node [latent] (U) [ below right = 1.2 cm and 0.5 cm of N] {$U_i$}
	edge [->, color=black, thick,dashed]		(N);
	\node [obs] [ right = 1.5 cm of N] (Y){$ Y_i$}
	edge [	<-,	 color=black, thick]		(N)
	edge [<-, color=black, thick, dashed]		(U);

	\node [fit={(N) (Z) (U) (Y)}] (DAG) {};
	\node[above left = -0.5 cm and 0 cm of DAG](DAGlab)
	{(d)};

	\node[inner sep=0pt] [above = 1 cm of DAG] (Trial){\includegraphics[width=.45\textwidth]{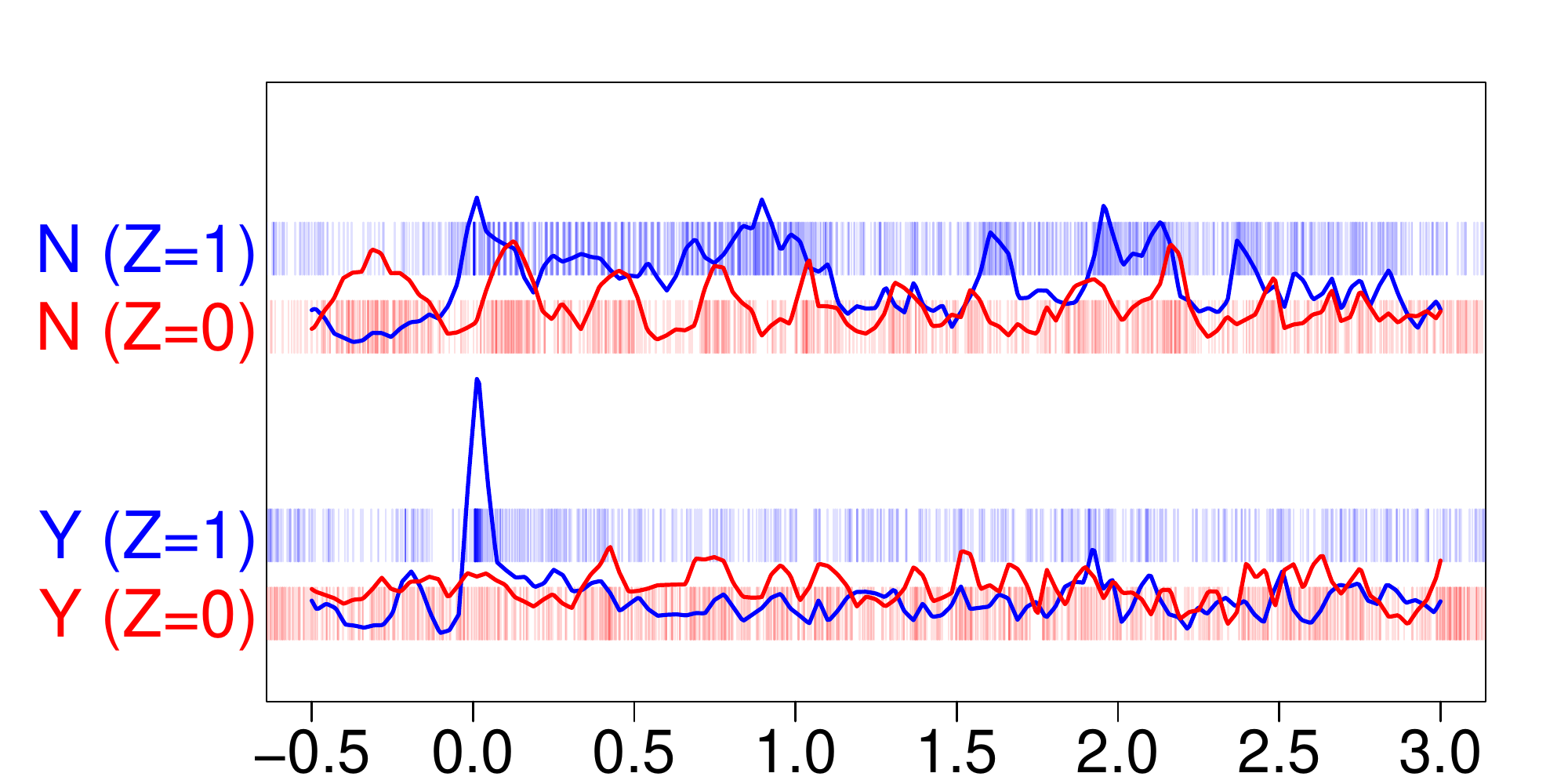}};
		\node[above = 3.8 cm of DAGlab](Tlab) {(c)};
		\node[below = 0 cm of Trial] (Taxis) {Time (s)};
	\node[inner sep=0pt] [left = 1 cm of Trial] (Ns){\includegraphics[width=.45\textwidth]{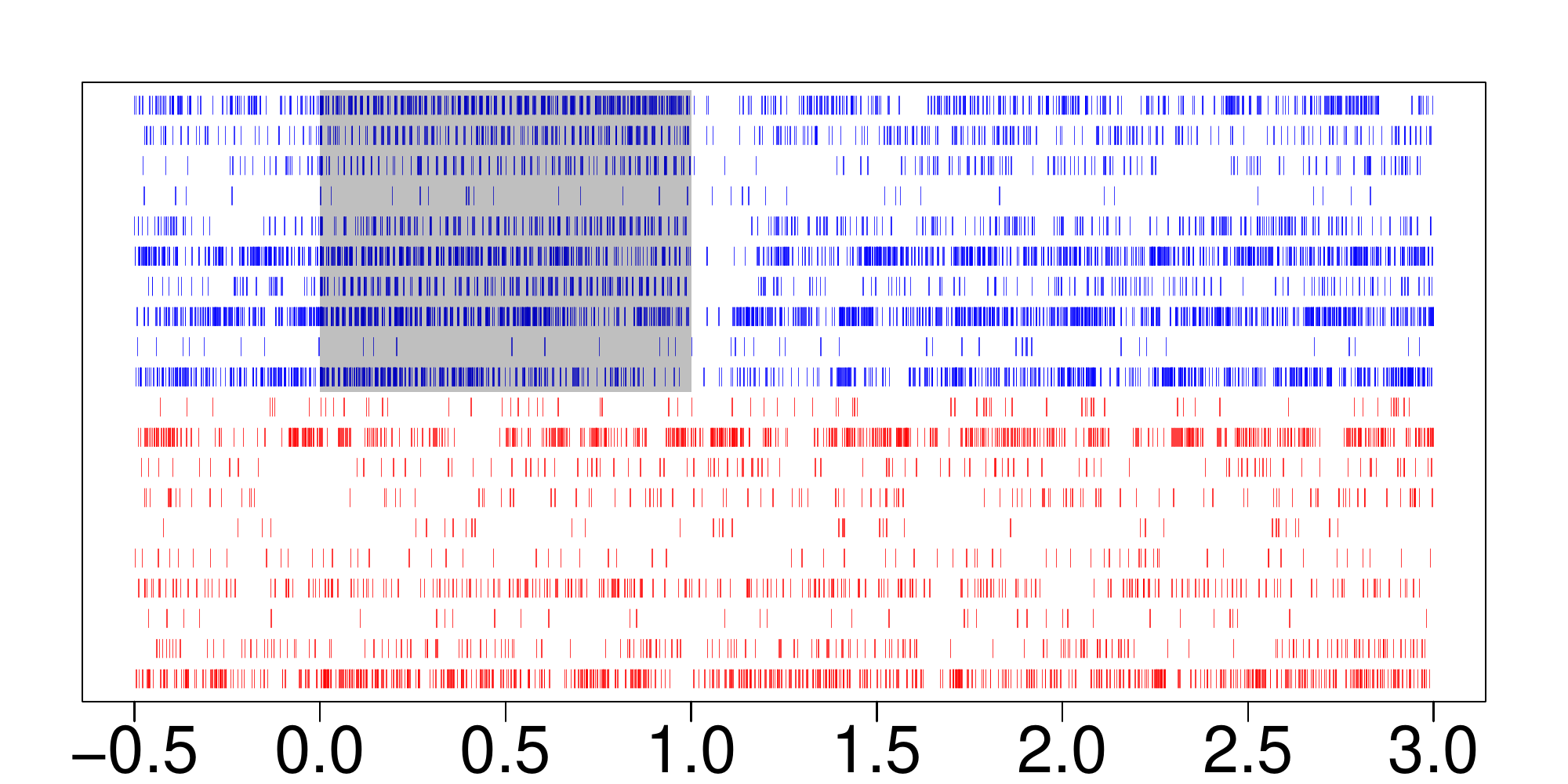}};
			\node[left =  8.5 cm of Tlab]( ) 	{(a)};
			\node[below = 0 cm of Ns] (Naxis) {Time (s)};
			\node[above left = -1.2 cm and -0.3 cm of Ns, rotate=90] () {Trials};
	\node[inner sep=0pt] [below = 0.5 cm of Ns] (Ys){\includegraphics[width=.45\textwidth]{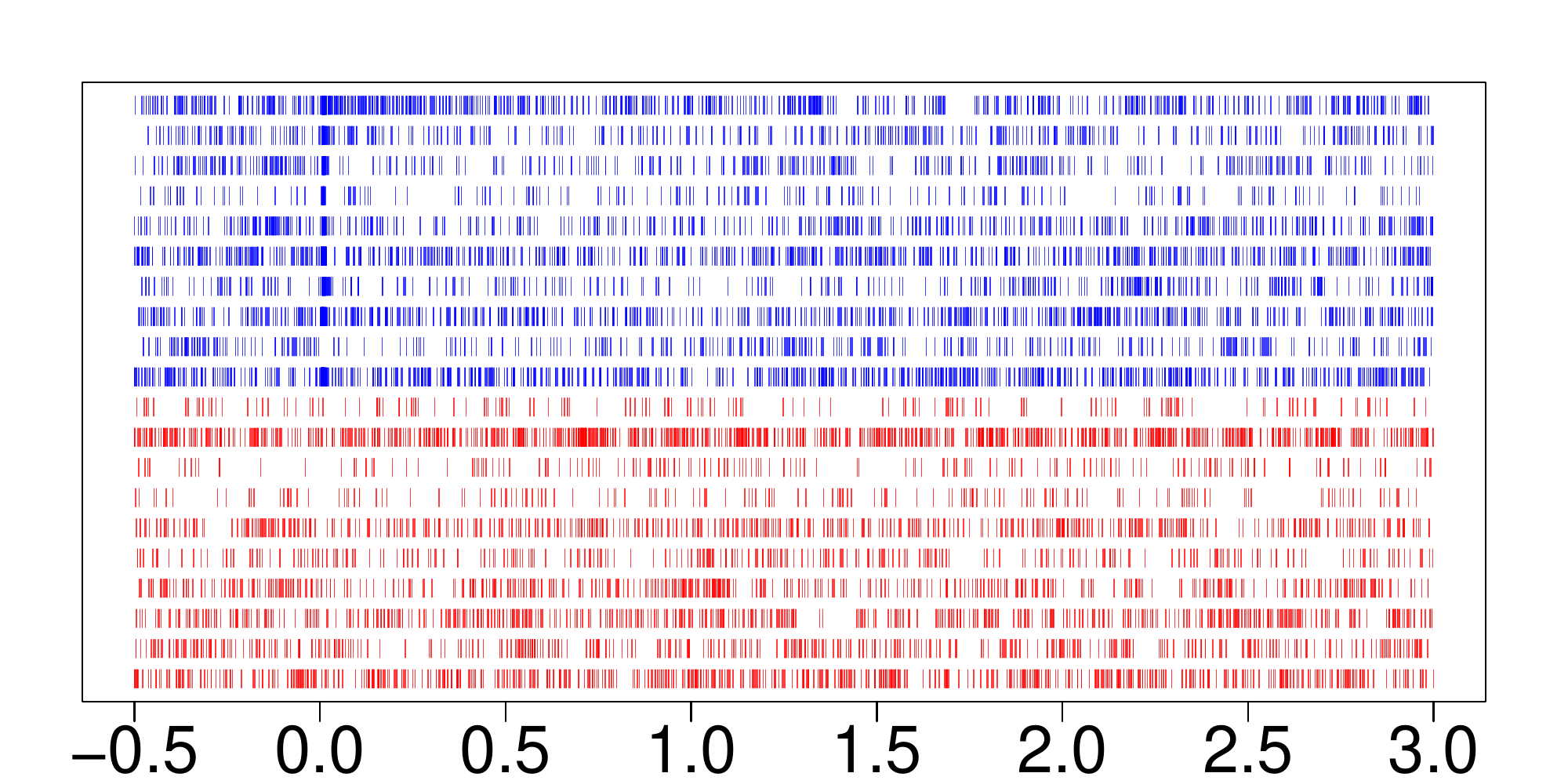}};
	\node[left = 8.5 cm of DAGlab]() {(b)};
			\node[below = 0 cm of Ys] (Yaxis) {Time (s)};
			\node[above left = -1.2 cm and -0.3 cm of Ys, rotate=90] () {Trials};

	\end{scope}

	\end{tikzpicture}
	\caption{Neural data from \cite{boldingfranks2018} and the causal diagram depicting the relationships among the variables in the instrumental variable framework. Panel (a) shows the spike trains collected in the mouse olfactory bulb in the stimulated (in blue) and unstimulated (in red) trials. Each row represents a spike train in the olfactory bulb in one trial ($N_i$). The shaded area depicts the duration of the light pulse. Panel (b) shows the spike trains collected in the mouse piriform cortex in the stimulated (in blue) and unstimulated (in red) trials. Each row represents a spike train in the piriform cortex in one trial ($Y_i$). Panel (c) zooms in on two randomly selected trials, where the solid curves are the smoothed intensities. Panel (d) shows the causal diagram for the relationship among the variables. Variables in the solid square and circles are observed, and the variable in the dashed circle is unobserved. The subscript $i$ represents the $i$th unit.}
	\label{fig:diagram}
\end{figure}
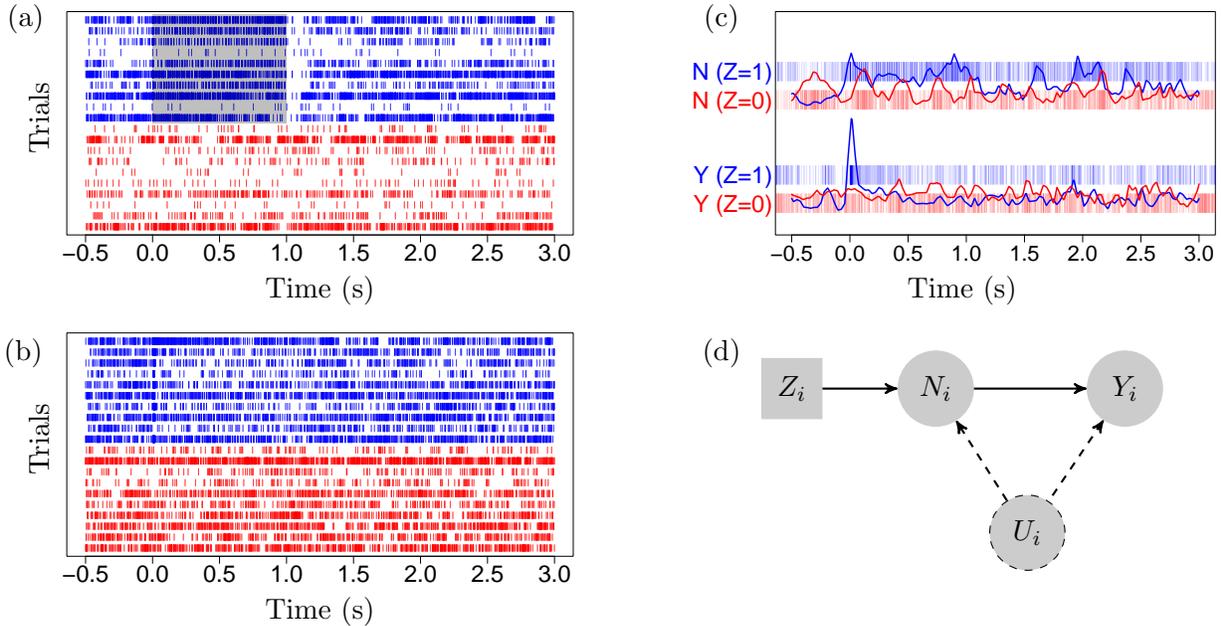

We now consider the setting when both the treatment $N_i$ and the outcome $Y_i$ are point processes. 
We will establish a similar ratio relationship as in~\eqref{eq::identify-cace} for the point process treatment and outcome, but in the frequency domain.
As a concrete example, we consider the neuroscience application from \cite{boldingfranks2018}. In this application, the treatment and the outcome are the neural activities of the mouse olfactory bulb and piriform cortex within each brain region, respectively. 
In experiments at the single-cell resolution, one can also model single-neuron activities as the treatment and outcome.
As shown in Figure~\ref{fig:diagram}(a)~and~(b), these data take the form of \emph{spike trains} that are commonly modeled as point processes. \cite{boldingfranks2018} apply light pulses to randomly selected trials to stimulate the olfactory bulb without affecting other brain regions. Therefore, the light pulse serves as an instrumental variable. To formally discuss causal inference, we need to generalise the model in \cite{angrist1996identification} to account for the point process treatment and outcome.

Let $i=1,\ldots,m$ index the units.
We use \textit{point processes} to describe the neuron activities (see, e.g., Chapter 2 in \citealp{cox1980point} or Chapter 3 in \citealp{daley2003introduction}). Define the treatment point process $N_i(\cdot)$ as a family of random non-negative integers $\{N_i(A)\}_{A \in \mathcal{B}(\mathbb{R})}$ counting the number of events in each set $A$. Let $\d N_i(t) \equiv N_i([t,t+\d t))$. Throughout this paper, we consider point processes that are simple, $\Pr \{ \d N_i(t)= 0 \ \textnormal{or}\ 1 \ \textnormal{for all} \ t\} = 1$, and with bounded intensity $\Pr \{ \d N_i(t)=1\} /\d t < \infty$. In a similar manner, we introduce the outcome point process $Y_i(\cdot)$.
We focus on a binary instrumental variable, $Z_i \in\{0,1\}$, and present the results on a discrete instrumental variable in \S\ref{sec:discrete} in the Supplementary Material.
Without loss of generality, we assume the instrumental variable onsets at time $0$, and $N_i(\cdot)$ and $Y_i(\cdot)$ are observed from $0$ to $T$. To avoid cumbersome bookkeeping, we constrain the processes to the observed period and ignore the history before time $0$, and we write $N_i([0,t])$ as $N_i(t)$.

For ease of discussion, we first consider the treatment $N_i(\cdot)$ being a point process with at most one event. We refer to $N_i(\cdot)$ as a \emph{single-point process} if $N_i(T)\leq 1$. We will extend the methodology to a general point process in \S\ref{sec:multi}.
We can characterize a single-point process $N_i(\cdot)$ using its event time: define $\mathcal{T}_i = T^+$ if $N_i(T)=0$ and $\mathcal{T}_i \equiv \tau$ if $N_i(t)=1$ for $t\geq \tau$ and $N_i(t)=0$ for $t <\tau$.

We adopt the potential outcomes framework under the following stable unit treatment value assumption \citep{rubin1980randomization}.
\begin{assumption}
There is no interference between units and there are no different versions of the instrument and the treatment process.
\label{asm::sutva}
\end{assumption}
Assumption~\ref{asm::sutva} rules out the spillover effect of other units' instrumental variable on one's treatment process and that of other units' instrumental variable and treatment process on one's outcome process. It also requires that there is only one version of the instrument and the treatment process.
In our motivating example, one unit corresponds to one trial, and trials conducted at different times might use the same mouse.
The no-interference assumption would be violated if the neural dynamics of a mouse adapt to stimulation over time, causing activities in one trial to depend on previous trials. This phenomenon is known as neural plasticity.   To restrict spillover between trials, adequate washout periods are incorporated to separate trials sufficiently apart. 
As a result, we can reasonably assume away spillover effects.  Furthermore, uniform stimulation is employed to ensure that there is only one version of the instrument. 
For the treatment process, we follow the common practice in neural data analysis to focus on the effect of the timings of spikes, ignoring the variation in the spike intensities \citep[cf.,][]{brillinger1988maximum, byron2009gaussian, zhao2017variational, wu2017gaussian}.

Assumption~\ref{asm::sutva} allows us to define the potential values as the function of a unit's own instrument and treatment process. Let
 $N_{iz}(\cdot)$ and $Y_{iz}(\cdot)$ be the potential processes of the treatment and outcome, and $\mathcal{T}_{iz}$ be the potential event time of the treatment process if the instrumental variable were set to $Z_i=z$. Also, define $Y_{iz\tau}(\cdot)$ as the potential process of the outcome if the instrumental variable were set to $Z_i=z$ and the event time were set to $\mathcal{T}_i=\tau$. By definition, the two versions of the potential outcome process satisfy
$Y_{iz}(\cdot)=Y_{iz,\mT_{iz}}(\cdot)$.
The observed treatment process is $N_i(\cdot) = Z_iN_{i1}(\cdot)+(1-Z_i)N_{i0}(\cdot)$, and the observed outcome process can be written as $Y_i(\cdot) = Z_iY_{i1}(\cdot)+(1-Z_i)Y_{i0}(\cdot)$ or
$Y_i(\cdot) = Z_iY_{i1\tau}(\cdot)+(1-Z_i)Y_{i0\tau}(\cdot)$ if $\mT_i=\tau$.
We assume $\{Z_i,N_{iz}(\cdot), Y_{iz\tau}(\cdot): z=0,1; \tau \in [0,T] \cup T^+\}_{i=1}^m$ are independently and identically distributed, and thus the observables $\{Z_i,N_i(\cdot), Y_i(\cdot)\}_{i=1}^m$ are also independently and identically distributed. We simplify $N_i(\cdot)$ as $N_i$ and $Y_i(\cdot)$ as $Y_i$ when no confusion arises.
In our motivating example, the experiment is carefully designed to ensure that the trials are independent and identically distributed.
For instance, the optical stimulation is targeted at the same location at the same chosen power to eliminate unintentional variability to ensure the identical distribution condition; the trials are separated with adequate washout periods to ensure the independence between units; the power of optical stimulation and the length of the experiment are limited to avoid physical damage to the neural circuits. 

Under Assumption~\ref{asm::sutva}, we impose the following three assumptions throughout the paper. 
First, we generalise the exclusion restriction assumption in \citet{angrist1996identification}.
\begin{assumption}[Exclusion restriction]
\label{asm::er}
 $Y_{iz'\tau}=Y_{iz\tau}$ for $z,z'=0,1$ and all $i$.
\end{assumption}
Assumption~\ref{asm::er} means that the instrumental variable affects the outcome only through the treatment. It holds in optogenetic experiments since only the targeted neurons respond to optical stimulation. Under Assumption~\ref{asm::er}, we can simplify $Y_{iz\tau}$ as $Y_{i\tau}$. There are two ways to describe the potential outcome processes under Assumption~\ref{asm::er}, i.e., $Y_{iz}$ and $Y_{i\tau}$. We will use $Y_{i1}$ and $Y_{i0}$ to represent the potential processes if the instrumental variable were set to $Z_i=1$ and $Z_i=0$, respectively, and $Y_{i\tau}$ to represent the potential process if the event time of $N_i$ were set to $\mT_i=\tau$.

Second, the following independence assumption holds automatically because trials are randomly selected for optical stimulation.
\begin{assumption}[Randomization]
\label{asm::random}
$Z_i \ind \{ N_{iz}(\cdot), Y_{i\tau}(\cdot) : z \in \{0,1\}, \tau \in [0,T]\cup T^+ \}$.
\end{assumption}
Assumption~\ref{asm::random} implies $Z_i \ind \{\mT_{iz}, Y_{iz}(t): z \in \{0,1\}, t \in [0,T] \}$ under Assumption~\ref{asm::er}.
It allows for the identification of the intention-to-treat effects of the instrumental variable on the treatment and the outcome. However, it is insufficient to identify the effect of the treatment on the outcome due to the possibility of unmeasured confounders.

Lastly, we invoke the following no anticipation assumption because the event time of $N_i$ at a later time point cannot reversely affect $Y_i$ at a previous time.
\begin{assumption}[No anticipation]
	\label{asm::no-anticipation}
	$Y_{i\tau}(t)=Y_{i\tau'}(t)$ for $\tau,\tau'\geq t$ and all $i$.
\end{assumption}
Assumption~\ref{asm::no-anticipation} is well-known in causal inference with time series data \citep[e.g.,][]{bojinov2019time}. The use of a non-strict inequality sign instead of a strict inequality in Assumption~\ref{asm::no-anticipation} indicates that
the effect of $N_i$ on $Y_i$ is not
instantaneous.
Replacing the non-strict inequality with a strict inequality allows for $Y_{i\tau}(\tau) \neq Y_{i\tau'}(\tau)$ for $\tau< \tau'$, i.e., the event at time $\tau$ has an effect on the outcome at the same time. The non-strict inequality in Assumption~\ref{asm::no-anticipation} also implies $Y_{iT^+} = Y_{iT}$ because the event at time $T$ does not have an effect on $Y_i$ in $[0,T]$.

Under Assumptions~\ref{asm::sutva}--\ref{asm::no-anticipation}, the relationships among $Z_i$, $N_i$, $Y_i$, and the unmeasured confounder $U_i$ can be illustrated by the causal diagram in Figure~\ref{fig:diagram}(d). The randomized stimulation $Z_i$ affects the treatment $N_i$, which in turn affects the outcome $Y_i$.
Because the treatment $N_i$ is not randomized, unmeasured confounders $U_i$ may exist between $N_i $ and $Y_i$.

\subsection{Definitions of causal effects with point process treatment and outcome}

We are now ready to define the causal quantities of interest.
First, we define the average causal effect (ACE) of the instrumental variable on the treatment and outcome processes at time $t$ as
$
\ACE_N(t)\ =\ E \{N_{i1}(t)-N_{i0}(t)\}$ and $\ACE_Y(t)\ =\ E \{Y_{i1}(t)-Y_{i0}(t)\},
$
respectively.
Although $\ACE_N(t)$ and $\ACE_Y(t)$  are possible quantities of interest in the experiment, they do not directly answer how the treatment $N_i$ affects the outcome $Y_i$.
Therefore, we define the ACE of the treatment process on the outcome process as
\begin{equation}\label{eqn::ACE}
\ACE(t;\tau_1,\tau_2)= E  \{Y_{i\tau_1}(t) - Y_{i\tau_2}(t)\}, \quad \tau_1 \geq \tau_2 \ \text{ and }\ \tau_1, \tau_2 \in [0,T] \cup T^+,
\end{equation}
which characterizes how the change in the event time of $N_i$ from $\tau_2$ to $\tau_1$ affects $Y_i$ at time $t$.
A positive $\ACE(t;\tau_1,\tau_2)$ with $\tau_1 \geq \tau_2$ implies that a later event in the treatment process increases the expected outcome process at time $t$.
This effect varies over time $t$ and depends on the two event times $\tau_1$ and $\tau_2$. Define the average causal effect rate (ACER) of $N_i$ on $Y_i$ as
\begin{equation}\label{eqn::ACER}
\ACER(t;\tau)\ =\ \lim_{ \Delta \tau \rightarrow 0+} \frac{\ACE(t;\tau+\Delta\tau, \tau)}{\Delta \tau}\ =\ \frac{\partial E  \{Y_{i\tau}(t)\}}{\partial \tau}.
\end{equation}
The ACER measures how fast $E \{Y_{i\tau}(t)\}$ changes given an infinitesimal change in the event time $\tau$.
This concept is similar to the infinitesimal shift function defined in \citet{lok2008statistical}.
Under Assumption~\ref{asm::no-anticipation}, we have
 \begin{eqnarray*}
\ACE(t;\tau_1,\tau_2) \ = \ \begin{cases}
\ACE(t;\tau_1,\tau_2), &\ \text{ if }\ \tau_2<\tau_1 < t\\
\ACE(t;t,\tau_2), & \ \text{ if } \ \tau_2 < t \leq \tau_1\\
0, & \ \text{ if } \ t \leq \tau_2 \leq \tau_1
\end{cases},
\end{eqnarray*}
and thus $\ACER(t;\tau) =0 $ if $t \leq \tau$.
When the treatment is a single-point process, we have the following relationship between the ACE and ACER of the treatment,
\begin{eqnarray}
\label{eqn::ace-acer}
\ACE(t;\tau_1,\tau_2) &=& \int_{\tau_2}^{\tau_1} \ACER(t;\tau) \d \tau.
\end{eqnarray}
Therefore,  we can focus on the ACER because it determines the ACE.

Under Assumption~\ref{asm::random}, the ACEs of the instrumental variable on the treatment and outcome processes can be identified by
the observed differences between the stimulated and unstimulated groups,
\begin{eqnarray}
\label{eqn::pp-aceN}
\ACE_N(t)\ =\ f(t)  & \text{ with }& f(t)\ =\  E \{N_i(t)\mid Z_i=1\}- E \{N_i(t)\mid Z_i=0\},\\
\label{eqn::pp-aceY}
\ACE_Y(t)\ =\ h(t) & \text{ with }& h(t)\ =\   E \{Y_i(t)\mid Z_i=1\}- E \{Y_i(t)\mid Z_i=0\}.
\end{eqnarray}
However, Assumption~\ref{asm::random} is insufficient for the identification of the ACE and the
ACER of the treatment process on the outcome process, because the treatment process is not randomized.

\section{Nonparametric identification and estimation}\label{sec:IV}
\subsection{Nonparametric identification with a single-point process treatment}
\label{sec:identification}

We begin by generalising the monotonicity assumption in \citet{angrist1996identification}.
 \begin{assumption}[Monotonicity]\label{asm::mon}
 For each $i$, the potential event times of $N_i$ satisfy $\mathcal{T}_{i1} \leq \mathcal{T}_{i0}. $
\end{assumption}
Assumption~\ref{asm::mon} requires that the potential event time of $N_i$ under stimulation will be no later than that without stimulation.
Under Assumption~\ref{asm::mon}, the ACE of the instrumental variable on the treatment process at time $\tau$ equals the proportion of a subpopulation defined by the joint potential event times of $N_i$, i.e.,
$$\ACE_N(\tau) \ =\ \Pr(\mathcal{T}_{i1} \leq \tau < \mathcal{T}_{i0}), \quad \tau \in [0,T].$$
Units in this subpopulation would have the event time of treatment process before or equal to $\tau$ with stimulation and after $\tau$ without stimulation. Thus, these can be viewed as the compliers whose treatment is positively affected by the stimulation. With a point process treatment, the definition of compliers is time-dependent. Similarly, the other three subpopulations, $\mathcal{T}_{i0} \leq \tau < \mathcal{T}_{i1}$, $\max(\mT_{i1},\mT_{i0}) \leq \tau $, and $\tau < \min(\mT_{i1},\mT_{i0})$, generalise the defiers, always-takers, and never-takers in the binary instrumental variable model, respectively.

We cannot validate Assumption~\ref{asm::mon} since it depends on unit-level potential outcomes. However, Assumption~\ref{asm::mon} implies a testable condition that can be checked using the observed data.
\begin{proposition}
\label{prop::testable}
Under Assumption~\ref{asm::random},
Assumption~\ref{asm::mon} implies, for all $\tau \in [0,T]$,
\begin{eqnarray*}
\Pr( \mathcal{T}_i > \tau \mid Z_i=1)\ \leq\ \Pr( \mathcal{T}_i > \tau \mid Z_i=0).
\end{eqnarray*}
\end{proposition}
Proposition~\ref{prop::testable} states the stochastic dominance of the survival function of $\mathcal{T}_i$ under stimulation over that without stimulation.
We can assess Assumption~\ref{asm::mon} by comparing the empirical survival functions of $\mT_i$ in the stimulated and unstimulated groups. If the two curves cross, then the testable condition in Proposition~\ref{prop::testable} is violated, which in turn falsifies Assumption~\ref{asm::mon}.
Therefore, our identification results will consider scenarios both with and without Assumption~\ref{asm::mon}.

\citet{angrist1996identification} show that the effect of the instrumental variable on the outcome equals the product of the effect of the instrumental variable on the treatment and the effect of the treatment on the outcome. The following theorem generalises their result to our setting.
 \begin{theorem}
\label{thm:ivformula}
Suppose that $N_i$ is a single-point process and Assumptions~\ref{asm::sutva}--\ref{asm::no-anticipation} hold.
For any $t \in [0,T]$, we have
\begin{eqnarray}
\nonumber\ACE_Y(t) &=& \int_{0}^T { E \{ \partial Y_{i\tau}(t)/ \partial \tau \mid \mT_{i0} \leq \tau < \mT_{i1} \}} \Pr (\mT_{i0} \leq \tau < \mT_{i1} )\d \tau\\
\label{eqn:ivformula_general} &&- \int_{0}^T { E \{ \partial Y_{i\tau}(t)/\partial \tau \mid \mT_{i1} \leq \tau < \mT_{i0} \}} \Pr (\mT_{i1} \leq \tau < \mT_{i0} )\d \tau.
\end{eqnarray}
If Assumption~\ref{asm::mon} holds in addition, then for any $t \in [0,T]$, we have
\begin{eqnarray}
\label{eqn:ivformula_monotone}\ACE_Y(t) &=& - \int_{0}^T E \{ \partial Y_{i\tau}(t)/\partial \tau \mid \mT_{i1} \leq \tau < \mT_{i0} \}\cdot \ACE_N(\tau) \d \tau.
\end{eqnarray}
\end{theorem}
 
In Theorem~\ref{thm:ivformula}, $\partial Y_{i\tau}(t)/\partial \tau $ is a generalised derivative that may consists of Dirac $\delta$ functions \cite[Appendix~B]{lax2002functional}. Since the conditional set $\mT_{i1} \leq \tau < \mT_{i0} $ depends on $\tau$, it is important to note that $ E \{ \partial Y_{i\tau}(t)/\partial \tau \mid \mT_{i1} \leq \tau < \mT_{i0} \} =  \partial E  \{Y_{i\tau'}(t) \mid \mT_{i1} \leq \tau < \mT_{i0}\} /\partial \tau' \mid_{\tau'=\tau},$ which is generally not equal to $\partial E \{ Y_{i\tau}(t) \mid \mT_{i1} \leq \tau < \mT_{i0} \}/\partial \tau$  that takes into account the change of the conditional set.

By rewriting $\ACER(t;\tau)$ as $ E \{ \partial Y_{i\tau}(t)/\partial \tau \}$, we can view $ E \{ \partial Y_{i\tau}(t)/\partial \tau \mid \mT_{i1} \leq \tau < \mT_{i0} \}$ as the ACER in the subpopulation $ \mT_{i1} \leq \tau < \mT_{i0}$. In a sense, $E \{ \partial Y_{i\tau}(t)/\partial \tau \mid \mT_{i1} \leq \tau < \mT_{i0} \}$ generalises the complier average causal effect in the binary instrumental variable model.
Similarly, $E \{ \partial Y_{i\tau}(t)/\partial \tau \mid \mT_{i0} \leq \tau < \mT_{i1} \}$ 
 generalises the average causal effect for the defiers. These conditional expectations might not equal the ACER because $Y_{i\tau}(t)$ and $(\mT_{i1},\mT_{i0})$ might not be independent due to the unmeasured confounding between $N_i$ and $Y_i$.

The formula in~\eqref{eqn:ivformula_general} shows that the average causal effect of the instrumental variable on the outcome process, $\ACE_Y(t)$, equals the difference between the weighted averages of the two subpopulation ACERs over the timeline. The weights rely on the joint distribution of $( \mT_{i1}, \mT_{i0} ) $. When Assumption~\ref{asm::mon} holds, the first term on the right hand side of~\eqref{eqn:ivformula_general} vanishes and the weight $ \Pr (\mT_{i1} \leq \tau < \mT_{i0} )$ is equal to $\ACE_N(\tau)$. As a result,~\eqref{eqn:ivformula_general} reduces to~\eqref{eqn:ivformula_monotone} under monotonicity.

Under Assumption~\ref{asm::random}, $\ACE_Y(t) $ and $\ACE_N(t) $ are identifiable. Thus, we can view~\eqref{eqn:ivformula_general}~and~\eqref{eqn:ivformula_monotone} as integral equations for the subgroup ACERs \citep{newey2003instrumental}.
Unfortunately, these subpopulation ACERs are not identifiable without additional assumptions.
To provide some intuition, consider~\eqref{eqn:ivformula_monotone} under monotonicity.
Based on the observed data,~\eqref{eqn::pp-aceN}~and~\eqref{eqn::pp-aceY} give the identification formulas for
 $\ACE_Y(t)$ and $\ACE_N(\tau)$ for all $t,\tau \in [0,T]$ under Assumption~\ref{asm::random}.
So~\eqref{eqn:ivformula_monotone} is an integral equation for the unknown quantity defined as $\gamma(\tau,t)= E \{\partial Y_{i\tau}(t) /\partial \tau \mid \mT_{i1} \leq \tau < \mT_{i0} \}$. Consider a discrete approximation of $\gamma(\tau,t)$ by evaluating its values over a $K_1\times K_2$ two-dimensional grid of $(\tau, t)$. Equation~\eqref{eqn:ivformula_monotone} generates only $K_2$ equations by considering the $K_2$ grid of $t$, which cannot sustain the identification of $K_1 \times K_2$ unknown values of $\gamma(\cdot,\cdot)$. Consequently, the identification of the ACERs is infeasible without additional assumptions.

To address this problem, we invoke the following identification assumption.
\begin{assumption}[Stationarity]
\label{asm::stationarity}
$\ACE(t;\tau_1,\tau_2)=\ACE(t-\tau_1;0, \tau_2-\tau_1)$ for $ \tau_1 \leq \tau_2 \leq t$.
\end{assumption}
Assumption~\ref{asm::stationarity} states that the ACE of the treatment on the outcome is invariant to timeline shifts. The left-hand side is the effect of the treatment when the event time is $\tau_1$ versus $\tau_2$ on the outcome at time $t$. In contrast, the right-hand side represents the same effect, but with the timeline shifted forward by $\tau_1$. Therefore, Assumption~\ref{asm::stationarity} means that the ACE of the treatment is invariant regardless of the absolute time.
Under Assumption~\ref{asm::stationarity}, we have $\ACER(t;\tau) = \ACER(t-\tau;0)$ and thus can simplify $ \ACER(t;\tau)$ as $ \ACER(t-\tau)$ with $\ACER(t-\tau)=0$ if $t\leq \tau$. We can show that, together with Assumption~\ref{asm::no-anticipation}, Assumption~\ref{asm::stationarity} leads to
\begin{eqnarray}\label{eqn::ACER_new}
	\ACER(t;0) \ = \ -\partial  E \{Y_{i0}(t)\}/\partial t.
\end{eqnarray}
The formula in~\eqref{eqn::ACER_new} offers a more natural interpretation of $\ACER$, that is, $-\ACER(t;0)$ describes the expected change rate in the potential outcome at time $t$ when the event in $N_i$ happens at time $0$.
Theorem \ref{thm::identification} below gives sufficient conditions for identifying the ACER.
\begin{theorem}
	\label{thm::identification}
	 Suppose that  $N_i$ is a single-point process and Assumptions~\ref{asm::sutva}--\ref{asm::no-anticipation}~and~\ref{asm::stationarity} hold.
Furthermore, if either 
(i) Assumption~\ref{asm::mon} holds and, for all $t, \tau \in [0,T]$, 
\begin{eqnarray}
\label{eqn::condition_mon}
E  \left\{ { \partial Y_{i\tau}(t)}/{\partial \tau} \mid \mT_{i1} \leq \tau < \mT_{i0} \right\}\ =\ { \partial E  \left\{ Y_{i\tau}(t)\right\}}/{\partial \tau},
\end{eqnarray}
or (ii) for all $t, \tau \in [0,T]$,
\begin{eqnarray}
\label{eqn::condition}
E  \{ { \partial Y_{i\tau}(t)}/{\partial \tau} \mid \mT_{i1} \leq \tau < \mT_{i0} \} = E \{ { \partial Y_{i\tau}(t)}/{\partial \tau} \mid \mT_{i0} \leq \tau < \mT_{i1} \} = { \partial E  \{ Y_{i\tau}(t) \}}/{\partial \tau},
\end{eqnarray}
then $\ACER$ satisfies
 \begin{eqnarray}
 \label{eqn::ACER-property1}
\ACER(t;\tau) \ = \ \begin{cases}
\ACER(t-\tau; 0), &\ \text{ if }\ t > \tau\\
0, & \ \text{ if } \ t \leq \tau
\end{cases},
\end{eqnarray}
 and
 \begin{eqnarray}
 \label{eqn::ACER-property2}
 h(t)\ = \ - \int_0^{T} \ACER(t-\tau; 0) f(\tau) \d \tau
\end{eqnarray}
 for $t \in [0,T]$. If further $(\Psi f) (\nu) \neq 0$ for all $\nu \in \mathbb{R}$, then the $\ACER$ is identified by
$\ACER(t;\tau)= -\Psi^{-1}\big( G \big)(t-\tau)$ for $t > \tau$ and $\ACER(t;\tau)= 0$ for $t \leq \tau$, where
\begin{equation}\label{eqn:generalized_wald}
G(\nu)\ =\ \frac{ (\Psi h) (\nu)}{(\Psi f) (\nu)}
\quad \text{ for all } \quad  \nu \in \mathbb{R}.
\end{equation}
\end{theorem}

The condition  in~\eqref{eqn::condition_mon} means that the ACERs are homogenous across subpopulations defined by $\mT_{i1} \leq \tau < \mT_{i0}$ with different values of $\tau$, generalising 
the homogeneity assumption in~\eqref{eqn::homogeneity}. Without monotonicity, the condition in~\eqref{eqn::condition} further requires that the ACERs are homogenous across subpopulations defined by $\mT_{i0} \leq \tau < \mT_{i1}$.
Similar to the instrumental variable methods in survival analysis \citep[e.g.][]{li2015instrumental,tchetgen2015instrumental},
these conditions are satisfied as long as the impact of the confounders on the outcome is additive.
With a binary treatment and a scalar outcome, \citet{wang2018bounded} also use a similar condition assuming no additive interaction between the treatment and the unmeasured confounders on the outcome.
We will study this condition in detail under several commonly-used outcome models in \S\ref{sec::model}.

The deconvolution problem \eqref{eqn::ACER-property2}
 belongs to the family of Wiener--Hopf equations (see, among others, \citealp{noble1959methods}). It is essentially the same as the well-studied deconvolution of densities in statistics \citep[e.g.,][]{fan1991optimal,diggle1993,pensky1999, johannes2009,dattner2011, dattner2016}. 
From the Paley--Wiener--Schwartz theorem, 
we know that $(\Psi f) (\nu) \neq 0$ for all $\nu \in \mathbb{R}$ if $f(t)= E \{N_i(t)\mid Z_i=1\}-E \{N_i(t)\mid Z_i=0\}$ is a non-zero function with bounded support.  This holds as long as the effect of the instrument $Z_i$ on the treatment process $N_i$ vanishes in finite time. 
The non-zero condition of $(\Psi f)(\nu)$ is also employed in the nonparametric deconvolution problem \citep[see, e.g.,][]{fan1991optimal}.

In the binary instrumental variable model with the homogeneity assumption, the effect of the treatment on the outcome equals the ratio of the effects of the instrumental variable on the treatment and the outcome. Theorem~\ref{thm::identification} shows that this ratio relationship also holds with point process treatment and outcome, but in the frequency domain. The well-known convolution theorem ensures that the Fourier transform of a convolution of two functions is equal to the product of their Fourier transforms. Therefore, by applying the Fourier transform on each term of the convolution equation in~\eqref{eqn::ACER-property2}, we can obtain the generalised Wald estimation formula \eqref{eqn:generalized_wald} in Theorem~\ref{thm::identification}.

\subsection{Treatment with multiple events }\label{sec:multi}

We generalise the identification result in \S\ref{sec:identification} to a treatment process with possibly multiple events.
We begin by generalising the definition of potential values and causal effects. Let $Y_{i,n(\cdot)}(\cdot)$ be the potential process of the outcome if the treatment were set to a fixed process $n(\cdot)$. The observed outcome process is $Y_i(\cdot)=Y_{i,n(\cdot)}(\cdot)$ if $N_i(\cdot) =n(\cdot)$. Then, we can define the ACE of the treatment $n(\cdot)$ versus $n'(\cdot)$ on the outcome as
\begin{equation}\label{eqn:ACE_path}
\ACE\{t; n(\cdot), n'(\cdot)\}\ =\ E \{Y_{i,n(\cdot)}(t)- Y_{i, n'(\cdot)}(t)\}.
\end{equation}
For single-point process treatments, ~\eqref{eqn:ACE_path} reduces to the definition in~\eqref{eqn::ACE}.
Using the linearity of expectation, we can write
\begin{align*}
\ACE\{t; n(\cdot), n'(\cdot)\}\ = & \ E \{Y_{i,n(\cdot)}(t)- Y_{i,n'(\cdot)}(t)\} \\
\ = & \ E \{Y_{i,n(\cdot)}(t)- Y_{iT+}(t)\} - E \{Y_{i,n'(\cdot)}(t)- Y_{iT+}(t)\},
\end{align*}
where $E \{Y_{i,n(\cdot)}(t)- Y_{iT+}(t)\}$ and $E \{Y_{i,n'(\cdot)}(t)- Y_{iT+}(t)\}$ are the effects of $n(\cdot)$ and $n'(\cdot)$ versus a null process with no events in $[0,T]$, respectively. Similar to \S\ref{sec:identification}, we can characterize the treatment process using event times.
Suppose that $n(\cdot)$ has $l$ events at  times $\tau_1,\ldots,\tau_l$. Then $Y_{i,n(\cdot)}(t)$ can be written as
$Y_{i, \tau_1,\ldots,\tau_l}(t)$, so its expectation decomposes as
\begin{eqnarray}
\label{eqn::decomposition}
E  \{Y_{i, \tau_1,\ldots,\tau_l}(t)\}\ =\ E \{Y_{iT+}(t)\}+\sum_{s=1}^{l} E \left\{Y_{i,\tau_1,\ldots,\tau_s}(t)-Y_{i,\tau_1,\ldots,\tau_{s-1}}(t)\right\}
\end{eqnarray}
with $Y_{i,\tau_1,\ldots,\tau_{s-1}}(t)=Y_{iT+}(t)$ for $s=1$. The following assumption simplifies the decomposition by assuming away the interactive effects of the event times in the potential outcome process.
\begin{assumption}[Additivity]
\label{asmp:additivity}
$E \left\{Y_{i, \tau_1,\ldots,\tau_s}(\cdot)-Y_{i,\tau_1,\ldots,\tau_{s-1}}(\cdot)\right\} = E \{Y_{i, \tau_s}(\cdot)-Y_{iT+}(\cdot)\}$ for any $s\geq 1$ and any event times $(\tau_1,\ldots,\tau_s)$ satisfying $\tau_1 < \tau_2 < \cdots < \tau_s$.
\end{assumption}
Point processes with event times $(\tau_1,\ldots,\tau_s)$ and $(\tau_1,\ldots,\tau_{s-1})$ have the same trajectory up to time $\tau_{s-1}$, where the former has an additional event at $\tau_s$.
Assumption~\ref{asmp:additivity} means that the effect of the process with event times $(\tau_1,\ldots,\tau_s)$ versus that with event times $(\tau_1,\ldots,\tau_{s-1})$ does not depend on their common trajectory up to time $\tau_{s-1}$. Hence, the causal effect remains the same when the first $s-1$ events are removed from both processes.
 Under Assumption~\ref{asmp:additivity},~\eqref{eqn::decomposition} simplifies to
$E \{Y_{i, \tau_1,\ldots,\tau_l}(t) -Y_{iT+}(t)\} \ =\ \sum_{s=1}^{l} E \left\{Y_{i,\tau_s}(t)-Y_{iT+}(t)\right\},$
which means that the effect of each event time on the outcome process is additive.
In \S\ref{sec:proof-model1} in the supplementary material, we show that Assumption~\ref{asmp:additivity} holds under the Hawkes process \citep{hawkes1971spectra} or Aalen's additive hazard model \citep{aalen1980} for the potential outcome process.
 Assumption~\ref{asmp:additivity} may be violated due to the interactive effect of the event times in the treatment process. 
For instance,  neural ensembles are famous for their neural plasticity in the long term --- the ability to reorganize themselves in response to stimulation, which clearly violates Assumption~\ref{asmp:additivity}.
Such violations of Assumption~\ref{asmp:additivity} are sometimes of scientific interest. We leave the investigation of such effects for future research.

Under Assumption~\ref{asmp:additivity}, we can separately study the effect of each event in $N_i$.
Proposition \ref{prop::ACE-ACER-multiple} below generalises \eqref{eqn::ace-acer} to treatment processes with multiple events.

\begin{proposition}
\label{prop::ACE-ACER-multiple}
Under Assumptions~\ref{asm::sutva},~\ref{asm::er},~\ref{asm::no-anticipation},~and~\ref{asmp:additivity}, we have
\begin{eqnarray*}
\label{eqn:ACE-ACER-multiple}
\ACE\{t; n(\cdot), n'(\cdot)\}\ =\ -\int_0^t \ACER(t;\tau) \{n(\tau) - n'(\tau)\} \d \tau.
\end{eqnarray*}
\end{proposition}
Based on Proposition~\ref{prop::ACE-ACER-multiple}, we can focus on the identification of ACER. Theorem~\ref{thm::identification-ex} below generalises Theorem~\ref{thm::identification} to treatment processes with multiple events.

\begin{theorem}
	\label{thm::identification-ex}
	 Suppose that Assumptions~\ref{asm::sutva}--\ref{asm::no-anticipation},~\ref{asm::stationarity},~and~\ref{asmp:additivity} hold.
If for all $t \in [0,T]$ and any fixed processes $n(\cdot)$ and $n'(\cdot)$,
\begin{eqnarray}
\label{eqn::condition-ex}
E  \{ Y_{i, n(\cdot)}(t) - Y_{i, n'(\cdot)}(t) \mid N_{i1}(\cdot)=n(\cdot), N_{i0}(\cdot)=n'(\cdot) \} 
= 
E \{ Y_{i, n(\cdot)}(t) - Y_{i, n'(\cdot)}(t) \},
\end{eqnarray}
then the $\ACER$ satisfies~\eqref{eqn::ACER-property1}~and~\eqref{eqn::ACER-property2}. If $(\Psi f) (\nu) \neq 0$ for all $\nu \in \mathbb{R}$, then the $\ACER$ is identified by
$\ACER(t;\tau)= -\Psi^{-1}\big( G \big)(t-\tau)$ for $t > \tau$ and $\ACER(t;\tau)=0$ for $t \leq \tau$, with $G(\nu)$ defined in \eqref{eqn:generalized_wald}. 
\end{theorem}
When $N_i$ is a single-point process, Theorem~\ref{thm::identification-ex} does not require Assumption~\ref{asmp:additivity}, and the condition in~\eqref{eqn::condition-ex} reduces to~\eqref{eqn::condition} in Theorem~\ref{thm::identification}. As a result, Theorem~\ref{thm::identification-ex} reduces to Theorem~\ref{thm::identification} when $N_i$ has at most one event. Similar to Theorem~\ref{thm::identification}, the condition in~\eqref{eqn::condition-ex}
means that the ACERs are homogenous across subpopulations defined by $N_{i1}$ and $N_{i0}$.

\subsection{Estimation}
\label{sec::estimation}
We consider the estimation of the ACER based on identification results from
Theorems~\ref{thm::identification}~and~\ref{thm::identification-ex}.
This is essentially the deconvolution problem commonly studied in the literature \citep[see][for more discussion]{diggle1993,pensky1999, johannes2009,dattner2011, dattner2016}. 
Since an optimal estimation procedure is not the focus of this paper, we only provide a simple regression-based procedure to estimate $\ACER$. To be specific, we use a two-step procedure by first obtaining the estimates of $f$ and $h$ and then solving the ACER from the empirical version of the convolution equation in~\eqref{eqn::ACER-property2}.

Let $\hat f$ and $\hat h$ denote the estimators of $f$ and $h$ defined in~\eqref{eqn::pp-aceN} and~\eqref{eqn::pp-aceY}, which equal the empirical mean differences in the treatment and outcome processes in the stimulated and unstimulated groups.
We approximate the true ${\ACER}$ with truncated basis expansions, for $\Delta \in [0,T]$,
\begin{equation}\label{eqn:bsplines}
{\ACER}(\Delta;0)\ \approx\ \sum_{j=1}^J \psi_j(\Delta) \beta_j,
\end{equation}
where $J$ is a tuning parameter for the number of bases and $\{\psi_j(\cdot): j=1,2,\ldots,J\}$ is a set of pre-specified basis functions.
Here the support of ${\ACER}(\cdot;0)$ can be determined by prior knowledge.
Then, we estimate $\beta = (\beta_1,\ldots,\beta_J)$ by minimizing the following penalized $\ell_2$-distance based on the convolution equation \eqref{eqn::ACER-property2}
\begin{equation}\label{eqn::optim}
\widehat{\beta}\ =\  \underset{\beta \in \mathbb{R}^J}{\arg \min} \left\|\hat{h}+
\sum_{j=1}^J (\psi_j*\hat{f}) \beta_j
\right\|_2^2 + \eta \|\beta\|_2^2,
\end{equation}
where $*$ denotes the convolution between two functions and introduce the ridge penalty to reduce boundary effects. 
An analytic solution for $\widehat{\beta}$ is available since the objective function in \eqref{eqn::optim} is quadratic in $\beta$.
Recalling that we consider independent trials, we can choose the tuning parameter $J$ and $\eta$ using cross-validation or based on prior knowledge such as the smoothness of the $\ACER$. 
Denoting the selected parameter by $\widehat{J}$ and $\hat{\eta}$, the final estimator is given as $\widehat{\ACER}(\cdot;0)=\sum_{j=1}^{\widehat{J}} \psi_j(\cdot) \hat{\beta}_{j,\hat{\eta}}$.
We can then construct the confidence band for the function $\widehat{\ACER}(\cdot;0)$ using the bootstrap. The asymptotic properties for $\widehat{\ACER}$ as the sample size $m$ increases follow from the standard theory assuming independent samples. We leave the rigorous discussion for future analysis, as it is not the main focus of this paper.

\section{The role of models: causal interpretability and identifiability}\label{sec::model}

\subsection{Conditional intensity}\label{sec::model-intensity}

In this section, we study several commonly-used models for point process outcomes in applied research when an instrumental variable is available, allowing for the presence of unmeasured confounders. We do not impose any distributional assumptions on the unmeasured confounders. Consequently, it is difficult to study the identifiability of the model parameters themselves. We take an alternative route by connecting the model parameters to the causal effects and considering the identifiability and estimation of the causal effects directly. With a binary instrumental variable, we show that the ACER is identifiable and can be estimated using the generalised Wald estimation under many commonly-used models.
This  estimation strategy does not rely on the identification or estimation of the model parameters, as long as the unmeasured confounding is additive in  the underlying outcome model.

We begin by introducing some additional notation to characterize a point process. Let $U_i(\cdot)$ denote the unmeasured confounding process on $\mathbb{R}$. We use $\mathcal{H}_{it-}$ to represent the $\sigma$-algebra induced by the history up to, but not including, time $t$. Define the  \emph{conditional intensity} of $Y_i$ as
\begin{eqnarray}
\label{eqn:intensity-def}
\lambda_{Y}\left(t\right)\ = \ E  \left\{ {\d Y_i(t)}/{\d t} \mid \mathcal{H}_{it-} \right\}.
\end{eqnarray}
The conditional intensity, or intensity, is the conditional mean of the event rate of $Y_i$ in an infinitesimal time interval $[t,t + \d t)$, which is analogous to the conditional mean of the outcome in the binary instrumental variable model. It fully characterizes the probabilistic structure of a point process and is closely related to the hazard function in survival analysis. See Chapter 7 in \citet{daley2003introduction} for more discussion of the intensity.

In \eqref{eqn:intensity-def}, the conditional intensity could depend on the history of $Y_i$. When the outcome $Y_i$ describes recurrent events, it is common to allow the conditional intensity to depend on past events of $Y_i$ (e.g., \citealp{hawkes1971spectra, brillinger1988maximum,lawrence2004gaussian, kulkarni2007common, byron2009gaussian, gao2015high, macke2015estimating, gao2016linear,  wu2017gaussian, zhao2017variational, pandarinath2018inferring}). As concrete examples, in the context of neural data, the dependence on past events captures the known phenomenon that a single neuron cannot fire consecutively in a very short period of time and
activities in a region may trigger inhibitory circuits to stabilize the activity on a longer time scale.

An inherent constraint on the intensity is that it must be non-negative for the probabilistic model to be well-defined.
A similar constraint is well acknowledged in modeling the hazard function in survival analysis. 
This constraint of the intensity creates a schism in the modeling of point processes --- whether to employ a linear working model \citep{aalen1980} or a non-negative generative model \citep{cox1972regression}.
In either model, since $U_i$ is unobserved, existing methods usually impose strong parametric assumptions on $U_i$ in order to estimate the parameters. A common assumption is that $U_i$ is a Gaussian process (see, e.g., \citealp{byron2009gaussian, zhao2017variational}), primarily due to its simplicity for the Bayesian computation. However, the analysis can be sensitive to these parametric assumptions.
In \S\ref{sec:model-para}, we will study both types of models.

Before specifying $\lambda_Y(t)$, we introduce the following assumption on the relationships among $N_i$, $Y_i$, and $U_i$, which are commonly used in instrumental variable methods when outcome models are employed \citep[see][for an example in survival analysis]{tchetgen2015instrumental}.
\begin{assumption}	\label{asmp:ignorability-self-latent}
(a) $Z_i \ind \{ N_{iz}(t), Y_{i\tau}(t), U_i(t): z \in \{0,1\}, t \in [0,T], \tau \in [0,T]\cup T^+ \}$;
(b) For $t \in [0,T]$,	$Y_{i, n (\cdot)}(t)\ \ind\ N_{i}(\cdot) \mid \{\mathcal{H}^{*\setminus N_i}_{it-}, N_i(s)=n(s), s \in [0,t) \}$ and any fixed point process $n(\cdot)$, where $\mathcal{H}^{*\setminus N_i}_{it-}$ denotes the $\sigma$-algebra induced by all potential processes including $U_i$, except for $N_i$ up to time $t$. 
\end{assumption}
See \citet{lok2008statistical} for the measure-theoretic description of the independence given histories of point processes.
Assumption~\ref{asmp:ignorability-self-latent}(a) is a restatement of Assumption~\ref{asm::random} with the additional notation of $U_i$. It holds because $Z_i$ is randomized.
Assumption~\ref{asmp:ignorability-self-latent}(b) generalises the latent sequential ignorability in \citet{ricciardi2020bayesian} to a continuous-time setting with point process treatment and outcome. It assumes that $U_i$ fully characterizes the confounding between the treatment and the outcome so the treatment is independent of the potential outcome at time $t$ given the histories of $N_i$, $Y_i$, and $U_i$.
Under Assumption \ref{asmp:ignorability-self-latent}, we have, for any $t \in [0,T]$,
\begin{eqnarray}
\label{eqn:intensity-po-ob}
	E  \left\{ { \d Y_{i,n(\cdot)}(t)}/{\d t} \mid \mathcal{H}_{it-} \right\} &=&
	E  \left\{ { \d Y_{i}(t)}/{\d t} \mid N_{i}(\cdot) = n(\cdot), \mathcal{H}^{*\setminus N_i}_{it-} \right\},
\end{eqnarray}
which links the potential processes to the conditional intensity of the observed outcome.
Therefore, the discussion in \S\ref{sec:model-para} will focus on the models for the observed outcome.
Under each of the models, we will connect the model parameters with the ACER and study its identification.

\subsection{Identification of causal effects with linear additive unmeasured confounding}\label{sec:model-para}

We start with linear models for the intensity. This type of models has been widely used in different contexts because of its mathematical tractability \citep[e.g.,][]{hawkes1971spectra,aalen1980,tchetgen2015instrumental,jiang2018twostep}.
In particular,
consider $Y_i(\cdot)$ to be a linear Hawkes process with the following intensity,
\begin{eqnarray}
\label{eqn:model1}
\lambda_{Y}\left(t \right) = \mu_Y+ \int_{0}^{t} g(t-s) \d N_i(s)+ \int_{0}^{t} \omega(t-s) \d Y_i(s) + \psi_{U_i}(t),
\end{eqnarray}
 where $g(\Delta)=\omega(\Delta)=0$ for $\Delta \leq 0$ and $\psi_{U_i}(t)$ represents any function of $\{U_i(s): s \in [0,t)\}$.
Proposition \ref{prop:model1} below connects the ACER with parameters in \eqref{eqn:model1} and shows the identification.

\begin{proposition}
	\label{prop:model1}
	Suppose that Assumptions~\ref{asm::sutva}--\ref{asm::random}~and~\ref{asmp:ignorability-self-latent} hold, and the underlying outcome model satisfies~\eqref{eqn:model1}. (a) We have
	\begin{eqnarray*}
		\label{eqn:model1-acer}
		\ACER(t;\tau) \ =\ \ACER(t-\tau;0) \ =\ - \big( \Psi^{-1} \widetilde{G} \big)(t-\tau),
	\end{eqnarray*}
	where $\widetilde{G}(\nu)=\big\{ 1+ (\Psi \omega)(\nu) \big\}^{-1} \big( \Psi g\big) (\nu)$ if $1+ (\Psi \omega)(\nu) \neq 0$ for all $\nu \in \mathbb{R}$. (b)
When $ (\Psi f)(\nu) \neq 0$ for all $\nu \in \mathbb{R}$, \ACER{} is identified by $\ACER(t; \tau) =- \big(\Psi^{-1} G \big) (t-\tau)$ for $t > \tau$ and $\ACER(t;\tau)=0$ for $t \leq \tau$, with $G(\nu) $ defined in \eqref{eqn:generalized_wald}. 
\end{proposition}

In practice, the function $g(\cdot)$ is often interpreted as the effect of an event in $N_i$ on the outcome $Y_i$ conditional on the history up to time $t$.
Proposition~\ref{prop:model1}(a) expresses the ACER in terms of the model parameters, showing that
$g(\cdot)$ and $\omega(\cdot)$ jointly characterize the $\ACER$ of $N_i$ on $Y_i$. When the dependence on past $Y_i$ does not exist, i.e., $\omega(\cdot)\equiv 0$, we have $\ACER(t;\tau)= -g(t-\tau)$. From Proposition~\ref{prop:model1}(a), we can obtain the ACER if we can estimate the model parameters in~\eqref{eqn:model1}. However, this requires specifying the distribution of $U_i(\cdot)$.
Fortunately, Proposition~\ref{prop:model1}(b) shows that we can identify the ACER without any distributional assumption on $U_i$ when a binary instrumental variable is available, and hence estimate it using the method in \S\ref{sec::estimation}. It broadens the applicability of Model~\eqref{eqn:model1} with fewer parametric assumptions. 

Proposition~\ref{prop:model1}(b) is an application of Theorem~\ref{thm::identification} under Model~\eqref{eqn:model1}.
The linearity in Model~\eqref{eqn:model1} plays a key role in the causal interpretation of the model parameters and nonparametric identification of the ACER.
The linear terms of $N_i$ and $Y_i$ connect the ACER with $g(\cdot)$ and $\omega(\cdot)$, and the linear term of $U_i$ implies Assumption~\ref{asm::stationarity} and the condition in~\eqref{eqn::condition}.

\subsection{Identification of causal effects with nonlinear additive unmeasured confounding}

We now consider the following nonlinear model that is similar to models in survival analysis with an instrument \citep[e.g.,][]{mackenzie2014using,li2015instrumental,tchetgen2015instrumental}:
\begin{eqnarray}
\label{eqn:model2}
\lambda_{Y}\left(t\right) = \phi\left\{ \mu_Y+\int_{0}^{t} g(t-s) \d N_i(s)\right\}+ \psi_{U_i}(t).
\end{eqnarray}
Model~\eqref{eqn:model2} generalises Model~\eqref{eqn:model1} by allowing for a nonlinear relationship between $N_i$ and $Y_i$ through the link function $\phi$ while requiring the unmeasured confounding effect to be additive. For Model~\eqref{eqn:model2}, the following proposition characterizes the causal effect and its identifiability.
\begin{proposition}
	\label{prop:model2}
	Suppose that Assumptions~\ref{asm::sutva}--\ref{asm::random}~and~\ref{asmp:ignorability-self-latent} hold, $N_i$ is a single-point process, and the underlying outcome model satisfies~\eqref{eqn:model2}. (a) We have, for $t, \tau \in [0,T]$,
		\begin{eqnarray*}
		\label{eqn::model2-acer}
		\ACER(t;\tau)\ =\ \phi(\mu_Y) - \phi\{\mu_Y+g(t-\tau)\}.
	\end{eqnarray*}
 (b) When $ (\Psi f)(\nu) \neq 0$ for all $\nu \in \mathbb{R}$, \ACER{} is identified by $\ACER(t; \tau) =- \big(\Psi^{-1} G \big) (t-\tau)$ for $t > \tau$ and $\ACER(t;\tau)=0$ for $t \leq \tau$, where $G(\nu) $ is defined in \eqref{eqn:generalized_wald}. 
\end{proposition}
From Proposition~\ref{prop:model2}(a), the causal interpretation of $g(\cdot)$ depends on $\mu_Y$ and the link function $\phi(\cdot)$ under Model~\eqref{eqn:model2}.
As a result, even with the same link function, the interpretation of $g(\cdot)$ differs in populations with different values of $\mu_Y$. This warns us of interpreting $g(\cdot)$ as some causal effects.
 Proposition~\ref{prop:model2}(b) is an application of Theorem~\ref{thm::identification}. Similar to Model~\eqref{eqn:model1},
 we can use the method in \S\ref{sec::estimation} to estimate the ACER without the knowledge of $\phi(\cdot)$ or $U_i$. Therefore, Proposition~\ref{prop:model2} suggests directly targeting the ACER instead of the model parameter $g(\cdot)$. This circumvents the daunting task to identify, estimate, and interpret the model parameters in Model~\eqref{eqn:model2}, broadening its applicability with fewer parametric assumptions. 
 
Critically, although Model \eqref{eqn:model2} allows for nonlinearity, it restricts the effect of unmeasured confounder to be additive. Relaxing this modeling assumption is challenging. 
In \S\ref{sec::difficulty-nonlinear-nonadditive-model}
 in the supplementary material, we show that when the confounding effect on $Y_i$ is non-additive, the ACER would depend on the distribution of the confounder, making the identification not possible without a distributional assumption on $U_i$.

\section{Numerical analysis}\label{sec:num}

\subsection{Simulation}\label{sec:sim}

We use simulation to illustrate the numerical performance of the proposed nonparametric estimation procedure. In this simulation study, we generate the treatment $N_i$ and outcome $Y_i$ from the following model
\begin{eqnarray}
\label{eqn:N-sim} \lambda_N(t) & =& \mu_N + \phi_{\beta_0}\big\{\alpha(t; a_N,b_N)Z_i + U_i(t)\big\}, \\
\label{eqn:Y-sim} \lambda_Y(t) & =& \phi_{\beta_2}\left[ \phi_{\beta_1}\left\{ \mu_Y +\int_0^t \alpha(\Delta; a_Y,b_Y) \d N_i(t-\Delta)\right\} +\phi_{\beta_1}\left\{ U_i(t-d_U)\right\} \right],
\end{eqnarray}
where $\alpha(\cdot; a,b)=b a^2 t \exp(-at )$ is the \emph{alpha function} (see, among others, Chapter 7 in \citealp{ermentrout2010mathematical}) and $\phi_{\beta}(x)=x^{\beta}$ is the link function. The confounding variable $U_i$ is generated as a Gaussian process with mean zero and a squared exponential kernel $\cov\{U_i(t), U_i(t+d)\}=\sigma^2_U \exp\{- d^2/(2l_U^2)\}$. The parameters in \eqref{eqn:N-sim} and \eqref{eqn:Y-sim} are set as $\mu_N=\mu_Y=0.2$, $a_N=10, b_N=0.5$, $a_Y=8,b_Y=1$, $d_U=0.5$, $\sigma_U=0.2$, and $l_U=0.1$.
We consider five scenarios in this simulation.\\
\noindent \textbf{Scenario 1a $(\beta_0=\beta_1=\beta_2=1)$:} A linear model  for $Y_i$ with a  single-point process $N_i$, which is achieved by suppressing the intensity \eqref{eqn:N-sim} to zero after the first event in $N_i$ is generated.

\noindent \textbf{Scenario 1b $(\beta_1=2, \beta_0=\beta_2=1)$:} An additive confounding model  for $Y_i$ with a single-point process $N_i$.

\noindent \textbf{Scenario 2a $(\beta_0=\beta_1=\beta_2=1)$:} A linear model  for $Y_i$ with multiple events in $N_i$.

\noindent \textbf{Scenario 2b $(\beta_0=3,\beta_1=2, \beta_2=1)$:} A linear model  for $Y_i$ with multiple events in $N_i$ and non-additive confounding effects on $N_i$.

\noindent \textbf{Scenario 3  $(\beta_0=\beta_1=1, \beta_2=3)$:} A non-additive confounding model for $Y_i$. 

For each scenario, we generate $m$ trials with $m$ ranging from $40$ to $800$. In each simulated dataset, half of the trials are set to have $Z_i=1$ and the other half $Z_i=0$. The processes $N_i$ and $Y_i$ are generated from $0$ to $T=3$ using thinning process, while the unmeasured confounding process $U_i$ is generated from $-1$ to $3$ to account for its delayed effect on $Y_i$.

For Scenario 3, the identification of ACER is difficult with nonlinear confounding effects, as illustrated in the supplementary material. So we use the Monte Carlo method to calculate the ACER  to show its dependence on the distribution of the unmeasured confounder. For Scenarios 1a to 2b, we apply the proposed generalised Wald estimation procedure in \S\ref{sec::estimation}.  We estimate the function $h(t)$ as the difference between the empirical cumulative intensities of $Y_i$ in the treatment group $(Z_i=1)$ and control group $(Z_i=0)$. The function $f(t)$ is estimated in a similar manner. We approximate the $\ACER$ using a cubic B-splines with 6 knots evenly-spaced in $[0,1]$ where the mass of $\alpha(\cdot;a_Y,b_Y)$ resides. The tuning parameter of the ridge penalty $\eta$ is set to be $m^{-1}$ to reduce boundary effects from the nonparametric approximation. To measure the performance, we calculate the proportion of integrated squared errors with respect to the true \ACER, that is
\begin{equation}\label{eqn:criterion}
r\ =\ \frac{\int_0^1 \left\{ \widehat{\ACER}(\Delta)-\ACER(\Delta;0) \right\}^2 \d \Delta }{ \int_0^1 \ACER^2(\Delta;0) \d \Delta},
\end{equation}
where the true $\ACER$ is calculated from \eqref{eqn:Y-sim} based on Propositions~\ref{prop:model1}~and~\ref{prop:model2}.  

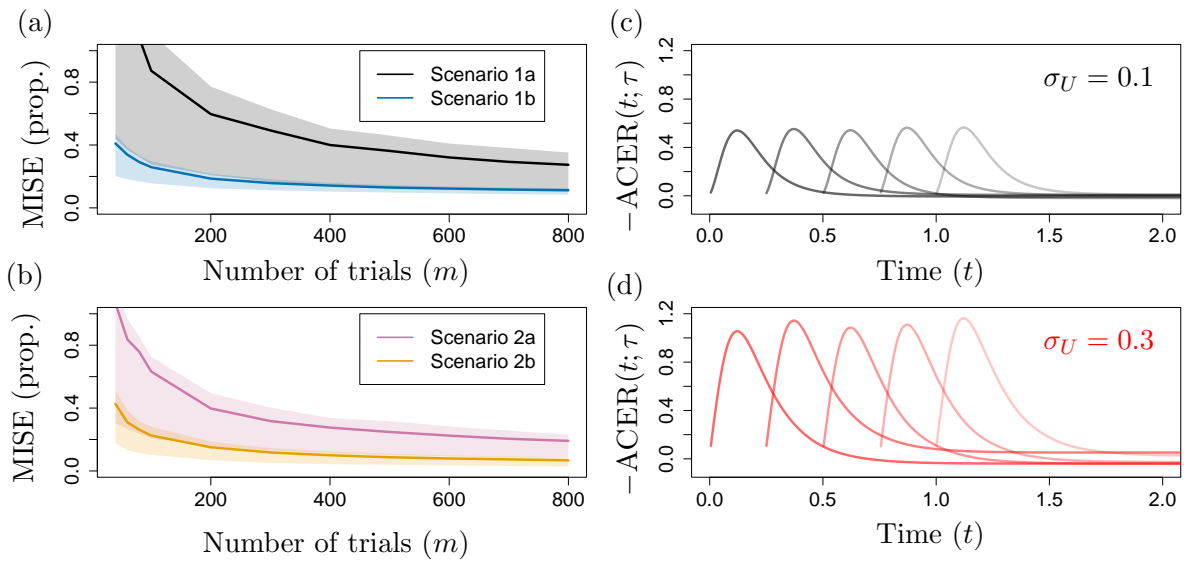
\begin{figure}[ht]
 \begin{tikzpicture}

\node[inner sep=0pt] (sim) at (-7.4,0)
{\includegraphics[width=.45\textwidth]{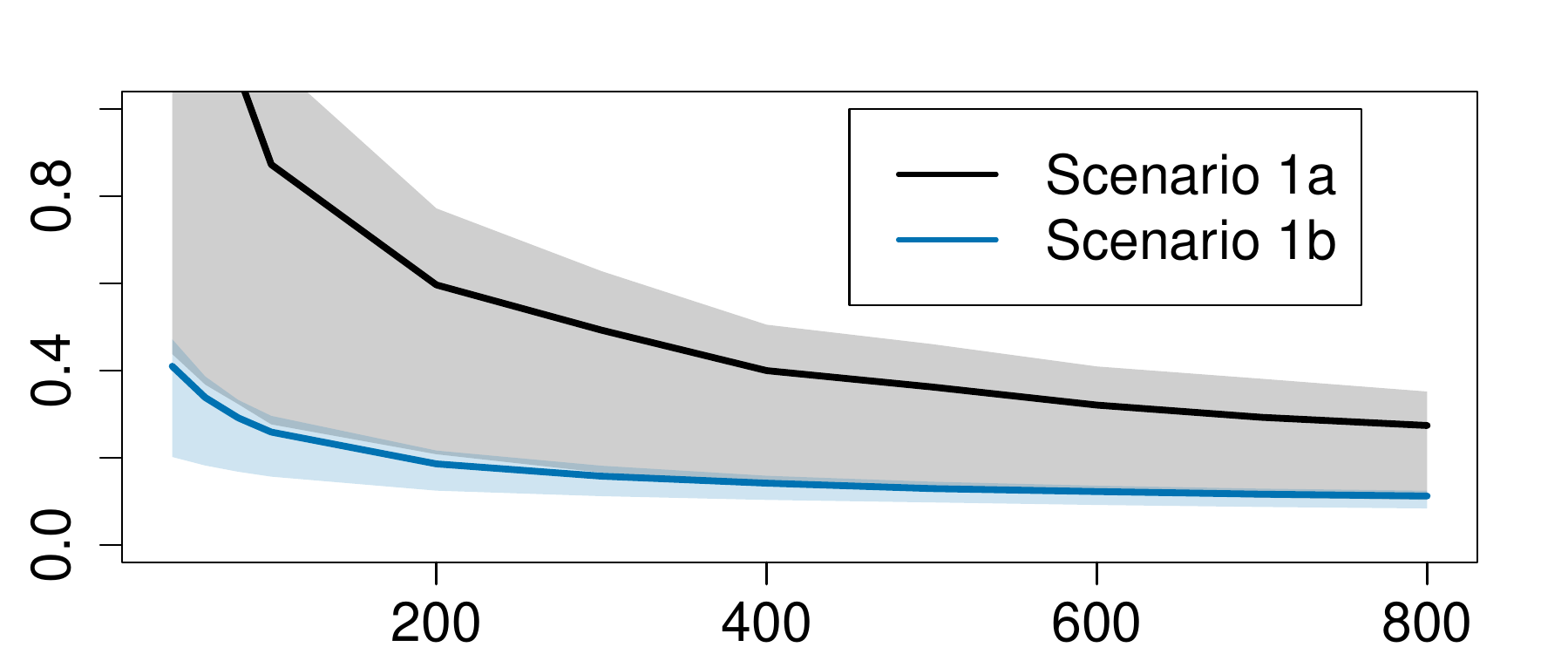}};
\node[below = 0 cm of sim](sim.lab)
{Number of trials ($m$)};
\node[above left = -0.5 cm and -0.1 cm of sim, rotate=90] () {MISE (prop.)};
\node[above left = -0.5 cm and -0.15 cm of sim] () {(a)};

\node (sim2) at (-7.4,-3.5)
{\includegraphics[width=.45\textwidth]{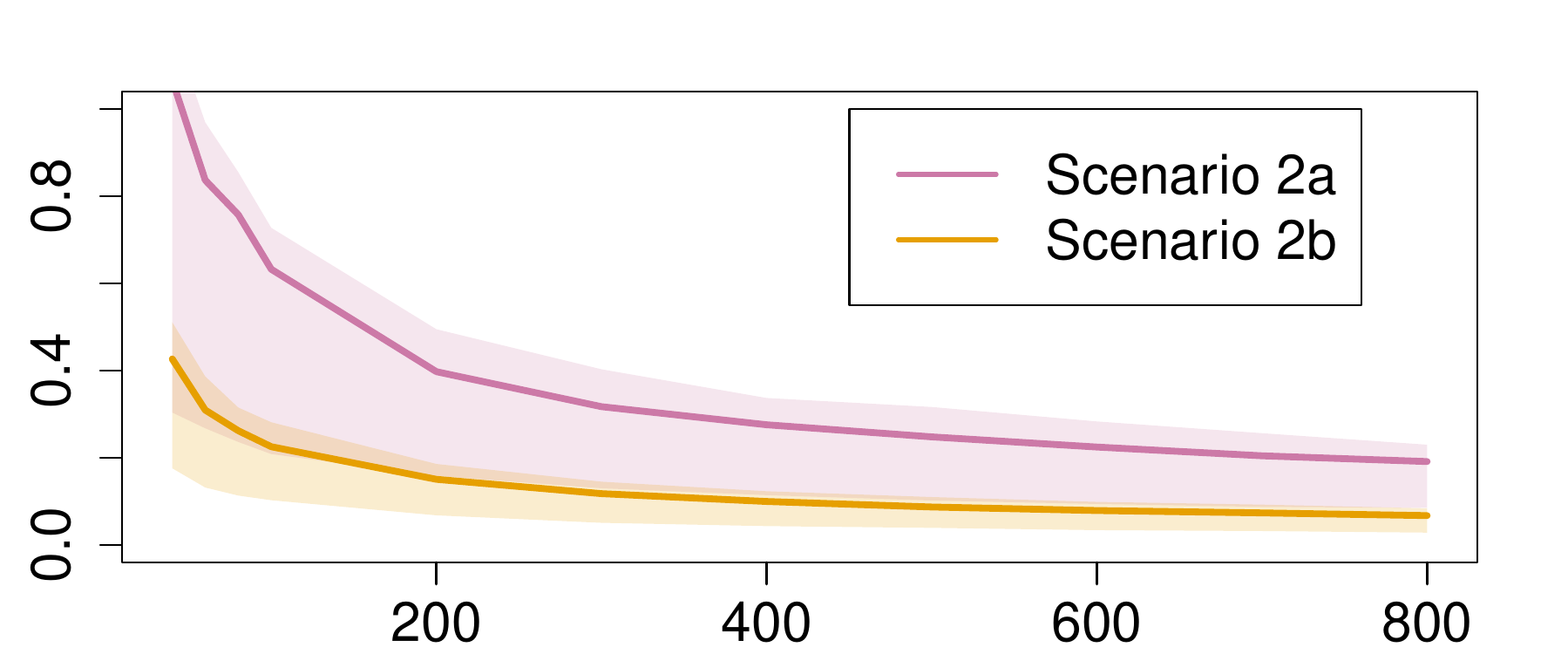}};
\node[below = 0 cm of sim2](sim2.lab)
{Number of trials ($m$)};
\node[above left = -0.5 cm and -0.1 cm of sim2, rotate=90] () {MISE (prop.)};
\node[above left = -0.5 cm and -0.15 cm of sim2] () {(b)};

\node[inner sep=0pt] (ACER) at (0.5,0)
 {\includegraphics[width=.45\textwidth]{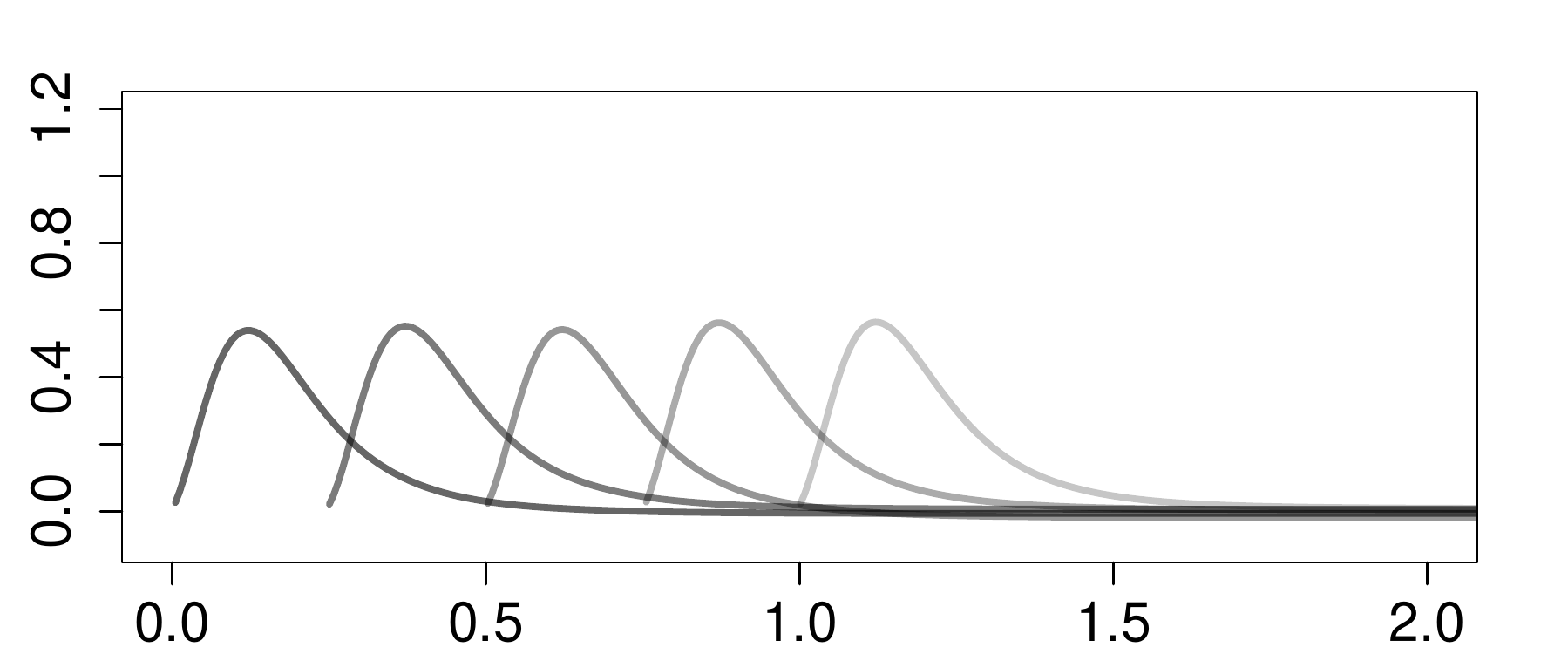}};
\node[below = 0 cm of ACER](){Time ($t$)};
\node[above left = -0.5 cm and -0.1 cm of ACER, rotate=90] () {$-\ACER(t;\tau)$ };
\node[above right = -1.2 cm and -2.4 cm of ACER] () {$\sigma_{U}=0.1$ };

\node[above left = -0.5 cm and -0.1 cm of ACER] () {(c) };

\node[inner sep=0pt] (ACER2) at (0.5,-3.5)
{\includegraphics[width=.45\textwidth]{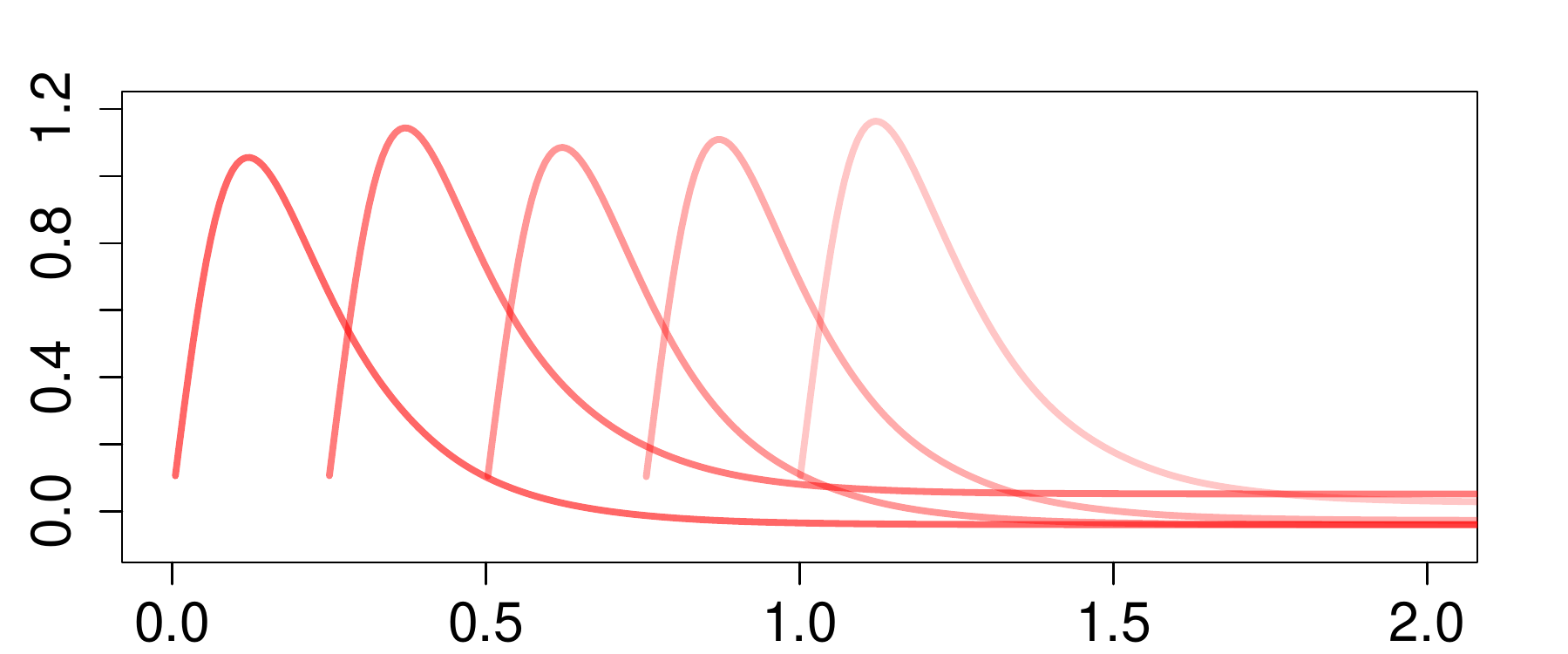}};
\node[below = 0 cm of ACER2](ACER2.lab){Time ($t$)};
\node[above left = -0.5 cm and -0.1 cm of ACER2, rotate=90] () {$-\ACER(t;\tau)$ };
\node[above right = -1.2 cm and -2.4 cm of ACER2] () {\textcolor{red}{$\sigma_{U}=0.3$} };
\node[above left = -0.5 cm and -0.1 cm of ACER2] () {(d) };

\end{tikzpicture}
\caption{Identification of \ACER\ and performance of the generalised Wald estimation averaged across 1000 replicates. Panels (a) and (b) illustrates the performance of the proposed estimation procedures in Scenarios 1a, 1b (Panel a),  2a, and 2b (Panel b). The x-axes are the numbers of trials ($m$), and the y-axes are the measure defined in \eqref{eqn:criterion}.  The shaded areas are the interquartile bands from the 1000 replicates. Estimation performances in the four scenarios are not directly comparable given the huge difference between the data generating mechanisms.
Based on the interpretation in~\eqref{eqn::ACER_new}, Panels (b) and (c) show the true value of $-\ACER(t;\tau)$ with non-additive confounding effects on $Y$ (Scenario 3). The spectrum of gray curves is calculated with $\sigma_U=0.1$, and the red curves with $\sigma_U=0.3$. Within each spectrum, the five curves correspond to $\tau$ being $0,0.25,0.5,0.75$ and $1$, respectively. Curves in each color spectrum are not shift-invariant. }
\label{fig:sim}
\end{figure}

Figure~\ref{fig:sim} shows the simulation results averaged over 1000 replicates. Figures~\ref{fig:sim}(a) and (b) show that the performances of estimators improve as the number of trials increases in Scenarios~1a, 1b, 2a, and 2b.  In particular, the proportion of integrated squared error in Scenario 1a is larger than that in Scenario 2a, despite having the same model for $Y_i$.
This reveals a feature for point process treatments that, given the same amount of trials, more events in $N_i$ contribute more information for recovering the causal effects of $N_i$ on $Y_i$. The four curves in Figures~\ref{fig:sim}(a)~and~(b) converge slowly towards zero due to the existence of approximation error in the basis expansion and the bias from penalization. 
Figures~\ref{fig:sim}(c)~and~(d) show the calculated true ACER in Scenario 3 under two different distributions of $U_i$. Within each of Figures~\ref{fig:sim}(c)~and~(d), the five curves correspond to $\tau$ being $0,0.25,0.5,0.75$, and $1$.
Even with the same $t$, the shape of $\ACER(t;\tau)$ varies as $\tau$ changes in both figures, implying that $\ACER(t;\tau)$ does not equal to $\ACER(t-\tau)$. Moreover, the contrast between Figure~\ref{fig:sim}(c) and (d) shows that \ACER \ depends on the distribution of the unmeasured process $U_i$, demonstrating the sensitivity of the ACER to distributional assumptions on $U_i$.  

\subsection{Empirical analysis}\label{sec:RDA}

\begin{figure}[ht]

	\begin{tikzpicture}[node distance=1.3cm,>=stealth',bend angle=45,auto]
	\tikzstyle{iv}=[draw=black!20,fill=black!20,minimum size=8mm]
	\tikzstyle{latent}=[circle,draw=black,fill=black!20,minimum size=10mm,dashed]
	\tikzstyle{obs}=[circle,draw=black!20,	fill=black!20,minimum size=10mm]

	\begin{scope}
	\node [obs] (N){OB};
	\node [iv] (Z) [left = 1 cm of N] {Light}
	edge [->, thick, color=black, thick] (N);
	\node [latent] (U) [ below right = 1.2 cm and 0.8 cm of N] {$U$}
	edge [->, color=black, thick,dashed]		(N);
	\node [obs] [ right = 2 cm of N] (Y){PN}
	edge [	<-,	 color=black, thick]		(N)
	edge [<-, color=black, thick, dashed]		(U);

	\node[above right = -0.4 cm and 0.3 cm of N] {$\ACER$};

	\node [fit={(N) (Z) (U) (Y)}] (DAG) {};

	\node[inner sep=0pt] [above = 0.5 cm of DAG] (Trial){\includegraphics[width=.45\textwidth]{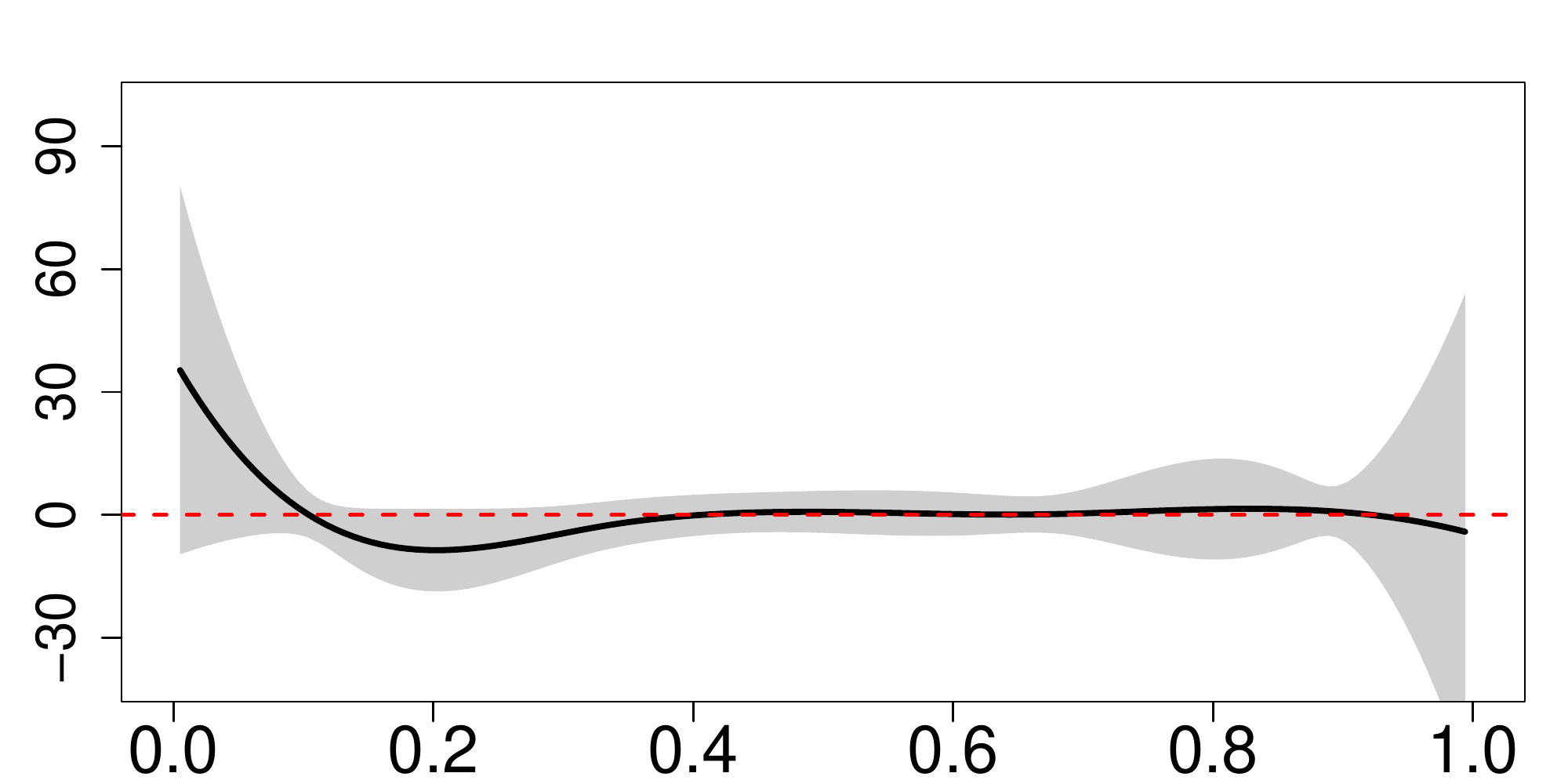}};

	\node[inner sep=0pt] [left = 0.6 cm of Trial] (Ns){\includegraphics[width=.45\textwidth]{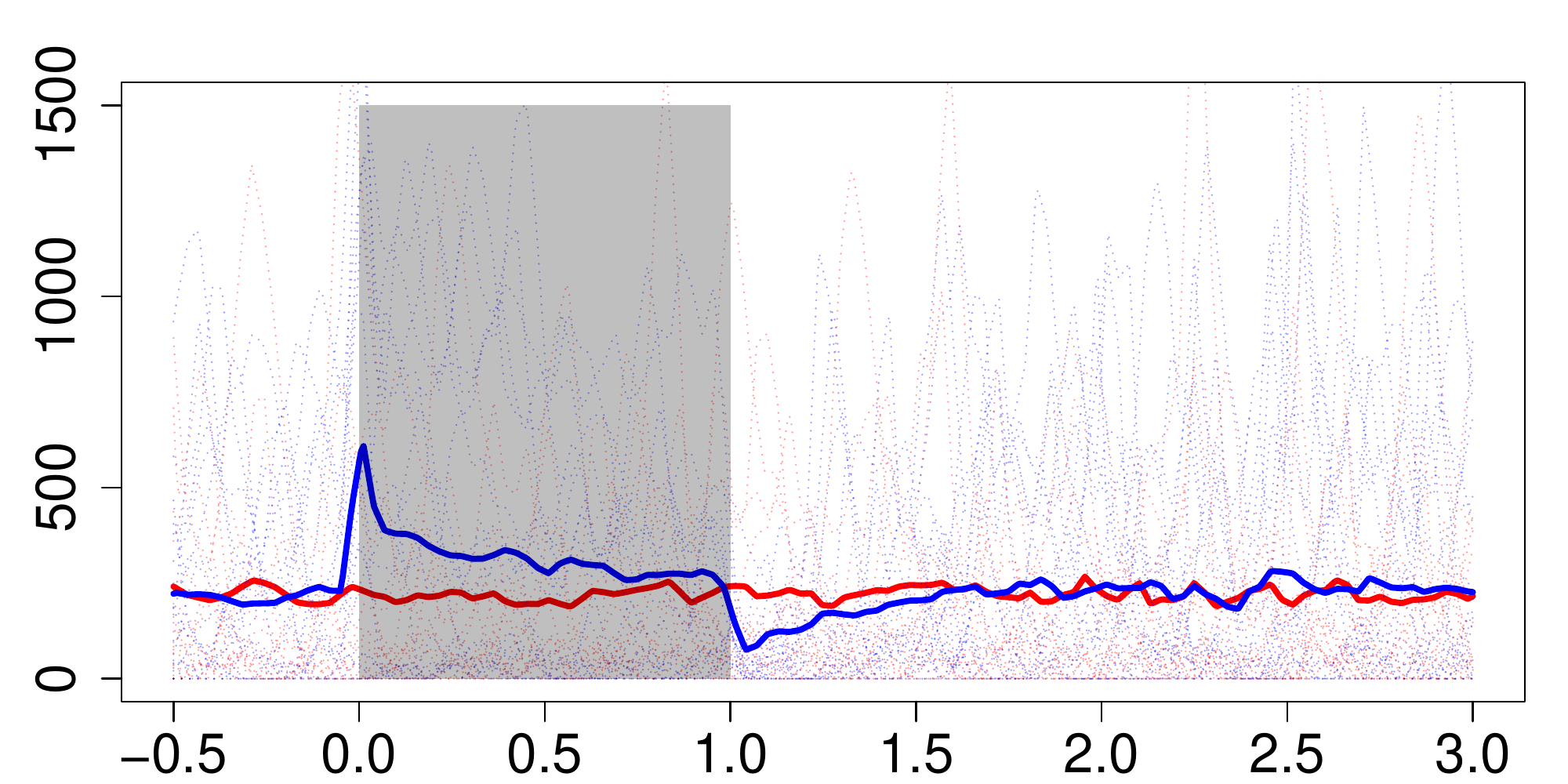}};

	\node[inner sep=0pt] [below = 0.2 cm of Ns] (Ys){\includegraphics[width=.45\textwidth]{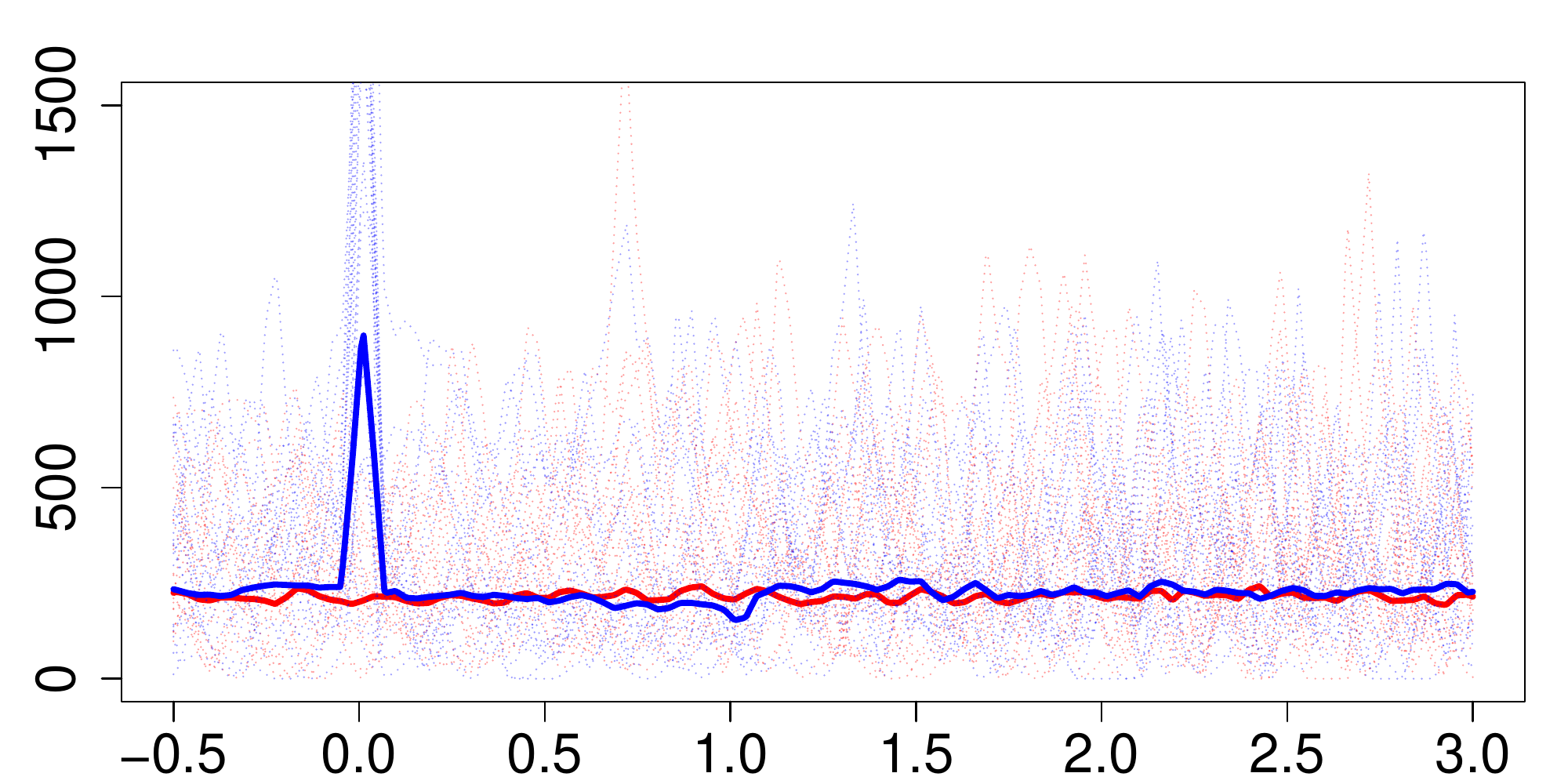}};
	
	\node[above left = -0.7 cm and -0.1cm of Ns] () {(a)};
	\node[above left = -1.2 cm and -0.05 cm of Ns, rotate=90] () {Trials};
	
	\node[above left = -0.7 cm and -0.1 cm of Ys] () {(b)};
	\node[below = -0.1 cm of Ys] (Yaxis) {Time (in seconds)};
	\node[above left = -1.2 cm and -0.05 cm of Ys, rotate=90] () {Trials};

	\node[below = -0.1 cm of Trial] (Taxis) {$\Delta$ (in seconds)};
	\node[above left = -0.7 cm and -0.1 cm of Trial, rotate=90] () {\small $-\ACER(\Delta;0)$};
	\node[above left = -0.7 cm and -0.1 cm of Trial](Tlab) {(c)};

			\node[below = -0.1 cm of Ns] (Naxis) {Time (in seconds)};
	\node[below = 3.35 cm of Tlab](DAGlab)
			{(d)};
	\end{scope}

	\end{tikzpicture}
	\caption{Empirical intensities and fitted $\ACER$ on data from \cite{boldingfranks2018data}.  Panels (a) and (b) show the empirical intensities of the neural activities of OB (Panel a) and PN (Panel b) in the stimulated (blue) and unstimulated (red) groups. The solid curves represent the average intensity over 80 trials, and the dashed curves demonstrate the empirical intensity from 20 randomly selected trials in each group. The shaded area in Panel (a) represents the duration of the light pulse. Based on the interpretation in~\eqref{eqn::ACER_new}, Panel (c) shows the estimated $-\ACER(\Delta;0)$ from the full data set. The shaded area represents a $90\%$ confidence band for visualizing the uncertainty of the estimates from 5000 bootstrap samples. 
Panel (d) shows the causal diagram for the relationship among the variables.}
	\label{fig:RDA}
\end{figure}
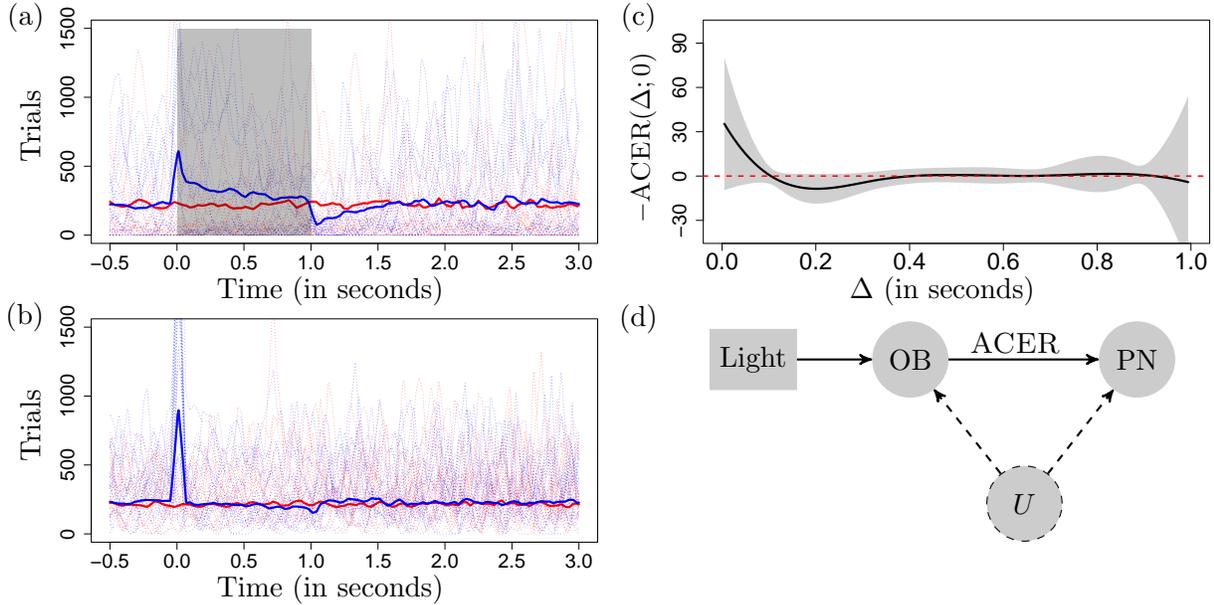

We now apply the proposed methodology to the neural data from \citet{boldingfranks2018}. We provide the basic background to help understand the experiment. For more details, see the supplementary material and \citet{boldingfranks2018}. \citet{boldingfranks2018} conduct an experiment to understand how a mouse brain maintains stationarity in odor detection regardless of odor concentration. To be specific, it is known that neural activities in the olfactory bulb (OB) increase in response to a higher concentration of odor particles, and that a spike in OB triggers neural activities of principal neurons (PN) in the piriform cortex, where the odor is perceived by the brain.
 To avoid other neural processes that normalize odor responses, \citet{boldingfranks2018} use optogenetics to stimulate neurons in OB with 1-s light pulses, which meet the requirements as an instrumental variable in our methodology. \cite{boldingfranks2018} also take an optogenetic  to circumventing the contribution of centrifugal inputs and other intrabulbar processes, which effectively cuts of the feedback from PN to OB.
 Figure~\ref{fig:RDA}(d) shows a causal diagram for the relationship among the stimulation, OB, and PN.

 The dataset contains spike trains recorded in OB and PN during the experiment. A total of 160 trials are conducted on 8 mice, where each mouse has 10 trials without stimulation ($Z_i=0$) and 10 trials with a one-second light pulse at 20 ${\rm mW/mm}^2$ ($Z_i=1$). The light pulse, if present, onsets at time $0$ and ends at $1$s. In our analysis, we consider the first 3.5 seconds of a trial, from $-0.5$ to $3$, as there are hardly any residual effects afterward.
 We consider the treatment $N_i$ as the process of events in OB, and the outcome $Y_i$ as the process of events in PN. Each recorded event in $N_i$ is a spike of one neuron in OB that may instigate a distinct group of PN in the piriform cortex.
  Given the vast amount of PN in the piriform cortex, the instigated groups may share few or no overlaps, limiting the interactive effect of the treatment process. Therefore,  the additivity in Assumption~\ref{asmp:additivity} is a plausible approximation to the true underlying mechanism. Figures~\ref{fig:RDA}(a) and (b) show the smoothed intensities of neural activities in the stimulated (blue) and unstimulated (red) groups. We can see that the stimulation triggers increased activities in OB in all trials. However, there are large variations in the neural activities across trials.

We first conduct a preliminary analysis assuming no unmeasured confounders between the treatment and outcome processes. In this case, the identification of the causal effects does not require the instrumental variable. We fit a model of $Y_i$ on $N_i$ and directly interpret the coefficient function of $N_i$ as the causal effect. However, the conclusion is inconsistent with the findings in \cite{boldingfranks2018}, implying possible unmeasured confounders or model misspecification.  See more details in \S\ref{sec:RDA-obs} in the supplementary material.

We then apply the generalised Wald estimation procedure to estimate the causal effects of neural activities in OB on the neural activities of PN.
From the observed data, we estimate the functions $h$ and $f$ using differences between empirical cumulative intensities in the stimulated ($Z_i=1$) and unstimulated ($Z_i=0$) groups. We approximate the unknown $\ACER$ using cubic B-splines with evenly-spaced knots in $[0,1]$, where two knots are selected by a 5-fold cross-validation. We set the tuning parameter for the ridge penalty to be $\eta=0.01$ to handle boundary effects. A $90\%$ confidence band is constructed using the bootstrap with details in the supplementary material. We use the bootstrap confidence band to approximate the uncertainty in the estimates. 

Figure~\ref{fig:RDA}(c) shows the estimated ACER. The curve shows that an event in OB elicits high activities in PN immediately after the event ($<0.1$ seconds), but the effect quickly turns negative for an extended duration (between 0.1 to 0.4 seconds) before it dies down. This is consistent with the findings in \cite{boldingfranks2018} that a temporal mechanism is in place to stabilize the neural activities of PN after the initial detection of odors. Additional analysis of the neural dataset can be found in \S\ref{sec:num-supp} in the supplementary material. The confidence band shows that the proposed generalised Wald estimation procedure yields high uncertainty near the boundaries, despite a large number of events and the ridge penalty. In this particular case, it appears that the \ACER\ vanishes after 0.5 seconds, but the boundary effect causes spurious estimates in the bootstrap samples. In practice, we recommend practitioners applying the generalised Wald estimation procedure, and then applying a suitable parametric form or shape constraint on the \ACER.

\pdfbookmark[1]{References}{References}
\spacingset{1.45}
\bibliographystyle{Chicago}
\bibliography{reference}

\newpage

\appendix

\setcounter{lemma}{0}
\renewcommand {\thelemma} {S\arabic{lemma}}
\setcounter{equation}{0}
\renewcommand {\theequation} {S\arabic{equation}}
\setcounter{section}{0}
\renewcommand\thesection{S\arabic{section}}
\renewcommand\thesubsection{S\arabic{section}.\arabic{subsection}}

\begin{center}
  \LARGE {\bf Supplementary Material}
\end{center}

\S\ref{sec:discrete} generalises the framework to discrete instrumental variables.

\S\ref{sec:proofs} provides the proofs of the theorems, propositions, and claims in the main text. 

\S\ref{sec:num-supp} provides more details about the empirical analysis.

Let  $1 (\cdot)$ denote the indicator function. Recall $\text{i} = \sqrt{-1}$.

\section{Generalisation to discrete instrumental variables}\label{sec:discrete}

In the main text, we discuss the case when the instrumental variable takes binary values. Here we briefly outline the generalisation to the case with discrete instrumental variables where there could be multiple levels of $Z$.  For a discrete instrumental variable, we can apply the proposed methodology by comparing two levels $z$ and $z'$ (instead of $0$ and $1$ in the binary instrumental variable case) under the monotonicity and homogeneity assumptions with respect to these two levels.
For simplicity, we only give the result with a single-point process $N_i$.
Define
\begin{eqnarray*}
\ACE_Y(t;z,z') & =& E\{ Y_{iz}(t)-Y_{iz'}(t) \}, \quad  \ACE_N(t;z,z')\ =\ E\{ N_{iz}(t)-N_{iz'}(t) \}.
\end{eqnarray*}
 We can obtain a similar result as Theorem 1 with respect to $z$ and $z'$:
\begin{eqnarray*}
\nonumber\ACE_Y(t;z,z') &=& \int_{0}^T { E \{ \partial Y_{i\tau}(t)/ \partial \tau \mid \mT_{iz'} \leq \tau < \mT_{iz} \}} \Pr (\mT_{iz'} \leq \tau < \mT_{iz} )\d \tau\\
 &&- \int_{0}^T { E \{ \partial Y_{i\tau}(t)/\partial \tau \mid \mT_{iz} \leq \tau < \mT_{iz'} \}} \Pr (\mT_{iz} \leq \tau < \mT_{iz'} )\d \tau,
\end{eqnarray*}
which, under monotonicity $\mathcal{T}_{iz} \leq \mathcal{T}_{iz'}$, simplifies as 
\begin{eqnarray*}
\ACE_Y(t;z,z') &=& - \int_{0}^T E \{ \partial Y_{i\tau}(t)/\partial \tau \mid \mT_{iz} \leq \tau < \mT_{iz'} \}\cdot \ACE_N(\tau;z,z') \d \tau.
\end{eqnarray*}
Define 
\begin{eqnarray*}
h^{z,z’}(t) &=& E\{Y_i(t)\mid Z_i=z\}-E [Y_i(t)\mid Z_i=z'\},\\
f^{z,z’}(t) &=& E\{N_i(t)\mid Z_i=z\}-E \{N_i(t)\mid Z_i=z'\}.
\end{eqnarray*}
Under randomization (Assumption~\ref{asm::random}) and stationarity (Assumption~\ref{asm::stationarity}), if 
\begin{eqnarray*}
E  \left\{ \frac{ \partial Y_{i\tau}(t)}{\partial \tau} \ \bigg | \ \mT_{iz} \leq \tau < \mT_{iz'} \right\}\ =\ E  \left\{ \frac{ \partial Y_{i\tau}(t)}{\partial \tau} \ \bigg |\ \mT_{iz'} \leq \tau < \mT_{iz} \right\}\ =\ \frac{ \partial E  \left\{ Y_{i\tau}(t)\right\}}{\partial \tau},
\end{eqnarray*}
then
$$
h^{z,z'}(t)\ =\ -\int_0^T \mathrm{ACER}(t-\tau;0)f^{z,z'}(\tau)\mathrm{d} \tau.
$$
This is similar to Theorem 2 with respect to $z$ and $z'$. Therefore, 
if $(\Psi f^{z,z'}) (\nu) \neq 0$ for all $\nu \in \mathbb{R}$, then the $\ACER$ is identified by
$$
\mathrm{ACER}(t;\tau)=-\Psi^{-1}(G)(t-\tau) \ \text{with} \ G(\nu)=\frac{\big( \Psi h^{z,z'}\big)(\nu)}{\big( \Psi f^{z,z'}\big)(\nu)}.
$$
For estimation, we can use the data with $Z_i=z,z'$ and apply the same method as in the binary instrumental variable case.

\section{Proofs}\label{sec:proofs}

\subsection{Proof of Theorem~\ref{thm:ivformula}}\label{sec:ivformula_proof}
First, we have
\begin{eqnarray}
\label{eqn::h-po} \ACE_Y(t)\ =\ E \{Y_{i1,\mT_{i1}}(t) \} - E \{Y_{i0,\mT_{i0}}(t) \}\ =\ E \{Y_{i,\mT_{i1}}(t) \} - E \{Y_{i,\mT_{i0}}(t) \},
\end{eqnarray}
where the second equality follows from Assumption~\ref{asm::er}.

Using the Stieltjes integral, we can write
\begin{eqnarray}
\nonumber
Y_{i,\mT_{i1}}(t) &=& \int_{[0,T]} Y_{i\tau}(t) \d N_{i1}(\tau) + Y_{iT+}(t) 1 {\{N_{i1}(T)=0\}}\\
\label{eqn:Stieltjes1} &=& \int_{[0,T]} Y_{i\tau}(t) \d N_{i1}(\tau) + Y_{iT}(t) 1 {\{N_{i1}(T)=0\}},
\end{eqnarray}
where the second equality follows from Assumption~\ref{asm::no-anticipation}.
Similarly, we have
\begin{eqnarray}
\label{eqn:Stieltjes0} Y_{i,\mT_{i0}}(t) &=& \int_{[0,T]} Y_{i\tau}(t) \d N_{i0}(\tau) + Y_{iT}(t) 1 {\{N_{i0}(T)=0\}}.
\end{eqnarray}
Plugging~\eqref{eqn:Stieltjes1}~and~\eqref{eqn:Stieltjes0} into~\eqref{eqn::h-po} yields
\begin{eqnarray}
\nonumber && \ACE_Y(t) \\
\nonumber &=& E \left\{ \int_{[0,T]} Y_{i\tau}(t) \d N_{i1}(\tau) + Y_{iT}(t) 1 {\{N_{i1}(T)=0\}}\right\}\\
\nonumber &&- E \left\{ \int_{[0,T]} Y_{i\tau}(t) \d N_{i0}(\tau) + Y_{iT}(t) 1 {\{N_{i0}(T)=0\}}\right\} \\
\nonumber &= & E \left [ \int_{[0,T]} Y_{i\tau}(t) \{\d N_{i1}(\tau) -\d N_{i0}(\tau)\} \right] + E  \left[ Y_{iT}(t) 1 \{N_{i1}(T)=0\} - 1 \{N_{i0}(T)=0\} \right] \\
\label{eqn::h-po-decomposition} &= & E \left [ \int_{[0,T]} Y_{i\tau}(t) \{\d N_{i1}(\tau) -\d N_{i0}(\tau)\} \right] + E  \left[ Y_{iT}(t) \{N_{i0}(T)- N_{i1}(T)\} \right],
\end{eqnarray}
where the last equality holds because $N_{i0}$ and $N_{i1}$ are single-point processes.

We then focus on the first term of~\eqref{eqn::h-po-decomposition}.
Let $\mathcal{H}^{N}_{izt}= \{N_{iz}(s): s \in [0,t]\}$ denote the history of $N_{iz}$ up to time $t$ for $z=0,1$.
By switching the order of the expectation and integral, we have
\begin{eqnarray}
\nonumber &&E \left [ \int_{[0,T]} Y_{i\tau}(t) \{\d N_{i1}(\tau) -\d N_{i0}(\tau)\} \right] \\
 \nonumber&=& \int_{[0,T]} E \left[ Y_{i\tau}(t) \{\d N_{i1}(\tau) -\d N_{i0}(\tau)\} \right]\\
 \nonumber&=& \int_{[0,T]} E \left[ E \left\{ Y_{i\tau}(t) \mid \mathcal{H}^{N}_{i1T}, \mathcal{H}^{N}_{i0T}\right\} \d \{N_{i1}(\tau) - N_{i0}(\tau)\}\right]\\
 \nonumber&=& E \left[\int_{[0,T]} E \left\{ Y_{i\tau}(t) \mid \mathcal{H}^{N}_{i1T}, \mathcal{H}^{N}_{i0T}\right\} \d \{N_{i1}(\tau) - N_{i0}(\tau)\}\right]\\
\nonumber &=& E \left[ E \left\{ Y_{i\tau}(t) \mid \mathcal{H}^{N}_{i1T}, \mathcal{H}^{N}_{i0T}\right\} \{N_{i1}(\tau) - N_{i0}(\tau)\}\Big |^{\tau=T}_{\tau=0} \right]\\
 \label{eqn::intergration-by-parts} &&- E \left[\int_{[0,T]} \frac{\partial E \left\{ Y_{i\tau}(t) \mid \mathcal{H}^{N}_{i1T}, \mathcal{H}^{N}_{i0T}\right\}}{\partial \tau} \{N_{i1}(\tau) - N_{i0}(\tau)\}\d \tau\right],
\end{eqnarray}
where the second equality follows from the law of total expectation and the last equality follows from integration by parts.
We can write the first term of~\eqref{eqn::intergration-by-parts} as
\begin{eqnarray*}
&& E \left[ E \left\{ Y_{i\tau}(t) \mid \mathcal{H}^{N}_{i1T}, \mathcal{H}^{N}_{i0T}\right\} \{N_{i1}(\tau) - N_{i0}(\tau)\}\Big |^{\tau=T}_{\tau=0} \right]\\
&=& E \left[ E \left\{ Y_{iT}(t) \mid \mathcal{H}^{N}_{i1T}, \mathcal{H}^{N}_{i0T}\right\} \{N_{i1}(T) - N_{i0}(T)\}-0\right]\\
&=& E \left[ Y_{iT}(t) \{N_{i1}(T) - N_{i0}(T)\}\right],
\end{eqnarray*}
where the first equality follows from $N_{i1}(0)-N_{i0}(0)=0$ and the second equality follows from the law of total expectation.
Therefore, the first term of~\eqref{eqn::h-po-decomposition} becomes
\begin{eqnarray}
\nonumber &&E \left [ \int_{[0,T]} Y_{i\tau}(t) \{\d N_{i1}(\tau) -\d N_{i0}(\tau)\} \right] \\
\nonumber &=& E \left[ Y_{iT}(t) \{N_{i1}(T) - N_{i0}(T)\}\right] \\
\label{eqn::intergration-by-parts2}&&- E \left[\int_{[0,T]} \frac{\partial E \left\{ Y_{i\tau}(t) \mid \mathcal{H}^{N}_{i1T}, \mathcal{H}^{N}_{i0T}\right\}}{\partial \tau} \{N_{i1}(\tau) - N_{i0}(\tau)\}\d \tau\right].
\end{eqnarray}
Plugging~\eqref{eqn::intergration-by-parts2} into~\eqref{eqn::h-po-decomposition}, we obtain
\begin{eqnarray}
\label{eqn::aceY-decomposition}
\ACE_Y(t)\ =\ - E \left[\int_{[0,T]} \frac{\partial E \left\{ Y_{i\tau}(t) \mid \mathcal{H}^{N}_{i1T}, \mathcal{H}^{N}_{i0T}\right\}}{\partial \tau} \{N_{i1}(\tau) - N_{i0}(\tau)\}\d \tau\right].
\end{eqnarray}
Because $N_i(\cdot)$ is a single-point process, $N_{i1}(\tau) - N_{i0}(\tau)$ takes values only in $\{-1,0,1\}$.
In addition, $N_{i1}(\tau) - N_{i0}(\tau)=1$ is equivalent to $\mT_{i1} \leq \tau < \mT_{i0}$ and $N_{i1}(\tau) - N_{i0}(\tau)=-1$ is equivalent to $\mT_{i0} \leq \tau < \mT_{i1}$. Therefore, by switching the order of the expectation, integral, and derivative in~\eqref{eqn::aceY-decomposition},
we have
\begin{eqnarray*}
\ACE_Y(t) &=& -\int_{[0,T]} E \left[ \frac{\partial E \left\{ Y_{i\tau}(t) \mid \mathcal{H}^{N}_{i1T}, \mathcal{H}^{N}_{i0T}\right\}}{\partial \tau}\ \Bigg |\ \mT_{i1} \leq \tau < \mT_{i0} \right] \Pr (\mT_{i1} \leq \tau < \mT_{i0}) \d \tau \\
&&+ \int_{[0,T]} E \left[ \frac{\partial E \left\{ Y_{i\tau}(t) \mid \mathcal{H}^{N}_{i1T}, \mathcal{H}^{N}_{i0T}\right\}}{\partial \tau}\ \Bigg |\ \mT_{i0} \leq \tau < \mT_{i1} \right] \Pr (\mT_{i0} \leq \tau < \mT_{i1}) \d \tau\\
&=& -\int_{[0,T]} E \left[ E \left\{ \frac{\partial Y_{i\tau}(t)}{\partial \tau} \ \bigg| \ \mathcal{H}^{N}_{i1T}, \mathcal{H}^{N}_{i0T}\right\}\ \Bigg |\ \mT_{i1} \leq \tau < \mT_{i0} \right] \Pr (\mT_{i1} \leq \tau < \mT_{i0}) \d \tau \\
&&+ \int_{[0,T]} E \left[E \left\{ \frac{\partial Y_{i\tau}(t)}{\partial \tau} \mid \mathcal{H}^{N}_{i1T}, \mathcal{H}^{N}_{i0T}\right\}\ \Bigg |\ \mT_{i0} \leq \tau < \mT_{i1} \right] \Pr (\mT_{i0} \leq \tau < \mT_{i1}) \d \tau\\
&=&-\int_{[0,T]} E \left\{ \frac{\partial Y_{i\tau}(t) }{\partial \tau} \ \bigg |\ \mT_{i1} \leq \tau < \mT_{i0} \right\} \Pr (\mT_{i1} \leq \tau < \mT_{i0}) \d \tau \\
&&+\int_{[0,T]} E \left\{ \frac{\partial Y_{i\tau}(t)}{\partial \tau}\ \bigg |\ \mT_{i0} \leq \tau < \mT_{i1} \right\} \Pr (\mT_{i0} \leq \tau < \mT_{i1}) \d \tau,
\end{eqnarray*}
 where the first equality follows from the law of total probability and the third equality follows from the law of total expectation. This proves~\eqref{eqn:ivformula_general}.

 When Assumption~\ref{asm::mon} holds, $N_{i1}(\tau) - N_{i0}(\tau)$ takes values only in $\{0,1\}$ and $\ACE_N(\tau) = \Pr (\mT_{i1} \leq \tau < \mT_{i0}) $,
Therefore, by switching the order of the expectation, integral, and derivative in~\eqref{eqn::aceY-decomposition},
we have
\begin{eqnarray*}
\ACE_Y(t) &=& -\int_{[0,T]} E \left[ \frac{\partial E \left\{ Y_{i\tau}(t) \mid \mathcal{H}^{N}_{i1T}, \mathcal{H}^{N}_{i0T}\right\}}{\partial \tau}\ \Bigg |\ \mT_{i1} \leq \tau < \mT_{i0} \right] \Pr (\mT_{i1} \leq \tau < \mT_{i0}) \d \tau \\
&=&-\int_{[0,T]} E \left\{ \frac{\partial Y_{i\tau}(t) }{\partial \tau}\ \bigg |\ \mT_{i1} \leq \tau < \mT_{i0} \right\} \Pr (\mT_{i1} \leq \tau < \mT_{i0}) \d \tau\\
&=& -\int_{[0,T]} E \left\{ \frac{\partial Y_{i\tau}(t) }{\partial \tau}\ \bigg |\ \mT_{i1} \leq \tau < \mT_{i0} \right\} \ACE_N(\tau) \d \tau.
\end{eqnarray*}
This proves~\eqref{eqn:ivformula_monotone}. \QEDB

\subsection{Proof of Theorem~\ref{thm::identification}}\label{sec:identification_proof}

First,
from Assumption~\ref{asm::stationarity} and the definition of ACER, we have
\begin{equation*}\label{eqn:ACER_shift_invariance}
 \ACER(t;\tau)\ =\ \lim_{ \Delta \tau \rightarrow 0+} \frac{E  \{Y_{i,\tau+\Delta \tau}(t) - Y_{i\tau}(t)\}}{\Delta \tau} = \lim_{ \Delta \tau \rightarrow 0+} \frac{E  \{Y_{i,\Delta \tau}(t-\tau) - Y_{i0}(t-\tau)\}}{\Delta \tau},
\end{equation*}
which depends only on $t-\tau$. Therefore, we can write $\ACER(t;\tau) = \ACER(t-\tau)$.

Second, under Assumption~\ref{asm::random}, we have
\begin{eqnarray*}
\ACE_Y(t) \ = \ E \{Y_i(t) \mid Z_i=1\} - E \{Y_i(t) \mid Z_i=0\} \ = \ h(t)
\end{eqnarray*}
and
\begin{eqnarray*}
\Pr (\mT_{i1} \leq \tau < \mT_{i0})-\Pr (\mT_{i0} \leq \tau < \mT_{i1})\ =\ \ACE_N(\tau)\ = \ f(\tau).
\end{eqnarray*}

Finally, we can write~\eqref{eqn:ivformula_general} in Theorem~\ref{thm:ivformula} as
\begin{eqnarray}
\nonumber h(t) &=& -\int_{[0,T]} E \left\{ \frac{\partial Y_{i\tau}(t) }{\partial \tau} \bigg | \mT_{i1} \leq \tau < \mT_{i0} \right\} \Pr (\mT_{i1} \leq \tau < \mT_{i0}) \d \tau \\
\nonumber&&+\int_{[0,T]} E \left\{ \frac{\partial Y_{i\tau}(t)}{\partial \tau} \bigg | \mT_{i0} \leq \tau < \mT_{i1} \right\} \Pr (\mT_{i0} \leq \tau < \mT_{i1}) \d \tau\\
\nonumber &=& \int_{[0,T]} \frac{\partial E \left\{ Y_{i\tau}(t) \right\}}{\partial \tau} \left\{\Pr (\mT_{i0} \leq \tau < \mT_{i1}) - \Pr (\mT_{i1} \leq \tau < \mT_{i0})\right\} \d \tau \\
\label{eqn::convolution} &=& -\int_{[0,T]} \ACER(t-\tau) f(\tau) \d \tau,
\end{eqnarray}
where the second equality follows from the condition in~\eqref{eqn::condition}.

Taking the Fourier transform on both sides of \eqref{eqn::convolution} yields, for all $\nu$, 
$$(\Psi h) (\nu)= -(\Psi \ACER) (\nu) (\Psi f)(\nu). $$
Therefore, if $(\Psi f) (\nu) \neq 0$ for all $\nu \in \mathbb{R}$, then 
$\ACER(t;\tau)= -\Psi^{-1}\big( G \big)(t-\tau)$ for $t > \tau$ and $0$ otherwise, where $G(\nu) = (\Psi h) (\nu) /(\Psi f) (\nu)$ for all $\nu \in \mathbb{R}$. 
\QEDB


\subsection{Proof of Proposition~\ref{prop::ACE-ACER-multiple}}
From Assumption~\ref{asmp:additivity}, we have
\begin{eqnarray*}
E \{Y_{i,n(\cdot)}(t)-Y_{iT+}(t)\}\ =\ \sum_{j=1}^{l} E \left\{Y_{i,\tau_j}(t)-Y_{iT+}(t)\right\}.
\end{eqnarray*}
Under Assumption~\ref{asm::no-anticipation}, we have
$Y_{i,\tau_j}(t) = Y_{iT+}(t) $ for $j > n(t)$ and $ Y_{iT+}(t) =Y_{it}(t) $.
Therefore,  for $j > n(t)$,  we have
\begin{equation}
\label{eqn::no-anti-eq}
 E \left\{Y_{i,\tau_j}(t)-Y_{iT+}(t)\right\}=0.
\end{equation}

Therefore, 
\begin{eqnarray*}
E \{Y_{i,n(\cdot)}(t)-Y_{iT+}(t)\}\ =\ \sum_{j=1}^{n(t)} E \left\{Y_{i,\tau_j}(t)-Y_{iT+}(t)\right\}\ =\  \sum_{j=1}^{n(t)} \ACE(t; \tau_j, T^+) \ = \ \sum_{j=1}^{n(t)} \ACE(t; \tau_j, t).
\end{eqnarray*}
As a result,
\begin{eqnarray*}
E \{Y_{i,n(\cdot)}(t)-Y_{iT+}(t)\}
 &=& - \sum_{j=1}^{n(t)}\int_{\tau_j}^{t} \ACER(t;\tau) \d \tau\\
 &=& - \sum_{j=1}^{n(t)} \left\{ \int_{\tau_{n(t)}}^{t} \ACER(t;\tau) \d \tau+\int_{\tau_j}^{\tau_{n(t)}} \ACER(t;\tau) \d \tau\right\}\\
 & =& - n(t)\cdot \int_{\tau_{n(t)}}^{t} \ACER(t;\tau) \d \tau - \sum_{j=1}^{n(t)-1} \left\{ \sum_{k=j}^{n(t)-1} \int_{\tau_{k}}^{\tau_{k+1}} \ACER(t;\tau) \d \tau \right\} \\
&=& - n(t)\cdot \int_{\tau_{n(t)}}^{t} \ACER(t;\tau) \d \tau - \sum_{k=1}^{n(t)-1} \sum_{j=1}^{k} \int_{\tau_{k}}^{\tau_{k+1}} \ACER(t;\tau) \d \tau\\
&=&- n(t)\cdot \int_{\tau_{n(t)}}^{t} \ACER(t;\tau) \d \tau - \sum_{k=1}^{n(t)-1} k\cdot \int_{\tau_{k}}^{\tau_{k+1}} \ACER(t;\tau) \d \tau.
\end{eqnarray*}
Because
\begin{eqnarray*}
n(\tau)\ = \ \begin{cases}
k,& \ {\rm if}\ \tau \in [\tau_{k},\tau_{k+1})\\
0,& \ {\rm if}\ \tau \in [0, \tau_1)\\
n(t),& \ {\rm if}\ \tau \in [\tau_{n(t)},t]
\end{cases},
\end{eqnarray*}
 we have
\begin{eqnarray*}
E \{Y_{i,n(\cdot)}(t)-Y_{iT+}(t)\} & =& -
 \int_{\tau_{n(t)}}^{t} \ACER(t;\tau)n(\tau) \d \tau - \sum_{k=1}^{n(t)-1} \int_{\tau_{k}}^{\tau_{k+1}} \ACER(t;\tau) n(\tau) \d \tau\\
&=& -\int_0^t \ACER(t;\tau) n(\tau) \d \tau.
\end{eqnarray*}
As a result, we have
\begin{eqnarray*}
E \{Y_{i,n(\cdot)}(t)-Y_{i,n'(\cdot)}(t)\}
\ =\ - \int_0^t \ACER(t;\tau) \left\{n(\tau) - n'(\tau)\right\} \d \tau. 
\end{eqnarray*}

\QEDB

\subsection{Proof of Theorem~\ref{thm::identification-ex}}
\label{sec:additivity_proof}
Under Assumptions~\ref{asm::er}~and~\ref{asm::random}, we can write, for any $t \in [0,T]$,
 \begin{eqnarray*}
h(t) & = & E  \{ Y_{i,N_{i1}}(t)\}-E  \{ Y_{i,N_{i0}}(t)\}\\
&=& \sum_{n(\cdot),n'(\cdot) \in \mathcal{N}} E  \{ Y_{i,n(\cdot)}(t)- Y_{i,n'(\cdot)}(t) \mid N_{i1}=n(\cdot), N_{i0}=n'(\cdot) \} \Pr \{N_{i1}=n(\cdot), N_{i0}=n'(\cdot)\},
\end{eqnarray*}
where $\mathcal{N}$ is the sample space of simple point process on $[0,T]$. Here we abuse the notation $\Pr(\cdot)$ to represent the density and take $\Sigma$ as the summation over all possible pairs of $n(\cdot)$ and $n'(\cdot)$ in $\mathcal{N}$. 

From the condition in~\eqref{eqn::condition-ex}, we know that 
$$ h(t) \ =\ \sum_{n(\cdot),n'(\cdot) \in \mathcal{N}} E  \{ Y_{i,n(\cdot)}(t)- Y_{i,n'(\cdot)}(t)\} \Pr \{N_{i1}=n(\cdot), N_{i0}=n'(\cdot)\}.$$
It follows from Proposition~\ref{prop::ACE-ACER-multiple} that 
$$ h(t)\ =\ -\sum_{n(\cdot),n'(\cdot) \in \mathcal{N}} \Pr \{N_{i1}=n(\cdot), N_{i0}=n'(\cdot)\} \int_0^t \ACER(t;\tau) \left\{n(\tau) - n'(\tau)\right\} \d \tau .$$
As a result, we have
\begin{eqnarray*}
	h(t) \ =\ - \int_0^t \ACER(t;\tau) E \{N_{i1}(\tau)-N_{i0}(\tau)\}\d \tau.
\end{eqnarray*}
From Assumption~\ref{asm::random}, we know that 
\begin{eqnarray*}
E \{N_{i1}(\tau)-N_{i0}(\tau)\}\ =\ E \{N_{i}(\tau) \mid Z_i = 1\} - E  \{ N_{i}(\tau) \mid Z_i = 0\} = f(\tau). 
\end{eqnarray*}
Finally, using that $\ACER(t;\tau)\ =\ \ACER(t-\tau)$ from Assumption~\ref{asm::stationarity}, we have 
\begin{eqnarray*}
	h(t)\ =\ - \int_0^t \ACER(t-\tau) f(\tau)\d \tau.
\end{eqnarray*}
 The rest follows from the same argument as in \S\ref{sec:identification_proof}.\QEDB

%

\subsection{Proof of Proposition~\ref{prop:model1}}
\label{sec:proof-model1}

We first prove Proposition~\ref{prop:model1}(a).
Consider a single-point treatment process $N_i$. 

From Assumption~\ref{asmp:ignorability-self-latent}, we know~\eqref{eqn:intensity-po-ob} holds that, for any $t \in [0,T]$, 
\begin{eqnarray*}
	E  \left\{ \frac{ \d Y_{i\tau}(t)}{\d t} \ \bigg |\ \mathcal{H}_{it-} \right\} &=&
	E  \left\{ \frac{ \d Y_{i}(t)}{\d t} \ \bigg |\ \mathcal{T}_i = \tau, \mathcal{H}^{*\setminus N_i}_{it-} \right\} = \lambda_Y(t),
\end{eqnarray*}
where the last equality follows from the definition of $\lambda_Y(t)$ in \eqref{eqn:intensity-def}. 

Plugging in Model~\eqref{eqn:model1}, we know that 
\begin{eqnarray}
\label{eqn:model1-po}
	E  \left\{ \frac{ \d Y_{i\tau}(t)}{\d t} \ \bigg |\ \mathcal{H}_{it-} \right\} = \mu_Y + g(t-\tau) +\int_0^{t} \omega(t-s) \d Y_{i\tau}(s) + \psi_{U_i}(t).
\end{eqnarray}

Taking the expectation on both sides of~\eqref{eqn:model1-po}, we obtain
\begin{eqnarray}
\label{eqn:model1-po2}
	E  \left\{ \frac{ \d Y_{i\tau}(t)}{\d t} \right\} & \ =\ & \mu_Y+g(t-\tau)+\int_0^{t} \omega(t-s) E \{ \d Y_{i\tau}(s) \} + E  \{\psi_{U_i}(t)\}.
\end{eqnarray}
Integrating both sides of~\eqref{eqn:model1-po2} from $0$ to $t$ yields 
\begin{eqnarray}
\nonumber
E  \left\{ Y_{i\tau}(t) \right\}\ =\ \mu_Y t+ \int_0^t g(s-\tau) \d s +\int_0^{t} \int_0^{s} \omega(s-\Delta) E  \left\{\d Y_{i\tau}(\Delta) \right\} \d s+\int_0^tE  \{\psi_{U_i}(s)\} \d s .\\
\label{eqn:model4-EY-one}
\end{eqnarray}
For the third term in~\eqref{eqn:model4-EY-one}, we have 
\begin{eqnarray*}
& & \int_0^{t} \int_0^{s} \omega(s-\Delta) E  \left\{\d Y_{i\tau}(\Delta) \right\} \d s\\
&= & \int_0^{t} \int_0^{s} \omega(s-\Delta) E  \left\{ \frac{\d Y_{i\tau}(\Delta)}{\d \Delta} \right\}\d \Delta \d s\\
&= & \int_0^{t} \int_0^{s} \omega(\Delta') E  \left\{ \frac{\d Y_{i\tau}(s-\Delta')}{\d (s-\Delta') } \right\}\d (s-\Delta') \d s \qquad (\Delta' = s-\Delta)\\
&= & \int_0^{t} \int_s^{0} \omega(\Delta') E  \left\{- \frac{\d Y_{i\tau}(s-\Delta')}{\d (s-\Delta') } \right\}\d \Delta' \d s \qquad ( \d(s-\Delta) = - \d \Delta' )\\
& = & \int_0^{t} \int_0^{s} \omega(\Delta') E  \left\{ \frac{\d Y_{i\tau}(s-\Delta')}{\d (s-\Delta') } \right\} \d \Delta' \d s \\
& = & \int_0^{t} \int_0^{s} \omega(\Delta') E  \left\{ \frac{\d Y_{i\tau}(s-\Delta')}{\d s} \right\} \d \Delta' \d s \\
& = & \int_0^{t} \int_{\Delta'}^{t} \omega(\Delta') E  \left\{ \frac{\d Y_{i\tau}(s-\Delta')}{\d s} \right\}  \d s\d \Delta'  \qquad  \text{(Fubini's Theorem)} \\
& = & \int_0^{t} \omega(\Delta') \left[\int_{\Delta'}^{t} E  \left\{ \frac{\d Y_{i\tau}(s-\Delta')}{\d s} \right\}  \d s \right] \d \Delta'\\
& = &\int_0^{t} \omega(\Delta) E \{ Y_{i\tau}(t-\Delta) \} \d \Delta - \left\{ \int_0^{t} \omega(\Delta) \d \Delta \right\}E \{ Y_{i\tau}(0)\}.
\end{eqnarray*}
Therefore, 
\begin{eqnarray*}
 E  \left\{ Y_{i\tau}(t) \right\} & = & \mu_Y t+ \int_0^t g(s-\tau) \d s+\int_0^{t} \omega(\Delta) E \{ Y_{i\tau}(t-\Delta) \} \d \Delta \\
 & & - \left\{ \int_0^{t} \omega(\Delta) \d \Delta \right\}E \{ Y_{i\tau}(0)\} + \int_0^tE  \{\psi_{U_i}(s)\} \d s.
\end{eqnarray*}
Taking the derivative with respect to $\tau$ on both sides of the above equation yields
\begin{eqnarray*}
	\ACER(t;\tau) &\ = \ & \frac{ \partial E  \{Y_{i\tau}(t)\}}{\partial \tau} \\ 
	&=& -\int_0^t \frac{\partial g(s-\tau)}{\partial s} \d s - \int_0^{t} \omega(\Delta) \frac{\partial}{\partial \tau }E \{ Y_{i\tau}(t-\Delta) \} \d \Delta \\
	& =  & -g(t-\tau) - \int_0^t \omega(\Delta) \ACER(t-\Delta;\tau) \d \Delta,
\end{eqnarray*}
where  $\partial E \{ Y_{i\tau}(0)\} / \partial \tau=0$ follows from~\eqref{eqn:model4-EY-one}. 
Using the property of the Dirac $\delta$ function, we obtain,
\begin{eqnarray}
\label{eqn:ACER_conv_eq}
\int_0^t \{\omega(\Delta) +\delta(\Delta) \}\ACER(t-\Delta;\tau) \d \Delta 	& = & -g(t-\tau). 
\end{eqnarray}
Taking the Fourier transform on both sides of~\eqref{eqn:ACER_conv_eq} yields, for all $\nu$, 
\begin{eqnarray*}
\{(\Psi \omega)(\nu) +1\} (\Psi \ACER_{\tau})( \nu) 	& \ = \ & - \exp(-\text{i} \nu \tau) (\Psi g)(\nu),
\end{eqnarray*}
where $\Psi \ACER_{\tau}$ is a short-hand notation for the Fourier transform of $\ACER(\cdot;\tau)$ and the right-hand side follows from the Fourier shift theorem that 
$\Psi g(t-\tau)= \exp(-\text{i} \nu \tau) \Psi g(\nu)$.
Therefore, 	we have 
\begin{eqnarray*}
(\Psi \ACER_{\tau})( \nu) 	= - \exp(-\text{i} \nu \tau) \frac{(\Psi g)(\nu)}{(\Psi \omega)(\nu) +1} = - \exp(-\text{i}\nu \tau) \widetilde{G}(\nu),
\end{eqnarray*}
where $\widetilde{G}(\nu) = \big\{ 1+ (\Psi \omega)(\nu) \big\}^{-1} \big( \Psi g\big) (\nu)$. 
By setting $\tau=0$, we obtain $\ACER(t;0) =- \big( \Psi^{-1} \widetilde{G} \big)(t)$. 
Therefore, we have
\begin{eqnarray*}
 (\Psi \ACER_{\tau})( \nu)\ = \  \exp(-\text{i} \nu \tau)  (\Psi \ACER(t;0) )( \nu),
\end{eqnarray*}
where $\Psi \ACER(t;0)$ is the Fourier transform of $\ACER(t;0)$.
Using the Fourier shift theorem, we obtain $\ACER(t;\tau) =\ACER(t-\tau;0)$. 


\bigskip
We now prove Proposition~\ref{prop:model1}(b). The result is an application of Theorem~\ref{thm::identification-ex}. In particular, we know that Assumption~\ref{asm::stationarity} holds from Part (a), we need to verify the condition in~\eqref{eqn::condition-ex}, Assumption~\ref{asmp:additivity}, and Assumption~\ref{asm::no-anticipation}. 

\noindent \textit{Verifying the condition in \eqref{eqn::condition-ex}:} 
Define
$$x(t;n(\cdot),n'(\cdot),n_1(\cdot), n_0(\cdot))\ \equiv \ E  \left\{ Y_{i,n(\cdot)}(t) - Y_{i, n'(\cdot)}(t) \mid N_{i1}=n_1(\cdot), N_{i0}=n_0(\cdot) \right\}. $$
By the law of total expectation, we have, for any fixed point processes $n_1(\cdot)$ and $n_0(\cdot)$,
\begin{eqnarray*}
	& & E  \left\{ Y_{i,n(\cdot)}(t) - Y_{i, n'(\cdot)}(t) \mid N_{i1}=n_1(\cdot), N_{i0}=n_0(\cdot) \right\} \\
		& = & E  \left[ E  \left\{ Y_{i,n(\cdot)}(t) -Y_{i, n'(\cdot)}(t) \mid \mathcal{H}_{it-},N_{i1}=n_1(\cdot), N_{i0}=n_0(\cdot)\right\} \big| N_{i1}=n_1(\cdot), N_{i0}=n_0(\cdot) \right] \\
	& = & E  \left[ E  \left\{ Y_{i,n(\cdot)}(t) \mid \mathcal{H}_{it-}\right\} - E  \left\{ Y_{i, n'(\cdot)}(t) \mid \mathcal{H}_{it-}\right\} \big| N_{i1}=n_1(\cdot), N_{i0}=n_0(\cdot) \right],
\end{eqnarray*}
where the last equality holds since the $N_{i1}$ and $N_{i0}$ are fixed up to time $t$ conditioning on the history $\mathcal{H}_{it}$.
Similar to the derivation of~\eqref{eqn:model4-EY-one}, we can obtain
\begin{eqnarray*}
& & E  \left\{ Y_{i,n(\cdot)}(t) \mid \mathcal{H}_{it-}\right\} - E  \left\{ Y_{i, n'(\cdot)}(t) \mid \mathcal{H}_{it-}\right\} \\
&=& \mu_Y t+ \int_0^t \int_0^s g(s-l) \d n(l) \d s +\int_0^t \psi_{U_i}(s) \d s +\int_0^{t} \int_0^{s} \omega(s-l) \d Y_{i, n(\cdot)}(l) \d s \\
& & - \mu_Y t - \int_0^t \int_0^s g(s-l) \d n'(l) \d s - \int_0^t \psi_{U_i}(s) \d s - \int_0^{t} \int_0^{s} \omega(s-l) \d Y_{i, n'(\cdot)}(l) \d s\\
&=& \int_0^t \int_0^s g(s-l) \d \{n(l)-n'(l)\} \d s +\int_0^{t} \int_0^{s} \omega(s-l) \d \{ Y_{i, n(\cdot)}(l) - Y_{i, n'(\cdot)}(l) \}\d s.
\end{eqnarray*}
%
%
Therefore, we have
\begin{eqnarray*}
	& & E  \left\{ Y_{i,n(\cdot)}(t) - Y_{i, n'(\cdot)}(t) \mid N_{i1}=n_1(\cdot), N_{i0}=n_0(\cdot) \right\} \\
	& = & E  \left[ E  \left\{ Y_{i,n(\cdot)}(t) \mid \mathcal{H}_{it-}\right\} - E  \left\{ Y_{i, n'(\cdot)}(t) \mid \mathcal{H}_{it-}\right\} \mid N_{i1}=n_1(\cdot), N_{i0}=n_0(\cdot) \right] \\
	&=& \int_0^t \int_0^s g(s-l) \d \{n(l)-n'(l)\} \d s +\int_0^{t} \int_0^{s} \omega(s-l) \d x(l;n(\cdot),n'(\cdot),n_1(\cdot), n_0(\cdot)) \d s,
\end{eqnarray*}
where
\begin{eqnarray*}
&&\int_0^{t} \int_0^{s} \omega(s-l) \d x(l;n(\cdot),n'(\cdot),n_1(\cdot), n_0(\cdot)) \d s\\
&=& \int_0^{t} \int_0^{s} \omega(s-l) \frac{ \d x(l;n(\cdot),n'(\cdot),n_1(\cdot), n_0(\cdot))}{\d l} \d l \d s\\
	&=&\int_0^{t} \int_0^{s} \omega(l') \frac{ \d x(s-l';n(\cdot),n'(\cdot),n_1(\cdot), n_0(\cdot))}{\d (s-l')} \d (s-l') \d s \qquad (l'=s-l)\\
		&=&\int_0^{t} \int_s^{0} \omega(l') \frac{ -\d x(s-l';n(\cdot),n'(\cdot),n_1(\cdot), n_0(\cdot))}{\d (s-l')} \d l' \d s \\
		&=&\int_0^{t} \int_0^{s} \omega(l') \frac{ \d x(s-l';n(\cdot),n'(\cdot),n_1(\cdot), n_0(\cdot))}{\d s} \d l' \d s \\
		&=&\int_0^{t} \int_{l'}^{t} \omega(l') \frac{ \d x(s-l';n(\cdot),n'(\cdot),n_1(\cdot), n_0(\cdot))}{\d s} \d s \d l'   \qquad  \text{(Fubini's Theorem)} \\
		&=& \int_0^{t} \omega(l') \left\{\int_{l'}^{t} \frac{ \d x(s-l';n(\cdot),n'(\cdot),n_1(\cdot), n_0(\cdot))}{\d s} \d s\right\} \d l' \\
	& = & \int_0^{t} \omega(l') x(t-l';n(\cdot),n'(\cdot),n_1(\cdot), n_0(\cdot)) \d l'\\
	& = &  \int_0^{t} \omega(s) x(t-s;n(\cdot),n'(\cdot),n_1(\cdot), n_0(\cdot))\d s.
\end{eqnarray*}
As a result, for $t \in [0,T]$, 
\begin{eqnarray}
\nonumber &&x(t;n(\cdot),n'(\cdot),n_1(\cdot), n_0(\cdot)) \\
\label{eqn:x-convo-eq} &=& \int_0^t \int_0^s g(s-l) \d \{n(l)-n'(l)\} \d s + \int_0^{t} \omega(s) x(t-s;n(\cdot),n'(\cdot),n_1(\cdot), n_0(\cdot))\d s.
\end{eqnarray}
Taking the derivative with respect to $t$ on both sides of~\eqref{eqn:x-convo-eq} leads to
\begin{eqnarray*}
&&\dot{x}(t;n(\cdot),n'(\cdot),n_1(\cdot), n_0(\cdot)) \\
&=& \int_0^t g(t-l) \d \{n(l)-n'(l)\}+ \omega(t) \dot{x}(0;n(\cdot),n'(\cdot),n_1(\cdot), n_0(\cdot))\\
&& +\int_0^{t} \omega(s) \dot{x}(t-s;n(\cdot),n'(\cdot),n_1(\cdot), n_0(\cdot))\d s,
\end{eqnarray*}
where $\dot{x}(t;n(\cdot),n'(\cdot),n_1(\cdot), n_0(\cdot))$ denotes the derivative of $x(t;n(\cdot),n'(\cdot),n_1(\cdot), n_0(\cdot))$ with respect to $t$.
We know that $\dot{x}(0;n(\cdot),n'(\cdot),n_1(\cdot), n_0(\cdot))=0$ by definition. Therefore, using the Fourier transform, we know that $\dot{x}(t;n(\cdot),n'(\cdot),n_1(\cdot), n_0(\cdot))= - \big( \Psi^{-1} \tilde{G} \big)(t)$, where $\tilde{G}(\nu) = \big\{ 1+ (\Psi \omega)(\nu) \big\}^{-1} \big( \Psi \tilde{g}\big) (\nu) $ and $\tilde{g}(t)\equiv \int_0^t g(t-l) \d \{n(l)-n'(l)\}$. 
As a result, $\dot{x}(t;n(\cdot),n'(\cdot),n_1(\cdot), n_0(\cdot))$ does not depend on $n_1(\cdot)$ and $n_0(\cdot)$. Combining this with the fact that 
$x(0;n(\cdot),n'(\cdot),n_1(\cdot), n_0(\cdot)) = 0$, we know that $x(t;n(\cdot),n'(\cdot),n_1(\cdot), n_0(\cdot)) $ does not depend on $n_1(\cdot)$ and $n_0(\cdot)$ and can write 
$x(t;n(\cdot),n'(\cdot),n_1(\cdot), n_0(\cdot)) = x(t;n(\cdot),n'(\cdot))$.

Taking $n_1(\cdot)=n(\cdot)$ and $n_0(\cdot)=n'(\cdot)$, we have,
 $$
E  \left\{ Y_{i,n(\cdot)}(t) - Y_{i, n'(\cdot)}(t) \mid N_{i1}=n(\cdot), N_{i0}\ =\ n'(\cdot) \right\}\ =\ x(t;n(\cdot),n'(\cdot)).
$$
From the law of total probability, we obtain
\begin{eqnarray*}
& &	E  \left\{ Y_{i, n(\cdot)}(t) - Y_{i, n'(\cdot)}(t) \right\}\\
 & = & \sum_{n_1(\cdot),n_0(\cdot) \in \mathcal{N}} E  \{ Y_{i,n(\cdot)}(t)- Y_{i,n'(\cdot)}(t) \mid N_{i1}=n_1(\cdot), N_{i0}=n_0(\cdot) \} \Pr \{N_{i1}=n_1(\cdot), N_{i0}=n_0(\cdot)\}\\
	& =& \sum_{n_1(\cdot),n_0(\cdot) \in \mathcal{N}} x(t;n(\cdot),n'(\cdot)) \Pr \{N_{i1}=n_1(\cdot), N_{i0}=n_0(\cdot)\}\\
& = & x(t;n(\cdot),n'(\cdot)).
\end{eqnarray*}
As a result, the condition in~\eqref{eqn::condition-ex} holds.\\

\noindent \textit{Verifying Assumption~\ref{asmp:additivity} (Additivity):} We have shown that, for any $n(\cdot)$ and $n'(\cdot)$,  
\begin{eqnarray*}
	E  \left\{ Y_{i, n(\cdot)}(t) - Y_{i, n'(\cdot)}(t) \right\}\ =\ x(t;n(\cdot),n'(\cdot)),
\end{eqnarray*}
where $x(t;n(\cdot),n'(\cdot))$ is the solution of \eqref{eqn:x-convo-eq}. To verify additivity, we only need to show that, for all $t$,  
\begin{equation*}
 x(t;n_1(\cdot),n'_1(\cdot))= x(t;n_2(\cdot),n'_2(\cdot)),
\end{equation*}
where, without loss of generality, $n_1(\cdot)$ has events $\{\tau_1,\tau_2,\ldots, \tau_l\}$, $n'_1(\cdot)$ has events $\{\tau_1,\tau_2,\ldots, \tau_{l-1}\}$, $n_2(\cdot)$ has a single event $\{\tau_l\}$, and $n'_2(\cdot)$ has no events. 

From~\eqref{eqn:x-convo-eq}, we have
\begin{eqnarray*}
	x(t;n(\cdot),n'(\cdot)) & = & \int_0^t \int_0^s g(s-l) \d \{n(l)-n'(l)\} \d s + \int_0^{t} \omega(s) x(t-s;n(\cdot),n'(\cdot)) \d s\\
	 & = & \sum_{j=1}^{n(t)} \int_{\tau_j}^t g(s-\tau_j) \d s - \sum_{j=1}^{n'(t)} \int_{\tau_j}^t g(s-\tau'_j) \d s + \int_0^{t} \omega(s) x(t-s;n(\cdot),n'(\cdot)) \d s.
\end{eqnarray*}
Therefore, $x(t;n_1(\cdot),n'_1(\cdot))$ is the solution of 
\begin{eqnarray*}
	x(t;n_1(\cdot),n'_1(\cdot))\ =\ \begin{cases}
		\int_{\tau_l}^t g(s-\tau_j) \d s + \int_0^{t} \omega(s) x(t-s;n_1(\cdot),n'_1(\cdot)) \d s & t > \tau_l \\
		0 & t \leq \tau_l
	\end{cases}.
\end{eqnarray*}
Similarly, $x(t;n_2(\cdot),n'_2(\cdot))$ is the solution of 
\begin{eqnarray*}
	x(t;n_2(\cdot),n'_2(\cdot)) \ = \ \begin{cases}
		\int_{\tau_l}^t g(s-\tau_j) \d s + \int_0^{t} \omega(s) x(t-s;n_2(\cdot),n'_2(\cdot)) \d s & t > \tau_l \\
		0 & t \leq \tau_l
	\end{cases}.
\end{eqnarray*}
Therefore, we have $ x(t;n_1(\cdot),n'_1(\cdot))= x(t;n_2(\cdot),n'_2(\cdot))$.

\noindent \textit{ Verifying Assumption~\ref{asm::no-anticipation}:} We only need to show that, for any $t\leq \tau$,$\partial E\{ Y_{i\tau}(t)\} /\partial \tau =0$. From~\eqref{eqn:ACER_conv_eq}, we have 
  \begin{eqnarray}
 \int_0^t \{\omega(\Delta) +\delta(\Delta) \}\ACER(t-\Delta;\tau) \d \Delta 	& = & 0
\end{eqnarray}
 for $t \leq \tau$,
where we use $g(t-\tau)=0$ for $t \leq \tau$. 
For fixed $\tau$, denote $m(t) = \ACER(t;\tau)$ for $t \leq \tau$ and $x(t) = 0$ for $t>\tau$. Then, for any $t$
  \begin{eqnarray}
 \int_0^t \{\omega(\Delta) +\delta(\Delta) \}x(t-\Delta) \d \Delta 	& = & 0.
\end{eqnarray}
Taking the Fourier transform on both sides yields, for all $\nu$, 
\begin{eqnarray*}
\{(\Psi \omega)(\nu) +1\} (\Psi m)( \nu) 	& \ = \ & 0.
\end{eqnarray*}
Recalling that $\{(\Psi \omega)(\nu) +1\}\neq 0$, we have $\Psi m = 0$, and thus $m(t) = 0$ for all $t$. Therefore, $\partial E\{ Y_{i\tau}(t)\} /\partial \tau =\ACER(t;\tau)  =0$ for $t \leq \tau$. 
\QEDB

\subsection{Proof of Proposition~\ref{prop:model2}}
\label{sec:proof-prop-model2}

Using a similar argument as in \eqref{eqn:model1-po}, we have
\begin{eqnarray*}
	E  \left\{ \frac{ \d Y_{i\tau}(t)}{\d t} \ \bigg |\ \mathcal{H}_{it-} \right\} \ =\ \phi \left\{ \mu_Y+g(t-\tau)\right\}+ \psi_{U_i}(t).
\end{eqnarray*}
Taking expectation and then integrating from $0$ to $t$ on both sides, we obtain
\begin{eqnarray}
\label{eqn::model2-one}
E  \left\{ Y_{i\tau}(t) \right\} \ =\ \int_0^t \phi \left\{ \mu_Y+g(s-\tau)\right\} \d s + \int_0^t E  \left\{\psi_{U_i}(s) \right\} \d s.
\end{eqnarray}

Taking the derivative with respect to $\tau$ yields Part (a) of the proposition that 
\begin{eqnarray}
\nonumber	\ACER(t;\tau) \ =\ -\int_0^t \phi' \left\{ \mu_Y+g(s-\tau)\right\} g'(s-\tau)\d s \ = \ \phi(\mu_Y) - \phi\{\mu_Y+g(t-\tau)\}.\\
\label{eqn::ACER-model2-proof}
\end{eqnarray}

Similar to the proof of Proposition~\ref{prop:model1}, it suffices to verify Assumption~\ref{asm::stationarity}, the condition in~\eqref{eqn::condition}, and Assumption~\ref{asm::no-anticipation}.

From~\eqref{eqn::model2-one}, we have
\begin{eqnarray*}
	E \{Y_{i\tau_1}(t) - Y_{i\tau_2}(t)\}
	&=& \int_{\tau_1}^t \phi \left\{ \mu_Y+ g(s-\tau_1) \right\} \d s- \int_{\tau_2}^t\phi \left\{ \mu_Y+ g(s-\tau_2) \right\} \d s\\
	&=& \int_0^{t-\tau_1} \phi \left\{ \mu_Y+ g(s) \right\} \d s- \int_{\tau_2-\tau_1}^{t-\tau_1}\phi \left\{ \mu_Y+ g(s-\tau_2+\tau_1) \right\} \d s\\
	&=&E \{Y_{0}(t-\tau_1) - Y_{\tau_2-\tau_1}(t-\tau_1)\}.
\end{eqnarray*}
for $\tau_1\leq \tau_2\leq t$.
Thus, Assumption~\ref{asm::stationarity} holds.

We then verify the condition in~\eqref{eqn::condition}. Similar to the proof of Proposition~\ref{prop:model1},
we can obtain
\begin{eqnarray*}
&&E  \left\{ \frac{\partial Y_{i\tau}(t)}{\partial \tau} \cdot 1 (\mathcal{T}_{i1} \leq \tau) \right\}\\
&=& E  \left[ \frac{\partial E \{Y_{i\tau}(t) \mid  \mathcal{H}_{it-} \}}{\partial \tau} \cdot 1 (\mathcal{T}_{i1} \leq \tau)\right]\\
&=&E  \left[ \frac{\partial E \{Y_i(t) \mid \mathcal{T}_i = \tau, \mathcal{H}^{*\setminus N_i}_{it} \}}{\partial \tau} \cdot 1 (\mathcal{T}_{i1} \leq \tau)\right],
\end{eqnarray*}
where, from Model~\eqref{eqn:model2},
\begin{eqnarray*}
	\frac{\partial E \{Y_i(t) \mid \mathcal{T}_i = \tau, \mathcal{H}^{*\setminus N_i}_{it} \}}{\partial \tau} &=& \frac{\partial}{\partial \tau} \left[ \phi(\mu_Y)\tau+\int_\tau^t \phi\{\mu_Y+g(s-\tau) \} \d s +\int_0^t \psi_{U_i}(s) \d s\right]\\
	&=& \phi(\mu_Y)-\phi(\mu_Y)-\int_\tau^t g'(s-\tau)\phi' \{\mu_Y+g(s-\tau) \} \d s \\
	&=& \phi(\mu_Y)-\phi(\mu_Y)- \left[ \phi\{\mu_Y+g(t-\tau) \}-\phi(\mu_Y) \right]\\
	&=& \phi(\mu_Y)-\phi\{\mu_Y+g(t-\tau) \}.
\end{eqnarray*}
Therefore, we obtain
\begin{eqnarray}
\label{eqn::model2-condition1}
E  \left\{ \frac{\partial Y_{i\tau}(t)}{\partial \tau} \cdot 1 (\mathcal{T}_{i1} \leq \tau) \right\}\ = \ \left[ \phi(\mu_Y)-\phi\{\mu_Y+g(t-\tau) \}\right] \Pr( \mathcal{T}_{i1} \leq \tau).
\end{eqnarray}
Similarly, we obtain
\begin{eqnarray}
\label{eqn::model2-condition2}
E  \left\{ \frac{\partial Y_{i\tau}(t)}{\partial \tau} \cdot 1 (\mathcal{T}_{i0} \leq \tau) \right\}\ = \ \left[ \phi(\mu_Y)-\phi\{\mu_Y+g(t-\tau) \}\right] \Pr( \mathcal{T}_{i0} \leq \tau).
\end{eqnarray}
Combining~\eqref{eqn::model2-condition1}~and~\eqref{eqn::model2-condition2} yields the condition in~\eqref{eqn::condition}. 

From~\eqref{eqn::ACER-model2-proof}, we have 
$$	\ACER(t;\tau) \ = \ \phi(\mu_Y) - \phi\{\mu_Y+g(t-\tau)\}\ = \ \phi(\mu_Y) - \phi(\mu_Y)\ =\ 0$$
for $t \leq \tau$, where we use $g(\Delta)=0$ for $\Delta \leq 0$. 
Therefore, Assumption~\ref{asm::no-anticipation} holds.

\QEDB

\subsection{Comment on the difficulties of the nonlinear non-additive model}\label{sec::difficulty-nonlinear-nonadditive-model}

 We now comment on the difficulties of the following nonlinear non-additive model
\begin{eqnarray}
\label{eqn:model3}
\lambda_{Y}(t) = \phi\left\{ \mu_Y+ \int_{0}^{t} g(t-s) \d N_i(s)+ \int_{0}^{t} \omega(t-s) \d Y_i(s) + \psi_{U_i}(t)\right\},
\end{eqnarray}
where $\phi(\cdot )$ is a non-negative link function and the rest are the same as in~\eqref{eqn:model1}.
Model \eqref{eqn:model3} is widely used in the analysis of neural data with different forms of $\phi$ (see among others \citealp{ lawrence2004gaussian, kulkarni2007common, byron2009gaussian, gao2015high, macke2015estimating, gao2016linear, sussillo2016lfads, wu2017gaussian, zhao2017variational, pandarinath2018inferring}). Compared with Models~\eqref{eqn:model1}~and~\eqref{eqn:model2}, it guarantees a non-negative intensity without additional restrictions on the unmeasured confounders. However, the impact of the unmeasured confounding is no longer additive under Model~\eqref{eqn:model3}. Even when the treatment is a single-point process, the following proposition shows that the causal effect depends on the choices of the link function and the distribution of $U_i$ in Model~\eqref{eqn:model3}.
\begin{proposition}
\label{prop:interpretation-model3}
Suppose that Assumptions~\ref{asm::sutva},~\ref{asm::er},~and~\ref{asmp:ignorability-self-latent} hold and the underlying outcome satisfies~\eqref{eqn:model3}. When $N_i$ is a single-point process, we have, for $t, \tau \in [0,T]$,
\begin{eqnarray*}
\label{eqn:model3-acer}
\ACER(t;\tau) \ =\ \int_0^t E  \left[ \frac{ \partial }{\partial \tau} \phi \left\{ \mu_Y+g(s-\tau)+\int_{0}^{s} \omega(t-\Delta) \d Y_{i\tau}(\Delta) + \psi_{U_i}(s)\right\}\right]\d s.
\end{eqnarray*}
\end{proposition}
In Proposition~\ref{prop:interpretation-model3}, the ACER under Model~\eqref{eqn:model3} does not satisfy the identification assumptions in Theorems~\ref{thm::identification}~or~\ref{thm::identification-ex}. Thus, we cannot use the identification result in \S\ref{sec:IV}. Under Model~\eqref{eqn:model3}, the identification of ACER is difficult. Although it is strenuous to formally study its identifiability due to the complexity of Model~\eqref{eqn:model3}, econometricians have obtained negative results for the
identification of non-separable models in the cross-sectional setting when both the treatment and the outcome are scalars.
In particular, \citet{chesher2003identification} gives sufficient conditions for nonparametric identification, which generally requires the instrumental variable to be continuous.
Moreover, when the outcome is discrete, \citet{chesher2010instrumental} shows that the identification is typically not achieved even under parametric models. Therefore, the identification of Model~\eqref{eqn:model3} is gloomy in a more complex context with point processes. Due to the dependence of the ACER on the model parameters and the distribution of $U_i$, its identification is also unpromising.

We end this subsection with the proof of Proposition \ref{prop:interpretation-model3}. 

{\it Proof of Proposition \ref{prop:interpretation-model3}.}
Using a similar argument as in \eqref{eqn:model1-po}, we have
\begin{eqnarray*}
 E  \left\{ \frac{ \d Y_{i\tau}(t)}{\d t} \ \bigg |\ \mathcal{H}_{it} \right\} \ =\ \phi \left\{ \mu_Y+g(t-\tau)+\int_{0}^{t} \omega(t-s) \d Y_i(s) + \psi_{U_i}(t)\right\}.
\end{eqnarray*}
Taking expectation and then integrating from $0$ to $t$ on both sides, we obtain
\begin{eqnarray}
\label{eqn::model2-one}
E  \left\{ Y_{i\tau}(t) \right\} \ =\int_0^t E  \left[ \phi \left\{ \mu_Y+g(s-\tau)+\int_{0}^{s} \omega(t-l) \d Y_{i\tau}(l) + \psi_{U_i}(s)\right\}\right]\d s.
\end{eqnarray}
Taking the derivative with respect to $\tau$ yields
\begin{eqnarray}
\label{eqn::acer-nonadditive}
\ACER(t;\tau) \ =\ \int_0^t E  \left[ \frac{ \partial }{\partial \tau} \phi \left\{ \mu_Y+g(s-\tau)+\int_{0}^{s} \omega(t-l) \d Y_{i\tau}(l) + \psi_{U_i}(s)\right\}\right]\d s.
\end{eqnarray}

\section{Supplement of the numerical analysis}\label{sec:num-supp}

\subsection{Derivation of $\ACER$ in simulation}\label{sec:sim-supp}

We know from \S\ref{sec:model-para} that the \ACER\ in the presence of non-additive confounding takes the form in~\eqref{eqn::acer-nonadditive}.
For Scenario 3 in \S\ref{sec:sim}, we can derive that
\begin{align*}
 & \frac{ \partial }{\partial \tau } \phi_{\beta_2} \left\{ \mu_Y+g(t-\tau)+ U_i(t-d_U)\right\}\\
= & \left\{\frac{ \partial}{\partial \tau } g(t-\tau)\right\} \beta_2 \{\mu_Y+g(t-\tau)+ U_i(t-d_U) \}^{\beta_2-1} \\
=& \beta_2 b_Y a_Y^2 \big\{a_Y^2 (t-\tau)-1\big\}\exp\{ -a_Y(t-\tau) \}\{\mu_Y+\alpha(t-\tau; a_Y,b_Y)+ U_i(t-d_U) \}^{\beta_2-1},
\end{align*}
where the last equality follows from 
\begin{align*}
\frac{ \partial}{\partial \tau } g(t-\tau) &= \frac{ \partial}{\partial \tau } \alpha(t-\tau; a_Y,b_Y) \\
& =\frac{ \partial}{\partial \tau } b_Y a_Y^2  (t-\tau)\exp\{ -a_Y(t-\tau)\}\\
&=  b_Y a_Y^2 [ -\exp\{ -a_Y(t-\tau)\} +a_Y(t-\tau)\exp\{ -a_Y(t-\tau)\} ].
\end{align*}

We can calculate the true value of $\ACER(t;\tau)$ using the Monte Carlo method by simulating the unmeasured confounding process $U_i$.

\subsection{Additional information on the real data analysis}\label{sec:RDA-supp}

In \S\ref{sec:RDA}, we apply our methodology on the neural data to estimate the causal effect of neural activities in the olfactory bulbs on those in the piriform cortex. In this section, we discuss more scientific backgrounds and conduct additional analysis.

\begin{figure}[ht!]
	\centering
	\begin{tikzpicture}[node distance=1.3cm,>=stealth',bend angle=45,auto]
	\tikzstyle{inhub}=[draw=black!75,fill=black!20,minimum size=6mm]
	\tikzstyle{outhub}=[circle,draw=blue!75,fill=blue!20,minimum size=10mm,dashed]
	\tikzstyle{cell}=[circle,draw=red!75,	fill=red!20,minimum size=10mm]

	\begin{scope}
	\node [cell ] (OB){OB};
	\node [inhub] (S) [left = 1 cm of OB] {Odor}
	edge [->, thick, color=red, thick] (OB);
	\node [outhub] (FFI) [ above right = 1.2 cm and 0.5 cm of OB] {FFI}
	edge [<-, color=red, thick,dashed]		(OB);
	\node [cell] (PN)[ right = 1.5 cm of OB] {PN}
	edge [	<-,	 color=red, thick]		(OB)
	edge [<-, color=blue, thick, dashed]		(FFI);
	\node [outhub] (FBI)[ right = 1.5 cm of PN] {FBI}
	edge [<-, color=red, thick, dotted, bend right]	(PN)
	edge [->, color=blue, thick, dashed, bend left]	(PN);
	\end{scope}

	\end{tikzpicture}
	\caption{Causal diagram depicting the relationships among the olfactory bulb mitral cells (OB), principal neurons (PN), feedforward interneurons (FFI), and feedback interneurons (FBI) in the piriform cortex. A red arrow represents an excitatory pathway while a blue arrow represents an inhibitory pathway. The dotted arrow from PN to FBI is disconnected when TeLC is expressed. In the experiment analyzed here and in \S\ref{sec:RDA}, the odor is replaced by light pulses, and only the neural activities of OB and PN are recorded. }
	\label{fig:RDA-DAG}
\end{figure}
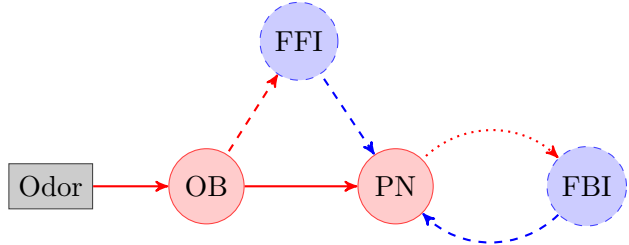

In addition to the olfactory bulb (OB) mitral cells and the principal neurons (PN) in the piriform cortex (PCx), layer 1 feedforward interneurons (FFI) and layer 2/3 feedback interneurons (FBI) play important roles in the neural circuits for odor perception \citep{boldingfranks2018}. They are inhibitory neurons that suppress the activities of PN. In particular, OB mitral cells excite both the PN and FFI in PCx, which may result in an immediate excitation and a slightly delayed inhibition in the PN. In addition, PN excites the FBI, which will in turn suppress future activities in PN. The causal pathways among the aforementioned neurons are illustrated in Figure~\ref{fig:RDA-DAG}.


The fitted \ACER\ in Figure~\ref{fig:RDA}(c) shows that, overall, a spike in OB causes an immediate excitation in PN in PCx, and causes a relatively long-term suppression till the effect vanishes. This finding corroborates the causal pathways in Figure~\ref{fig:RDA-DAG} that the immediate excitation may be due to the direct effects of OB cells on the PN, while the long-term suppression results from the induced activities from FFI and FBI. As a result, the stationarity of odor detection is maintained, as shown in Figures~\ref{fig:RDA}(a)~and~(b), that the PN return to normal activity level quickly after a sharp increase in activities.

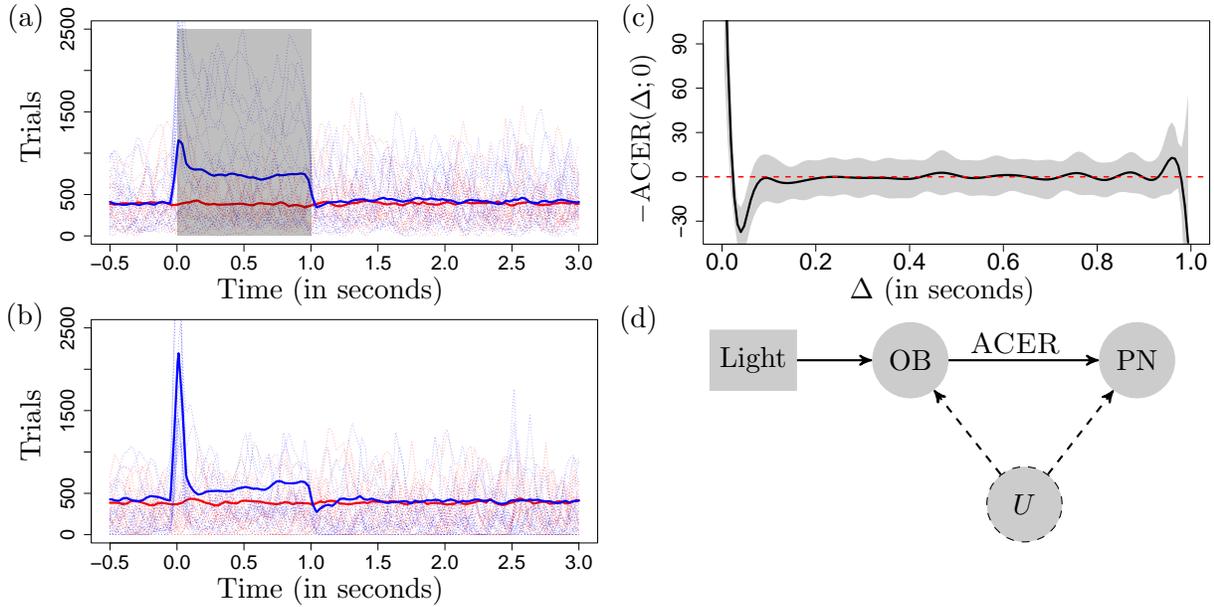
\begin{figure}[htpb]
\begin{tikzpicture}[node distance=1.3cm,>=stealth',bend angle=45,auto]
	\tikzstyle{iv}=[draw=black!20,fill=black!20,minimum size=8mm]
	\tikzstyle{latent}=[circle,draw=black,fill=black!20,minimum size=10mm,dashed]
	\tikzstyle{obs}=[circle,draw=black!20,	fill=black!20,minimum size=10mm]

	\begin{scope}
	\node [obs] (N){OB};
	\node [iv] (Z) [left = 1 cm of N] {Light}
	edge [->, thick, color=black, thick] (N);
	\node [latent] (U) [ below right = 1.2 cm and 0.8 cm of N] {$U$}
	edge [->, color=black, thick,dashed]		(N);
	\node [obs] [ right = 2 cm of N] (Y){PN}
	edge [	<-,	 color=black, thick]		(N)
	edge [<-, color=black, thick, dashed]		(U);

	\node[above right = -0.4 cm and 0.3 cm of N] {$\ACER$};

	\node [fit={(N) (Z) (U) (Y)}] (DAG) {};

	\node[inner sep=0pt] [above = 0.5 cm of DAG] (Trial){\includegraphics[width=.45\textwidth]{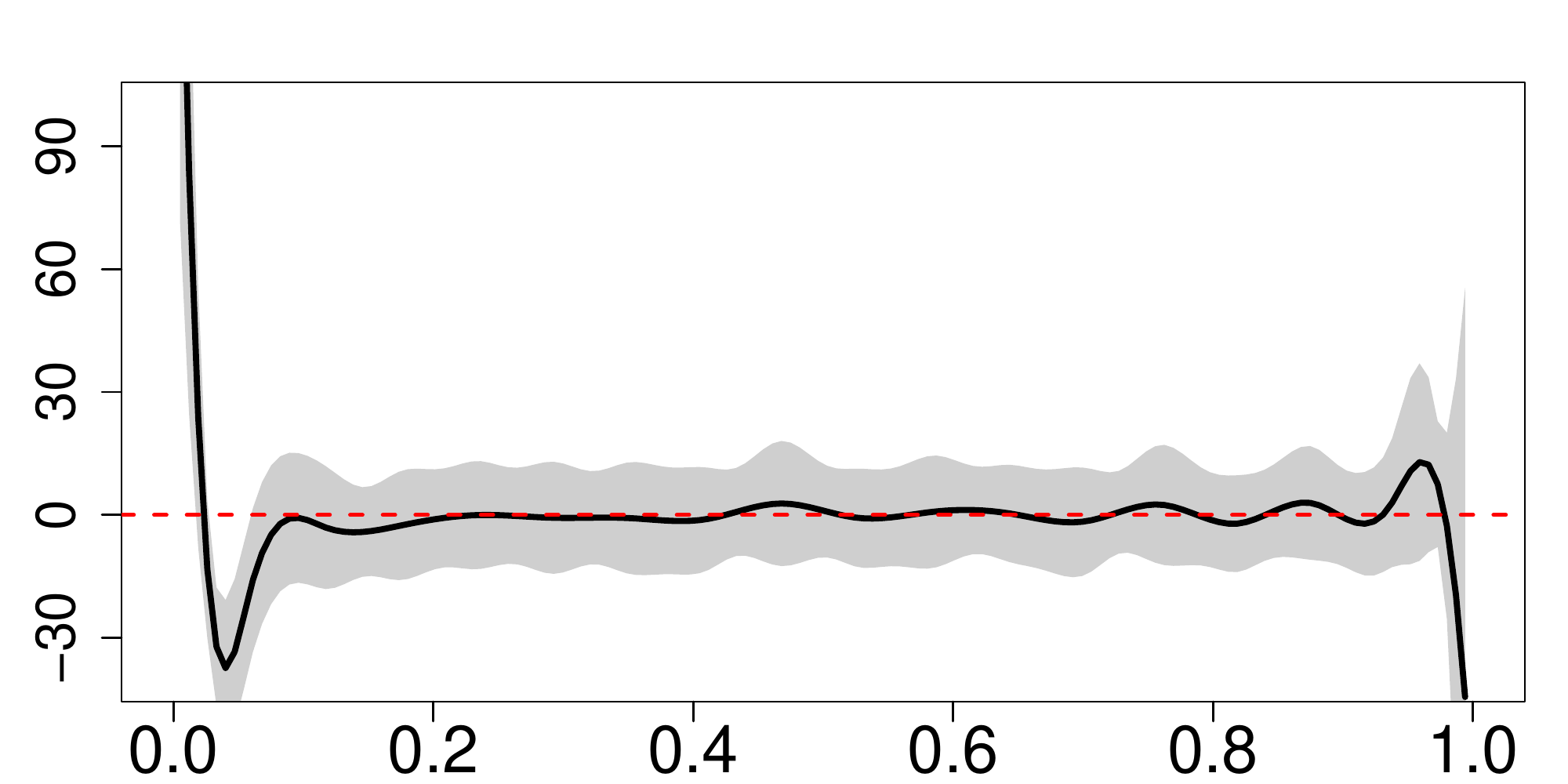}};

	\node[inner sep=0pt] [left = 0.6 cm of Trial] (Ns){\includegraphics[width=.45\textwidth]{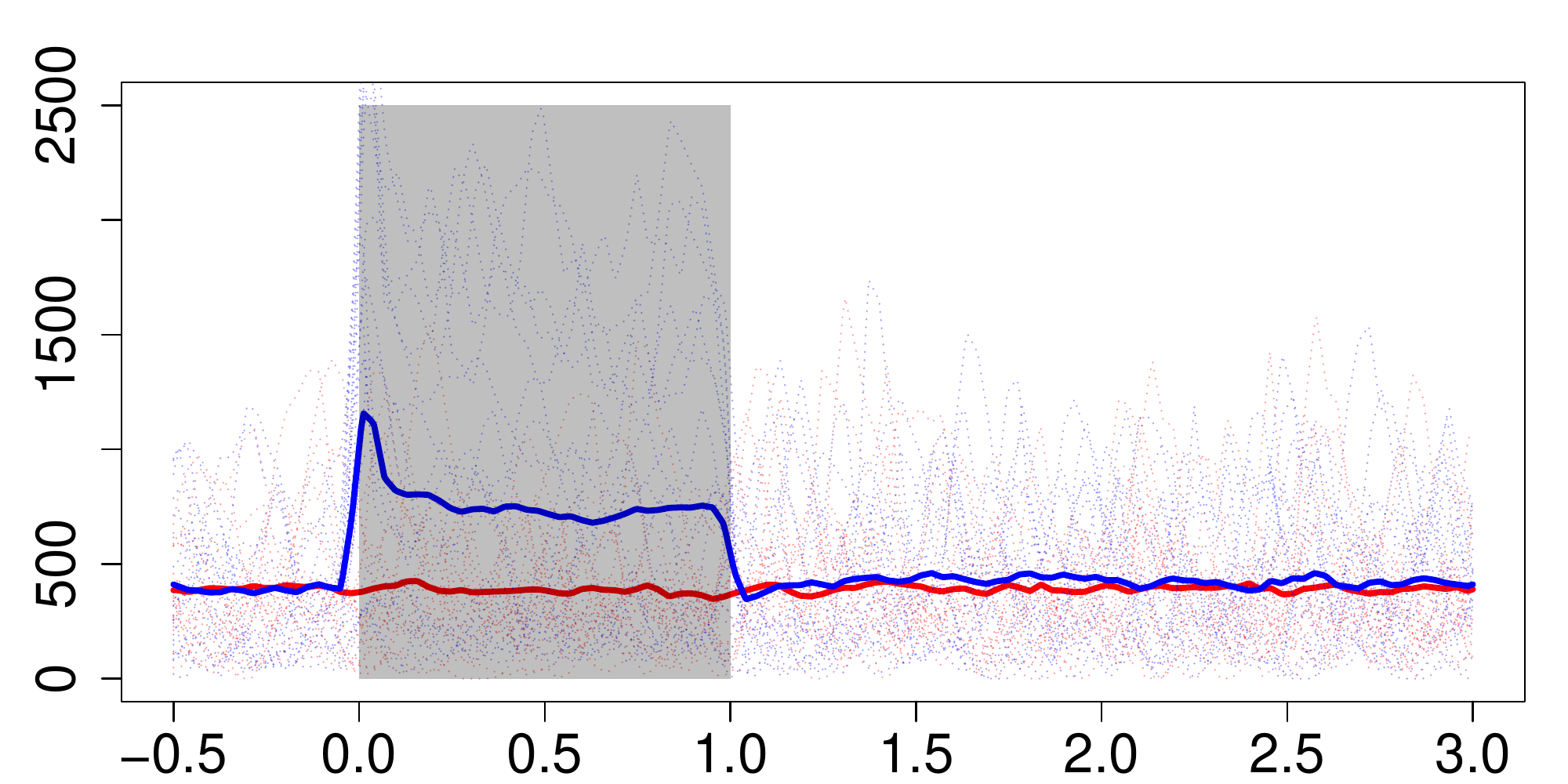}};

	\node[inner sep=0pt] [below = 0.2 cm of Ns] (Ys){\includegraphics[width=.45\textwidth]{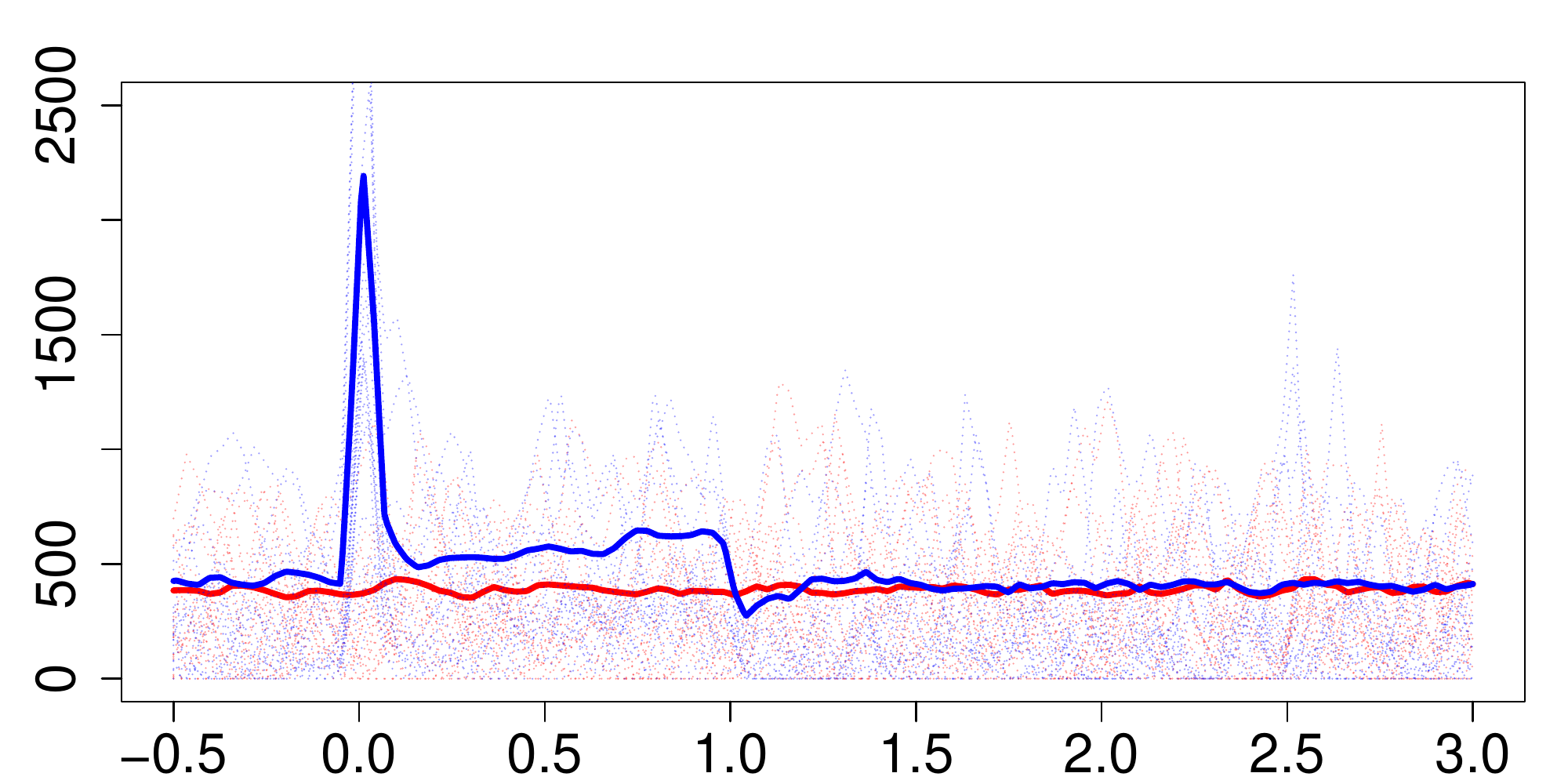}};
	
	\node[above left = -0.7 cm and -0.1cm of Ns] () {(a)};
	\node[above left = -1.2 cm and -0.05 cm of Ns, rotate=90] () {Trials};
	
	\node[above left = -0.7 cm and -0.1 cm of Ys] () {(b)};
	\node[below = -0.1 cm of Ys] (Yaxis) {Time (in seconds)};
	\node[above left = -1.2 cm and -0.05 cm of Ys, rotate=90] () {Trials};

	\node[below = -0.1 cm of Trial] (Taxis) {$\Delta$ (in seconds)};
	\node[above left = -0.7 cm and -0.1 cm of Trial, rotate=90] () {\small $-\ACER(\Delta;0)$};
	\node[above left = -0.7 cm and -0.1 cm of Trial](Tlab) {(c)};

			\node[below = -0.1 cm of Ns] (Naxis) {Time (in seconds)};
	\node[below = 3.35 cm of Tlab](DAGlab)
			{(d)};
	\end{scope}

	\end{tikzpicture}
	
	\caption{
		Empirical intensities and fitted $\ACER$ on data from \cite{boldingfranks2018data} on TeLC-expressed mice. Panels (a) and (b) show the empirical intensities of the neural activities of OB (Panel a) and PN (Panel b) in the stimulated (blue) and unstimulated (blue) groups. The solid curves represent the average intensity over 150 trials, and the shaped dashed curves demonstrate the empirical intensity from 20 randomly selected trials in each group. The shaded area in Panel (a) represents the duration of the light pulse. Panel (c) shows the estimated $-\ACER(\Delta;0)$ from the full data set. The shaded area represents a $90\%$ confidence band for visualizing the uncertainty of the estimates from 5000 bootstrap samples. Panel (d) shows the causal diagram for the relationship among the variables.}
	\label{fig:RDA-TeLC}
\end{figure}

To further investigate the roles of inhibitory pathways, \cite{boldingfranks2018} selectively expressed tetanus toxin light chain (TeLC) in PN. The expression of TeLC prevents a principal neuron to excite any other neurons including the FBIs, while retaining PN's excitability. In other words, in a TeLC-expressed mouse, the causal pathway between PN to FBI no longer exists in Figure~\ref{fig:RDA-DAG}. Figure~\ref{fig:RDA-TeLC} shows the resulting intensities. We can see the consequence of the absence of the FBI, where the activities of PN remain at a higher-than-normal level till the end of the stimulation.

Applying the same estimation procedure to the data from the TeLC-expressed mice reveals a very different \ACER. In this case, there are 15 TeLC-expressed mice. On each mouse, there are 10 trials in the treatment group and 10 in the control group. The confidence band is constructed as follows. Let $\{\hat{g}_{b}: b=1,\ldots, B\}$ be the estimates from the bootstrap samples, where trials are sampled with replacement in each treatment group. For each $t \in [0,T]$, we estimate the bootstrap mean $\bar{g}(t)$ and the bootstrap standard deviation $\widehat{\sigma}_g(t)$ from the bootstrap samples. We can then construct a $(1-\alpha)\%$ confidence band for $\hat{g}$ as $\{(\hat{g}(t)-q_{\alpha} m^{-1/2}\hat{\sigma}_g(t), \hat{g}(t)+q_{\alpha} m^{-1/2}\hat{\sigma}_g(t)): t\in [0,T] \} $, where $q_{\alpha}$ is the $(1-\alpha)\%$ bootstrap quantile of $Q_b = \max_{t} |\hat{g}_b(t)-\bar{g}(t)|/\hat{\sigma}_g(t)$. The number of trials $m$ is 300 in the TeLC experiment and $160$ in the analysis in the main text.

We can see that the causal effect lasts for a much shorter period, in the absence of self-excitation among PN and the inhibitions from FBI. 
Note that an inhibition period still exists, which is likely the effect through the FFI. 
Furthermore, the 5-fold cross-validation selected 18 knots in the TeLC-expressed mice, in sharp contrast to the 2 knots selected in Figure~\ref{fig:RDA}.
By comparing the fitted \ACER\ in Figure~\ref{fig:RDA-TeLC} and Figure~\ref{fig:RDA} in the main text, we can see that the causal pathway between PN and FFI is a key mechanism in maintaining stationarity in odor perception. We also notice that the nonparametric estimation procedure suffers from boundary effects, where a spurious inhibitory effect is estimated towards the right boundary of the support.
The estimated effect from the observational analysis does not reflect the sharp inhibition following an event in OB.

\subsection{Analysis assuming no unmeasured confounding}\label{sec:RDA-obs}

In this section, we conduct an observational analysis of the real data assuming there is no unmeasured confounding between the treatment $N_i$ and the outcome $Y_i$. In particular, we assume the outcome $Y_i$ follows the linear model
\begin{equation}\label{eqn:model-obs}
\lambda_Y(t) \ =\ E  \left\{ \frac{ \d Y_{i}(t)}{\d t} \ \bigg |\ \mathcal{H}_{it-} \right\}\ = \ \mu + \int_0^t g(t-s)\mathrm{d} N_i(s),
\end{equation} 
where $g(\cdot)$ is commonly interpreted as the effect of $N_i$ on $Y_i$. Here we do not include the self-dependence of $Y_i$ because we have shown in Proposition~\ref{prop:model1} that the function $g$ is not comparable to $\ACER$ when the self-dependence is included. 

\begin{figure}[htpb]
\begin{tikzpicture}[node distance=1.3cm,>=stealth',bend angle=45,auto]

	\begin{scope}
	\node[inner sep=0pt](Trial){\includegraphics[width=.45\textwidth]{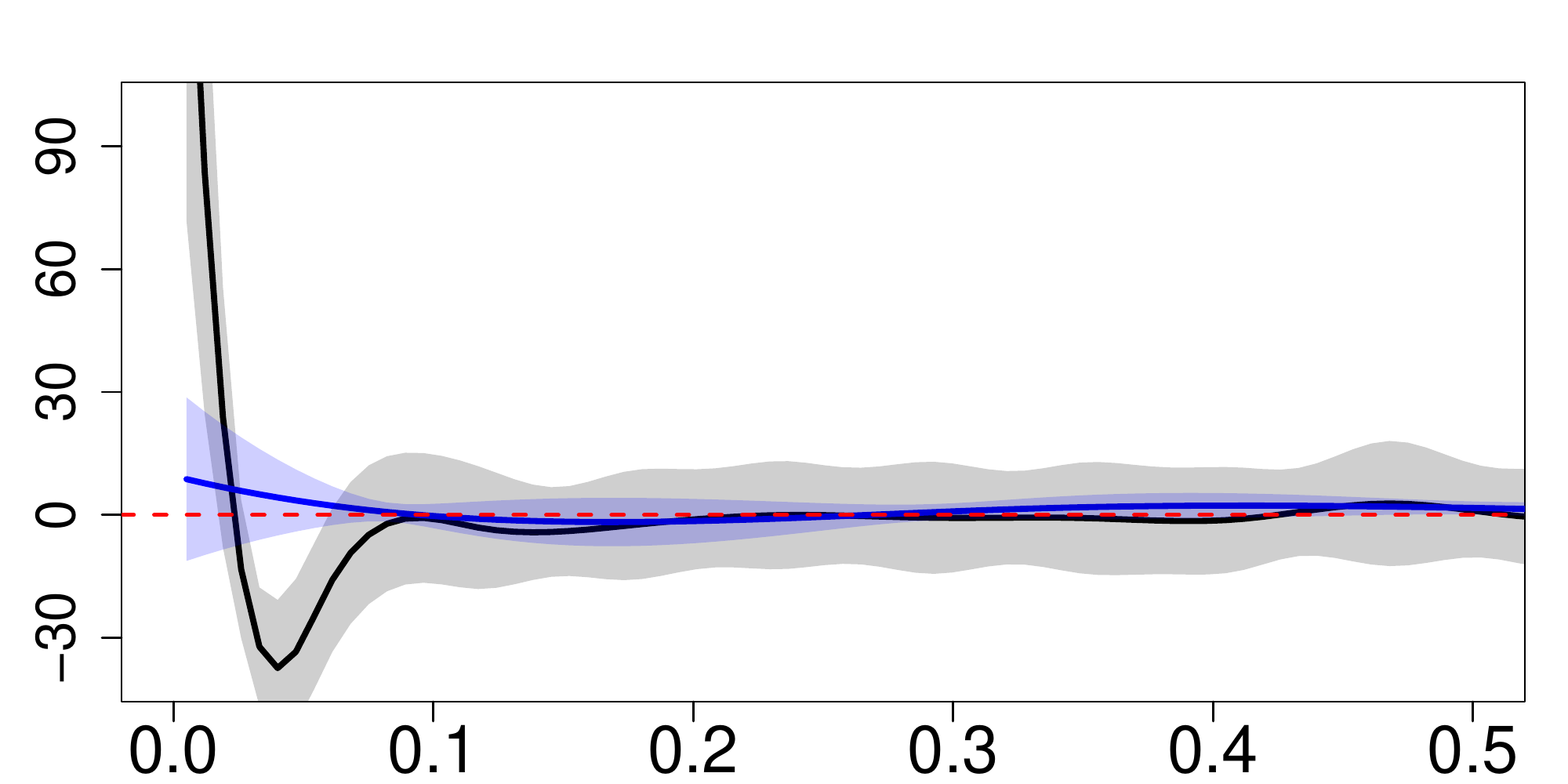}};

\node[below = -0.1 cm of Trial] (Taxis) {$\Delta$ (in seconds)};
	\node[above left = -0.7 cm and -0.1 cm of Trial, rotate=90] () {\small $-\ACER(\Delta;0)$};
	\node[above left = -0.7 cm and -0.1 cm of Trial](Tlab) {(b)};

	\node[inner sep=0pt] [left = 0.6 cm of Trial] (obs){\includegraphics[width=.45\textwidth]{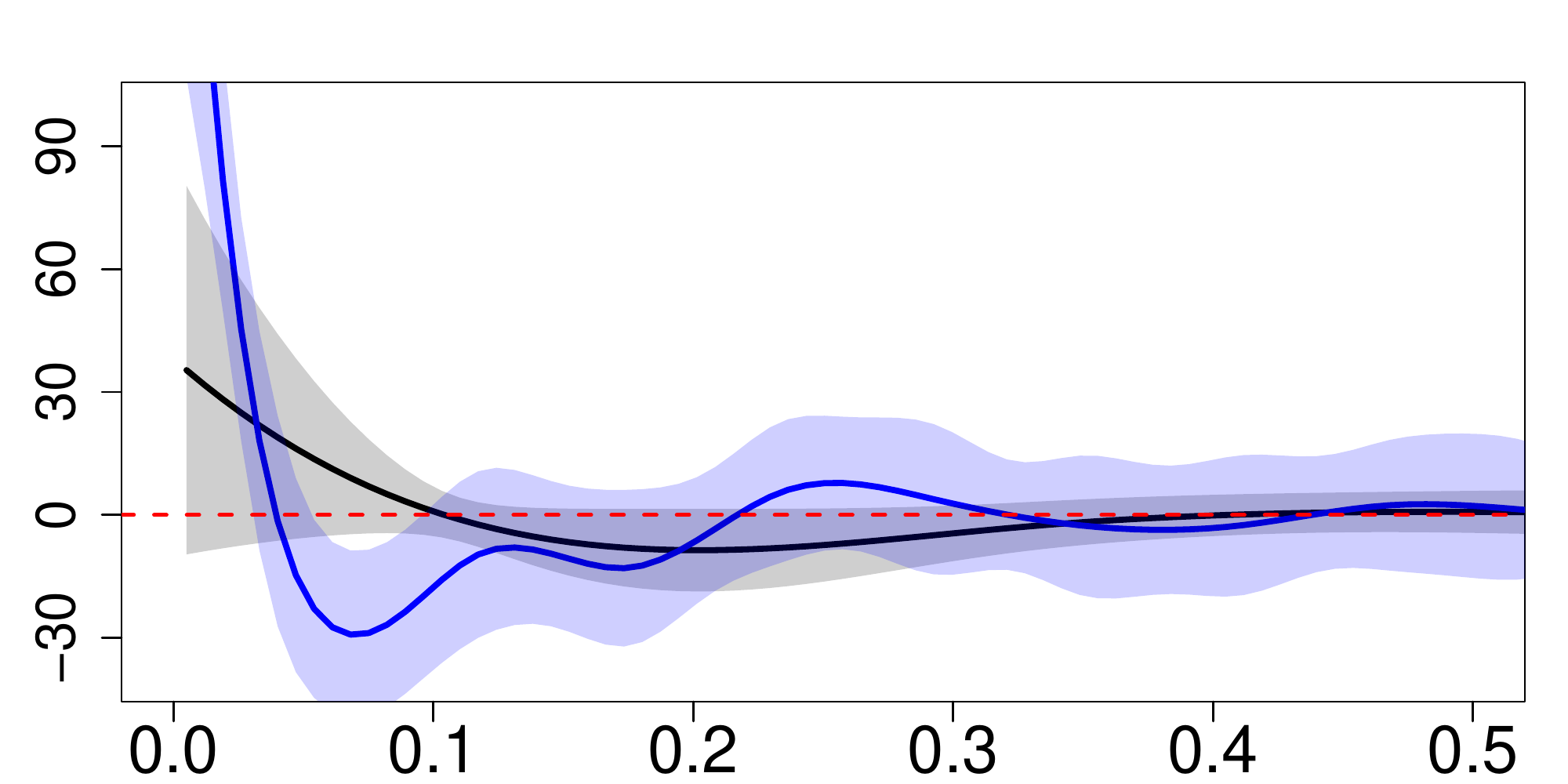}};
	\node[below = -0.1 cm of obs] (Taxis) {$\Delta$ (in seconds)};
	\node[above left = -0.7 cm and -0.1 cm of obs, rotate=90] () {\small $-\ACER(\Delta;0)$};
	\node[above left = -0.7 cm and -0.1 cm of obs](Tlab) {(a)};

		\end{scope}

	\end{tikzpicture}
	
	\caption{Estimated $-\ACER(\Delta;0)$ using the proposed instrumental variable method (black) and the estimated effect from the observational analysis (blue) using data from \cite{boldingfranks2018data} on normal mice (Panel a) and TeLC-expressed mice (Panel b). The shaded area represents a $90\%$ confidence band for visualizing the uncertainty of the estimates from 5000 bootstrap samples.}
	\label{fig:obs}
\end{figure}
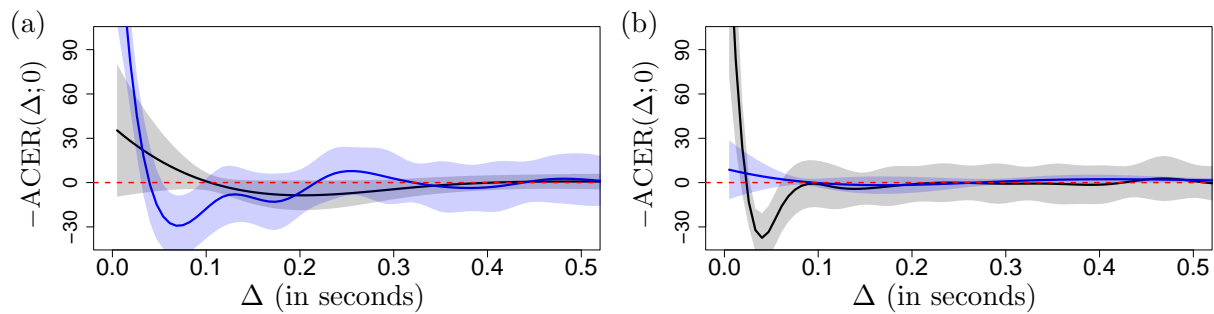

Taking the expectation on both sides of~\eqref{eqn:model-obs}, we have 
\begin{eqnarray}
\label{eqn::model-obs1}  E  \left\{ \frac{ \d Y_{i}(t)}{\d t} \right\}\ = \   \mu + \int_0^t g(t-s)E\{\mathrm{d} N_i(s)\}.
\end{eqnarray}
Integrating both sides of~\eqref{eqn::model-obs1} from $0$ to $t$ yields
\begin{eqnarray*}
E\{Y_i(t)\} &=&\int_0^t E  \left\{ \frac{ \d Y_{i}(l)}{\d l} \right\}  \mathrm{d} l \\
&=&\mu t + \int_0^t \int_0^l g(l-s)E\{ \d N_i(s)\} \mathrm{d} l \\
&=&\mu t + \int_0^t \int_0^l g(l-s)E\left \{ \frac{\d N_i(s)}{\d s }\right \}\d s \mathrm{d} l \\
&=&\mu t + \int_0^t \int_0^l g(\Delta)E\left \{ \frac{\d N_i(l-\Delta)}{\d (l-\Delta)}\right \}\d (t-\Delta) \mathrm{d} l  \qquad  (s=l-\Delta) \\
&=&\mu t + \int_0^t \int_l^0 g(\Delta)E\left \{ -\frac{\d N_i(l-\Delta)}{\d (l-\Delta)}\right \}\d \Delta \mathrm{d} l \\
&=&\mu t + \int_0^t \int_0^l g(\Delta)E\left \{ \frac{\d N_i(l-\Delta)}{\d l }\right \}\d \Delta \mathrm{d} l \\
&=&\mu t + \int_0^t \int_\Delta^t g(\Delta)E\left \{ \frac{\d N_i(l-\Delta)}{\d l }\right \}  \mathrm{d} l  \d \Delta   \qquad  \text{(Fubini's Theorem)} \\
&=&\mu t + \int_0^t g(\Delta) \left[ \int_\Delta^t E\left \{ \frac{\d N_i(l-\Delta)}{\d l }\right \}  \mathrm{d} l  \right] \d \Delta\\
&=&\mu t + \int_0^t g(s) E\{N_i(t-s)\} \d s\\
&=&\mu t + \int_0^t g(t-s) E\{N_i(s)\} \d s.
\end{eqnarray*}
Define $h'(t)=E\{Y_i(t)\}$ and $f'(t)=E\{N_i(t)\}$. We have
\begin{equation}\label{eqn:ee-obs}
h'(t)\ =\ \mu t + \int_0^t g(t-s) f'(s)\d s.
\end{equation}
We take a similar estimation procedure as in \S\ref{sec::estimation}. First, we estimate $f'(t)$ and $h'(t)$ using empirical cumulative intensities from all trials, denoted as $\hat{f}'$ and $\hat{h}'(t)$. Second, we approximate $g$ with truncated bases $\{\psi_j: j=1,\ldots, J\}$. Finally, we obtain the estimator from 
\begin{equation}\label{eqn::optim-obs}
\hat{\beta} = \underset{\beta \in \mathbb{R}^{J+1} }{\arg \min} \left\|\hat{h}'-\beta_{J+1} t - 
\sum_{j=1}^J (\psi_j*\hat{f}') \beta_j
\right\|_2^2 + \eta \|\beta\|_2^2.
\end{equation}
As with the proposed procedure, we set $\eta=0.01$ and use cubic B-splines with evenly-spaced knots. The number of knots is chosen using 5-fold cross-validation. We construct $90\%$ confidence bands using the bootstrap to approximate the uncertainty as in the main text. Results are shown in Figure~\ref{fig:obs}. We can see that not using the instrumental variable method yields estimates inconsistent with the findings in \cite{boldingfranks2018}. In particular, in Figure~\ref{fig:obs}(a), the observational analysis fails to capture the long-term inhibition between 0.1 to 0.3 seconds that contributes to the stable response in PN,  while displaying a false excitatory effect between 0.2 to 0.3 seconds. In Figure~\ref{fig:obs}(b), the observational analysis estimates a close-to-zero effect from OB to PN, contradicting the findings in  \cite{boldingfranks2018} on TeLC-expressed mice.

\end{document}